\documentclass[12pt]{article}
\usepackage{amsmath,amssymb,epsfig,tikz}
\usepackage[mathscr]{euscript}
\usepackage[affil-it]{authblk}
\usepackage{graphicx} 
\usepackage{color} 
\usepackage{enumitem}
\usepackage{amsthm}
\allowdisplaybreaks

\renewcommand\topfraction{.95}   
\renewcommand\bottomfraction{.6} 
\renewcommand{\theequation}{\thesection.\arabic{equation}}
\newlength{\extraspace}
\newlength{\extraspaces}
\newcommand{\be}{\begin{equation}
\addtolength{\abovedisplayskip}{\extraspaces}
\addtolength{\belowdisplayskip}{\extraspaces}
\addtolength{\abovedisplayshortskip}{\extraspace}
\addtolength{\belowdisplayshortskip}{\extraspace}}
\newcommand{\ee}{\end{equation}}
 
\newcommand{\ba}{\begin{eqnarray}
\addtolength{\abovedisplayskip}{\extraspaces}
\addtolength{\belowdisplayskip}{\extraspaces}
\addtolength{\abovedisplayshortskip}{\extraspace}
\addtolength{\belowdisplayshortskip}{\extraspace}}
\newcommand{\ea}{\end{eqnarray}}

\newcommand{\bas}{\begin{eqnarray*}
\addtolength{\abovedisplayskip}{\extraspaces}
\addtolength{\belowdisplayskip}{\extraspaces}
\addtolength{\abovedisplayshortskip}{\extraspace}
\addtolength{\belowdisplayshortskip}{\extraspace}}
\newcommand{\eas}{\end{eqnarray*}}

\newcommand{\btheo}{\begin{theorem}
		\addtolength{\abovedisplayskip}{\extraspaces}
		\addtolength{\belowdisplayskip}{\extraspaces}
		\addtolength{\abovedisplayshortskip}{\extraspace}
		\addtolength{\belowdisplayshortskip}{\extraspace}}
	\newcommand{\etheo}{\end{theorem}}

\newcommand{\blemma}{\begin{lemma}
		\addtolength{\abovedisplayskip}{\extraspaces}
		\addtolength{\belowdisplayskip}{\extraspaces}
		\addtolength{\abovedisplayshortskip}{\extraspace}
		\addtolength{\belowdisplayshortskip}{\extraspace}}
	\newcommand{\elemma}{\end{lemma}}

\newcommand{\bdefn}{\begin{definition}
		\addtolength{\abovedisplayskip}{\extraspaces}
		\addtolength{\belowdisplayskip}{\extraspaces}
		\addtolength{\abovedisplayshortskip}{\extraspace}
		\addtolength{\belowdisplayshortskip}{\extraspace}}
	\newcommand{\edefn}{\end{definition}}

\newcommand{\bproof}{\begin{proof}
		\addtolength{\abovedisplayskip}{\extraspaces}
		\addtolength{\belowdisplayskip}{\extraspaces}
		\addtolength{\abovedisplayshortskip}{\extraspace}
		\addtolength{\belowdisplayshortskip}{\extraspace}}
	\newcommand{\eproof}{\end{proof}}
 
\newcounter{subequation}[equation]
\makeatletter

\expandafter\let\expandafter
\reset@font\csname reset@font\endcsname

\def\subeqnarray{\arraycolsep1pt
    \def\@eqnnum\stepcounter##1{\stepcounter{subequation}%
        {\reset@font\rm(\theequation\alph{subequation})}}
\jot5mm     \eqnarray}

\def\subarray{\arraycolsep1pt
    \def\@eqnnum\stepcounter##1{\stepcounter{subequation}%
        {\reset@font\rm(\alph{subequation})}}
\jot5mm     \eqnarray}

\makeatother

\newcommand{\newsection}[1]{
\vspace{15mm}
\pagebreak[3]
\addtocounter{section}{1}
\setcounter{equation}{0}
\setcounter{subsection}{0}
\addcontentsline{toc}{section}
{\protect\numberline{\arabic{section}}{#1}}
 
\begin{flushleft}
{\large\bf \thesection. #1}
\end{flushleft}
\nopagebreak
\medskip
\nopagebreak}
 
\newcommand{\newsubsection}[1]{
\vspace{1cm}
\pagebreak[3]
\addtocounter{subsection}{1}
 
\addcontentsline{toc}{subsection}
{\protect\numberline{\thesection.\arabic{subsection}}{#1}}

\noindent{ \bf \thesection.\arabic{subsection} #1}
\nopagebreak
\vspace{2mm}
\nopagebreak}

\newcommand{\newappendix}[1]{
\vspace{15mm}
\pagebreak[3]
\addtocounter{section}{1}
\setcounter{equation}{0}
\setcounter{subsection}{0}

\addcontentsline{toc}{section}
{\protect\numberline{\thesection}{#1}}

\renewcommand{\theequation}{\Alph{section}.\arabic{equation}}
\begin{flushleft}
{\large\bf \Alph{section}: #1}
\end{flushleft}
\nopagebreak
\medskip
\nopagebreak}

\newcommand{\N}{\mathbb{N}}

\newcommand{\R}{\mathbb{R}}

\newcommand{\1}{\mbox{1\hspace{-.8ex}1}}

\newcommand{\ra}{\rightarrow}

\newcommand{\rra}{\ \longrightarrow \ }

\newcommand{\is}{ &\! =\! & }
\newcommand{\nonum}{\nonumber \\[1.5mm]}
\newcommand{\sspace}{\makebox[1cm]{ }}

\newcommand{\nspace}{\!\!\!\!\!\!\!\!\!\!}

\newcommand{\Tr}{{\rm Tr}}

\newcommand{\inv}{^{-1}}

\renewcommand{\th}{{\theta}}
\newcommand{\eps}{\epsilon}
\newcommand{\lb}{\lambda}
\newcommand{\om}{\omega}

\newcommand{\Gm}{\Gamma}

\newcommand{\vp}{\varphi}
\newcommand{\vph}{\varphi}

\newcommand{\dd}{{\partial}}

\newcommand{\cC}{{\cal C}}
\newcommand{\cD}{{\cal D}}
\newcommand{\cE}{{\cal E}}

\newcommand{\cL}{{\cal L}}

\newcommand{\cR}{{\cal R}}
\newcommand{\cS}{{\cal S}}

\newcommand{\cV}{{\cal V}}

\newcommand{\pf}{\mathfrak{p}}
\newcommand{\rf}{\mathfrak{r}}

\newcommand{\wg}{{\rm g}}

\newcommand{\FL}{{Friedmann-Lema\^{i}tre}$\,$}  
\newcommand{\nnU}[1]{\,\,^{#1}U}
\newcommand{\nnu}[1]{\,\,^{#1}u}
\newcommand{\nnZ}[1]{\,\,^{#1}Z}
\newcommand{\nnz}[1]{\,\,^{#1}z}
\newcommand{\nbU}[1]{\,\,^{#1}\bar{U}}
\newcommand{\nbu}[1]{\,\,^{#1}\bar{u}}
\newcommand{\nnv}[1]{\,\,^{#1}v}

\newcommand{\mcG}[0]{\mathcal{G}}

\newcommand{\mcR}[0]{\mathcal{R}}

\newcommand{\mcU}[0]{\mathcal{U}}
\newcommand{\scriptr}{\mathfrak{r}}
\newcommand{\mfp}{\mathfrak{p}}

\newcommand{\exx}[1]{e^{#1}}
\newcommand{\sigfac}[0]{\frac{1}{\sqrt{\eps_g}\hbar}}
\newcommand{\ddf}[2]{\frac{d^{#2} #1}{(2\pi)^{#2}}}

\newcommand{\reg}[1]{r^{(#1)}(\varrho^2)}

\begin{document}

\begin{titlepage}

\renewcommand{\thefootnote}{\fnsymbol{footnote}}
\makebox[1cm]{}
\vspace{1cm}

\begin{center}
\mbox{{\Large \bf The spatial Functional Renormalization Group}}\\[2mm]
\mbox{{\Large \bf and Hadamard states}}\\[2mm] 
\mbox{{\Large \bf on cosmological spacetimes}} 
\vspace{2.8cm}

{\sc R.~Banerjee}\footnote{email: {\tt rub18@pitt.edu}} 
{\sc and} 
{\sc M.~Niedermaier}\footnote{email: {\tt mnie@pitt.edu}}
\\[8mm]
{\small\sl Department of Physics and Astronomy}\\
{\small\sl University of Pittsburgh, 100 Allen Hall}\\
{\small\sl Pittsburgh, PA 15260, USA}
\vspace{18mm}

{\bf Abstract} \\[1mm]
\begin{quote}
A spatial variant of the Functional Renormalization Group (FRG) is
introduced on (Lorentzian signature) globally hyperbolic spacetimes. 
Through its perturbative expansion it is argued that such a FRG 
must inevitably be state dependent and that it should be based 
on a Hadamard state. A concrete implementation is presented for 
scalar quantum fields on flat Friedmann-Lema\^{i}tre spacetimes. 
The universal ultraviolet behavior of Hadamard states allows the flow to 
be matched to the one-loop renormalized flow (where strict 
removal of the ultraviolet cutoff requires a tower of potentials, 
one for each power of the Ricci scalar). The state-dependent infrared 
behavior of the flow is investigated for States of Low Energy, which are 
Hadamard states deemed to be viable vacua for a pre-inflationary 
period. A simple time-dependent infrared fixed point equation 
(resembling that in Minkowski space) arises for any scale factor, 
with analytically computable corrections coding the non-perturbative 
ramifications of the Hadamard property in the infrared. 
\end{quote} 
\end{center}

\vfill
\setcounter{footnote}{0}
\end{titlepage}


\setlength\paperheight {305mm}   %
\renewcommand\topfraction{.99}   

\thispagestyle{empty}
\makebox[1cm]{}

\vspace{-23mm}
\begin{samepage}

\tableofcontents
\end{samepage}

\setlength\paperwidth  {210mm}   %
\renewcommand\bottomfraction{.6} 
\nopagebreak

\renewcommand\topfraction{.95}   
\setlength\paperheight {297mm}   

\newpage

\newpage
\newsection{Introduction} 

The Functional Renormalization Group ({\bf FRG}) is a work-horse 
for non-perturbative Quantum Field Theory ({\bf QFT}) and has found 
applications in areas as diverse as solid-state physics, QCD, and 
quantum gravity; see \cite{FRGreview,Percbook,RSbook,Kopbook} for 
recent accounts. 
It is primarily used in Euclidean signature and in combination with 
heat kernel methodology. For certain applications, however, this 
standard framework does not convincingly capture the physics situation 
one seeks to model, 
the one in focus here being QFT on cosmological backgrounds.  
Cosmological spacetimes in general do not admit a satisfactory 
notion of Wick rotation and the formal application of the 
(pseudo-)heat kernel expansion to non-elliptic operators is 
mathematically dubious. More importantly, many cosmological 
signatures refer to the infrared dynamics (super-Hubble 
wavelengths) of the quantum fields and these are not 
convincingly captured by a resummation of the structurally 
unique heat kernel expansion. In fact, vacuum-like states in 
QFTs on curved backgrounds are inherently non-unique and this 
non-uniqueness should manifest itself on the level of the FRG.   
For the widely studied de Sitter background (where the Bunch-Davies 
vacuum provides a unique group invariant vacuum) the issue does 
not arise \cite{Kaya,Serreau1,Serreau2}, and of course neither 
does it in Minkowski space \cite{FRGWick1,FRGWick2}, but for generic 
\FL backgrounds it will. 
   
The main goal of the present note is to develop a 
spatial variant of the FRG on globally hyperbolic spacetimes  
where the issue of state dependence can successfully 
be addressed. A schematic summary is presented in Figure 1.
By a spatial FRG we mean one where the mode modulation affects only
the eigenvalues of the spatial Laplacian, while the temporal 
dynamics is left unaffected. In this way Lorentzian signature 
can be maintained, while implementing a mode suppression 
where nearly homogeneous configurations rather than near-null 
configurations correspond to small spectral values. In addition, 
in a perturbative solution of the FRG at each order  
the Green's function of a well-defined hyperbolic wave equation  
enters, a feature that would be spoiled by a temporal mode 
modulation. The Cauchy data then fix a Green's function 
but the admissible data themselves are insufficiently constrained
by physics requirements so that the ambiguity is best attributed to 
the Green's function.

It is through the non-uniqueness of such Green's functions that 
one can make contact to the selection criteria independently 
developed in the framework of perturbative QFT in curved backgrounds.     
Recall that for perturbatively defined QFTs on globally 
hyperbolic spacetimes the free state on which perturbation theory 
is based must be a Hadamard state. This is because by-and-large the
Hadamard property is necessary and sufficient for the existence of 
Wick powers of arbitrary order and hence for the perturbative series 
to be termwise well-defined, see e.g.~\cite{Moretti,Hackbook}. The 
Hadamard property entails that the 
associated Green's function has a universal ultraviolet behavior, 
the same for all Hadamard states. In contrast, the infrared behavior 
of the Green's function will be specific for the Hadamard state under 
consideration and whatever construction principle has been used to 
obtain it. In the perturbative expansion of the FRG a similar 
Green's function will occur, just with its spatial mode content 
modulated. By the very nature of a FRG modulator it should however 
leave an ultraviolet asymptotic expansion of the Green's function 
unaffected. 
This leads to the one-to-one correspondence depicted in Figure 1,
as far the FRG's perturbative expansion is concerned.

\begin{figure}[htb]
\hspace{20mm}
\epsfxsize=11cm
\epsfysize=10cm
\epsfbox{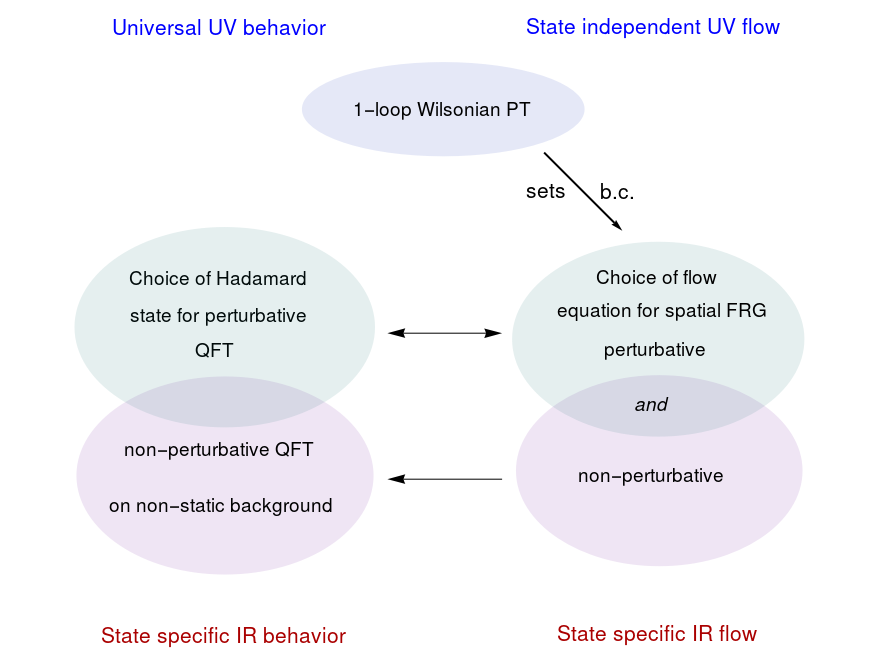}
\caption{\small Correspondence between state choice in perturbative QFT 
and choice of non-perturbative spatial FRG.}
\end{figure}
\medskip

Next, we consider the Effective Potential Approximation ({\bf EPA}) 
of the spatial FRG. Apart from the now scale dependent potential the same 
Green's function will occur and the correspondence carries over to 
the non-perturbative aspects covered by the EPA. Beyond the EPA it 
will of course become increasingly difficult to construct a Green's 
function for the (modified) Hessian of the effective action, but 
computationally this inversion is also difficult in the covariant 
Euclidean case. Further, at present no systematic non-perturbative    
formulation of QFTs on non-static backgrounds and the associated 
vacuum-like states exists. Our proposal is to use the 
right hand side of Figure 1 to explore the left hand side, 
i.e.~to use the spatial FRG to define and explore 
Lorentzian signature QFTs on non-static backgrounds beyond
perturbation theory. 

Importantly, the infrared aspects of the spatial FRG flow will 
always be state dependent. This is manifest in the perturbative 
expansion and on the level of the EPA. We expect it to be true     
generally whenever the full spatial FRG can be rendered well-defined.  
A systematic proposal to do so based on a spatial hopping expansion 
\cite{Graphpaper,Mainzproc} will be presented elsewhere. The state 
dependence of the spatial EPA flow will be studied in Section 4.  

As with other FRGs, the spatial FRG requires the specification of a 
boundary condition in the ultraviolet. The boundary functional 
is normally identified with the renormalized action at 
arbitrarily high renormalization scale. Assuming that the 
theory is renormalizable the latter coincides with the bare 
action and thus is known. In the case at hand, the renormalization 
of a scalar field theory with a {\it quartic} self-interaction in 
curved background is well understood 
\cite{ParkerTomsbook,Phi4onFRWPT1,Phi4onFRWPT2,Phi4onFRWPT3}. At the 
one-loop level the familiar ``tracelog'' of the action's Hessian needs 
to be evaluated, which can efficiently be done via heat kernel 
(for Euclidean signature) or pseudo-heat kernel (for Lorentzian 
signature) techniques \cite{ParkerTomsbook}. 
We shall find that even at one-loop order such a Wilsonian 
renormalization of a scalar field theory (with strict removal of 
the UV cutoff) demands the inclusion of an infinite tower of Ricci 
scalar terms $R(g)^n$, each coming with its own potential
\begin{equation} 
\label{i0} 
U(\phi) \mapsto {\cal U}(\phi,g) = \sum_{n \geq 0} \nnU{n}(\phi) R(g)^n\,.
\end{equation} 
Here $\nnU{0}(\phi) = U(\phi)$ is the original potential 
and the usual non-minimal coupling would invoke only  
a quadratic $\nnU{1}(\phi) = \xi \phi^2$. As soon as $U(\phi)$ 
contains a sextic term,  an infinite tower of additional 
interactions is  required to absorb the additional divergences 
generated. This holds for any non-trivial background metric,
though we shall focus on \FL backgrounds. The scalar field 
theory with the generalized potential (\ref{i0}) will be shown 
to be one-loop renormalizable in Section 3 and the beta functions for 
the relevant and marginal couplings will be computed. This will be 
done by re-integrating the one-loop spatial FRG flow itself,  
which ensures that it can consistently be used to set boundary 
conditions for the spatial FRG in the ultraviolet, see Figure 1.

With this framework in place, we elaborate the EPA for 
the spatial FRG in \FL backgrounds in Section 4. Here we 
focus on the so-called States of Low Energy ({\bf SLE}). These are 
Hadamard states that arise by minimizing a suitable time averaged 
energy functional \cite{Olbermann}. The function $f$ used for the 
time averaging is chosen to have support in the cosmological 
epoch in focus, e.g.~a pre-inflationary period of non-accelerated 
expansion. Importantly, the SLE admit an infrared expansion 
\cite{BonusSLE} where the dependence on the potential 
is rendered {\it explicit}. This allows one to extract analytically 
the dependence of the EPA's right hand side on the dimensionless 
version $\cV_k(\vp_0, t)$ of the generalized potential (\ref{i0}). 
Here $\vp_0$ is the constant dimensionless background and, 
as the Ricci scalar may appear in a temporally nonlocal form in 
the infrared, we merely indicate the time dependence. In addition, 
the flow equation carries an explicit time dependence. We analyze 
both by projecting onto a complete, dimensionless, orthonormal system 
of polynomials $\mfp_l[R](t)$ built from the Ricci scalar $R$. 
In particular, one has the expansion  
$\cV_k(\vp_0,t) = \sum_{l \geq 0} {}_l\cV''_k(\vp_0) 
\pf_l[R](t)$. 
The scale dependent `infrared' potentials ${}_l\cV_k(\vp_0)$, $l \geq 0$,    
then obey the following small $k$ flow equation (with suitable 
normalizations and in $1\!+\!d$ dimensions)
\ba 
\label{i1} 
& \nspace & k \frac{d}{dk} \,{}_l\!\cV_k(\vp_0) + (1\!+\!d) {}_l\!\cV_k(\vp_0) 
- \frac{d\!-\!1}{2} \vp_0 \,{}_l\!\cV'_k(\vp_0) = \int_0^{\infty}\! d\wp \wp^{d-1} \frac{{}_l\rho(\wp)}%
{\sqrt{{}_0\cV''_k(\vp_0) + {}_0 \rho(\wp)}} + O(k^2) \,, 
\nonum
&\nspace & \quad {}_l\rho(\wp) := \int\! dt N(t) a(t)^d \pf_l(t) 
\big[ r(\wp^2/a^2) - (\wp^2/a^2) r'(\wp^2/a^2) \big]\,, \quad l \geq 0\,.
\ea 
Here $r:\R_+ \ra \R_+$, $r(0) =1$,  is a mode modulator of rapid decay, 
$a(t)$ is the cosmological scale factor, $N(t)$ is the lapse function, 
the integration is over a dimensionless  
spatial momentum modulus $\wp = |p|/k$, and $'$ on $\cV_k$ denotes 
a derivative with respect to $\vp_0$. The $O(k^2)$ and higher order 
corrections are analytically computable. For $k \ra 0$ the flow 
(\ref{i1}) has a well-defined fixed point equation where all 
infrared fixed point potentials ${}_l \cV_*(\vp_0)$, $l \geq 1$, 
are determined by ${}_0\cV_*(\vp_0)$ (which is proportional to 
the temporal average of $\cV_*(\vp_0, t)$). Moreover the
$l=0$ fixed point equation resembles that in Minkowski space.   

By construction, the independently computed one-loop flow is 
consistent with the Hadamard property of the SLE and can be used 
to set boundary conditions for the dimensionless `ultraviolet' potentials 
${}^nV_k(\vp_0)$ for large $k$. Solving the flow numerically into 
the infrared fixed point regime governed by (\ref{i1}) will allow 
one to explore the non-perturbative ramifications of the Hadamard 
property for a generic self-interacting scalar field 
in the very early universe.  

The paper is organized as follows. In Section 2 we introduce the 
spatial FRG on a generic globally hyperbolic background. Via its 
perturbative expansion we establish the one-to-one correspondence 
between a choice of Hadamard state and a choice of spatial FRG,
see Fig 1. In Section 3 we specialize to spatially flat 
\FL backgrounds with generic scale factor $a$. The  SLE are 
introduced as the Hadamard states of choice and analytically 
controllable expansions for the ultraviolet and infrared regimes 
are prepared. The ultraviolet asymptotics is used in Section 3 
to extract the divergent part of the spatially regularized 
one-loop effective action. For a beyond quartic potential the successful 
absorption of these terms mandates the generalized potential 
(\ref{i0}). In Section 4 the spatial EPA flow equation associated 
with the SLE is introduced, leading to (\ref{i1}) with explicitly 
computable subleading terms. Appendix A contains computational 
details for Section 3. Appendix B presents the ``instantaneous 
limit'' of the SLE based flow equation, which resembles the 
spatial EPA flow in Minkowski space. In Appendix C the $O(k^2)$ 
corrections in (\ref{i1}) are computed.

\newpage 
 
\newsection{Spatial FRG on globally hyperbolic backgrounds} 

Here we derive the spatial FRG on generic globally 
hyperbolic manifolds and discuss some of its basic 
properties. It admits a perturbative expansion in powers 
of $\hbar$ in which the correspondence to the choice 
of a modified Hadamard state is transparent.

\newsubsection{Basics} 

For comparison's sake we initially consider foliated background 
manifolds with both signatures and line element $ds^2 = 
\eps_g N^2 dt^2 + \wg_{ij} (N^idt + dx^i) (N^j dt + dx^j)$. 
Here $N, N^i, \wg_{ij}$, $i,j =1,\ldots,d$ are the lapse, shift,
spatial metric as usual and $\eps_g = \pm 1$ is the signature 
parameter, and we take $d \geq 3$ throughout. On such a foliated 
background we consider the action 
of a self-interacting scalar field non-minimally coupled to it
\ba
\label{action1} 
S[\tilde{\chi};g] \is \eps_g\!\int d^Dy \sqrt{\eps_g g} \Big\{ 
\frac{1}{2} g^{\mu\nu} \dd_{\mu} \tilde{\chi} \dd_{\nu} \tilde{\chi}
+ {\cal U}(\tilde{\chi}, R) \Big\} 
\nonum
\is \int_{t_i}^{t_f} \! dt \int_{\Sigma}\! dx \Big\{ 
\frac{1}{2 n} e_0(\tilde{\chi})^2 + \frac{\eps_g}{2} n \wg \wg^{ij} 
\dd_i \tilde{\chi} \dd_j \tilde{\chi} + \eps_g n \wg 
U(\tilde{\chi}, R) \Big\}\,.
\ea 
Here, the metric $g_{\mu\nu}$ is treated as a background field 
and $R(g)$ is its Ricci scalar. In the second line 
$e_0 = \dd_t - \cL_{\vec{N}}$ is the derivation transversal to the 
leaves of the foliation, $n = N/\sqrt{\wg}$ is the lapse anti-density. 
Further, $R(g)$ is to be read as its $1+d$ decomposition  
$R(g) = R(\wg) + 2 \eps_g N^{-1} e_0(K) - \eps_g (K^2 + K_i^k K_j^i) 
- 2 N^{-1} \nabla^2 N$, 
where $K_{ij} = - e_0(\wg_{ij})/(2N)$ is the extrinsic curvature. 
Finally, ${\cal U}(\tilde{\chi}, R)$ is a generalized potential of 
the form 
\begin{equation} 
\label{pot}  
{\cal U}(\tilde{\chi}, R) = \sum_{n \geq 0} \nnU{n}(\tilde{\chi}) R(g)^n\,,
\quad \nnU{n}(\tilde{\chi}) = 
\sum_{j \geq 0} \nnu{n}_{2j} \,\frac{\tilde{\chi}^{2j}}{(2j)!}\,,
\end{equation}
with real coupling constants $\nnu{n}_{2j},\, n,j \geq 0$. The 
potentials $\nnU{n}(\tilde{\chi})$ have mass dimension $d\!+\!1 \!-\!2 n$, 
the field has $(d\!-\!1)/2$, giving $\nnu{n}_{2j}$, a mass dimension 
$d\!+\!1 - 2 n -(d\!-\!1)j$. Despite the proliferation of couplings 
only a small number have non-negative mass dimension, 
$6$ for $d=3$ and $\lfloor (d\!+\!1)/2 \rfloor +3$ for $d \geq 4$.

The derivation of the FRG with a covariant modulator is by now 
standard \cite{FRGreview}. In order to set conventions, trace 
the occurrence of the signature parameter, and highlight 
the step where a purely spatial modulator enters, we quickly run 
though the basic steps. In the schematic functional integrals
both metric signatures are treated in parallel with conventions 
\be
\label{signature} 
\text{Euclidean:}\quad \epsilon_g=1\,,\;\;\sqrt{\epsilon_g}=1\,,
\quad \quad
\text{Lorentzian:}\quad \epsilon_g=-1\,,\;\;\sqrt{\epsilon_g}=i\,.
\ee 
We stress the difference between a geometrically meaningful ``Wick flip'' 
$(\epsilon_g\mapsto -\epsilon_g)$ and a problematic ``Wick rotation'' 
($t\mapsto -it$) that complexifies the line element. 
A meaningful notion of a Wick rotation on a generic globally hyperbolic 
manifold would have to meet a number of desiderata, see e.g.~\cite{Wicked}. 
Any known notion fails to meet one or more of these criteria. In contrast, 
the Wick flip is a weaker notion that simply maps a given Lorentzian 
manifold onto a Riemannian one, with no claims to analyticity or bijectivity.

Adopting the background field formalism we write $\tilde{\chi} = \vp + \chi$,
where $\vp $ is an arbitrary background field, and add a modulator term  
\begin{eqnarray}
\label{sfrg1}
S_k[\chi,\vp;g]=S[\vp+\chi;g]+\Delta S_k[\chi;g]\,,
\quad \Delta S_k[\chi;g]=\frac{\epsilon_g}{2}\chi\cdot 
\mathcal{R}_k(g) \cdot\chi\,.
\end{eqnarray}
The ``$\cdot$'' shorthand represents an integration with 
respect to the standard (pseudo) Riemannian volume density.  
The regulator kernel $\mcR_k$ is normally taken to suppress the 
eigenvalues of the full Laplacian $\nabla^2$ in Euclidean signature 
(possibly shifted by a mass-like term) below a scale $k^2$.  
In the Riemannian setting, the ellipticity of $\nabla^2$ in principle 
allows for a well-defined spectral resolution with a clear discrimination 
between modes associated with large and small eigenvalues. Hence, 
modulating the mode content according to the eigenvalues of $\nabla^2$ 
leads to a well defined covariant modulation procedure which for 
$k \ra \infty$ suppresses almost all of the modes. On the other hand, 
for Lorentzian signature $\nabla^2$ is hyperbolic and the modes 
associated with vanishing spectral values are associated with field 
propagation along lightcones. A mode modulation $\cR_k$ of the same 
form would leave such null degrees of freedom (equivalent to a 
$d$-dimensional field theory) unaffected. The associated functional 
integral would still be intractable, rendering a naive transcription of 
the Euclidean procedure untenable. While in Minkowski space the simple 
form of the wave front sets allows for practically usable Lorentzian 
signature modulators ${\cal R}_k$ \cite{FRGWick1,FRGWick2}, such an 
adaptation seems both mathematically problematic and physically obscure 
for generic pseudo-Riemannian manifolds. A local Wick rotation might 
suffice for the ultraviolet aspects captured by a (pseudo-)heat 
kernel expansion \cite{Moretti1}, but for the infrared aspects 
needed to recover the unmodulated QFT results one encounters an impasse.

The route pursued here is to fix a foliation and to  replace the 
covariant modulation with a merely spatial one through 
$\mcR_k (t,x;t',x')=N(t,x)^{-1/2}N(t,x')^{-1/2}  \delta(t-t')R_k(t,x,x')$, 
while leaving the temporal dynamics unaffected. Moreover, the equal 
time kernel $R_k$ is taken to suppress modes associated with 
eigenvalues of the {\it spatial} part of the Laplacian $\nabla_{\!s}^2$
(see \ref{Lapl1}) at time $t$ below a scale $k^2$. The modes unaffected 
in the limit $k \ra \infty$ are then equivalent to those of a 
$1$-dimensional field theory, i.e.~quantum mechanics. Since 
quantum mechanics typically does not require renormalization 
in itself this should lead to a legitimate mode modulation that 
mirrors correctly the scale dependence of the full functional 
integral and its divergences. This extends the use of a spatial regulator employed in \cite{Kaya,Serreau1,Serreau2} for the de Sitter background to generic foliated manifolds.

Having introduced the mode modulation, the background field generating 
functional $W_k[J;\vp]$ and effective average action $\Gamma_k[\phi;\vp]$ 
are defined as usual. For notational simplicity we momentarily suppress 
the parametric dependence on the metric. One then has  
\begin{eqnarray}
\label{sfrg2}
\exx{\sigfac W_k[J;\vp]}\is \int\! \mathcal{D} 
\chi\,\exx{-\sigfac (S[\vp+\chi]+\Delta S_k[\chi])+\sigfac J\cdot\chi  }\,,
\nonum
\Gamma_k[\phi;\vp] &:=&  J_k[\phi;\vp]\cdot \phi -W_k\big|_{J=J_k[\phi;\vp]}
-\Delta S_k[\phi]\,,
\quad \frac{\delta W_k}{\delta J} \Big|_{J=J_k[\phi;\vp]} = \phi\,.
\end{eqnarray}
In general the regulator $\mathcal{R}_k$ may depend on the background 
field $\vp$, in which case the action $S[\vp + \chi;g] + 
\Delta S_k[\chi,\vp;g]$ depends the fluctuation field and the background field 
separately, i.e.~not just through their sum. On the level of the effective 
action this means that the splitting symmetry 
$\Gamma_k[\phi+\zeta;\vp-\zeta]=\Gamma_k[\phi;\vp]$ is violated. 
For the present purposes it suffices to take $\mcR_k$ independent 
of the background field $\vp$, in which case splitting symmetry  
is maintained, i.e. $\Gamma_k[\phi;\vp]=\Gamma_k[\phi+\vp]$ 
(by slight abuse of notation).

Throughout we use weighted functional derivatives adapted to 
the `$\cdot$' convolutions. A covariant source coupling 
$J \cdot \chi = \int\! dy\,\sqrt{g(y)} J(y)\chi(y)$ 
$= \int \!dt dx N \sqrt{\wg} J(t,x) \chi(t,x)$ suggests to define 
\begin{equation}
\label {Fderiv1}  
\lim_{\eps \ra 0} \frac{1}{\eps} 
( F[J + \eps H] - F[J]) = 
\int \! dt dx \,(N \sqrt{\wg})(t,x) \, H(t,x) 
\frac{\delta F[J]}{\delta J(t,x)} =: H \cdot \frac{\delta F}{\delta J}\,,
\end{equation}
where the measure contribution is taken out of the functional 
derivative. Whenever $d$ is inessential we write $dx$ for $d^dx$, 
the $d$-dimensional flat Euclidean measure.  Then $\delta F/\delta J(t,x)$ 
transforms as 
a temporal scalar and has length dimension $[\delta F/\delta J] 
= -(d\!+\!1) - [J] = -(d\!-\!1)/2$, for any dimensionless covariant 
functional $F[J]$. The covariant delta function is normalized 
by $\int \!dt dx (N \sqrt{\wg})(t,x) J(t,x) \delta(t,x,t',x') = J(t',x')$. 

In this setting we quickly run through the familiar steps of deriving 
the FRG. Differentiating  $W_k[J;\vp]$ with respect to $k$ yields the 
Polchinski equation
\begin{eqnarray}
\label{sfrg3}
k\dd_k W_k[J;\vp]=-\frac{\eps_g}{2}
\Tr \bigg\{k\dd_k \mathcal{R}_k\bigg[\frac{\delta W_k}{\delta J}
\frac{\delta W_k}{\delta J}+\sqrt{\epsilon_g}\hbar 
\frac{\delta^2 W_k }{\delta J \delta J}\bigg]\bigg\}\,,
\end{eqnarray}
where we note that the trace over spacetime indices is with respect 
to the usual volume density. Implementing the modified Legendre 
transformation in \eqref{sfrg2} one has 
\ba
\label{sfrg4}
k\dd_k \Gamma _k[\phi;\vp] \is  
\frac{\eps_g \sqrt{\epsilon_g}\hbar }{2}\Tr \bigg\{k\dd_k 
\mathcal{R}_k\frac{\delta^2 W_k }{\delta J \delta J}
\bigg|_{J=J_k[\phi;\vph]}\bigg\}\,,
\nonum
\frac{\delta^2 W_k }{\delta J \delta J}\Big|_{J=J_k[\vp;\phi]}
\is 
\bigg[\frac{\delta^{2}\Gamma_k}{\delta \phi \delta \phi }  
+\epsilon_g \mcR_k\bigg]\inv \,.
\ea
For our purposes a background independent $\mathcal{R}_k$ suffices 
and we can set the mean field $\phi$ to zero, writing 
$\Gamma_k[\vp]=\Gamma_k[0;\vp]$. Further, splitting symmetry 
entails $\frac{\delta^{2}\Gamma_k}{\delta \phi \delta \phi }=
\frac{\delta^{2}\Gamma_k}{\delta \vp \delta \vp }$, yielding 
the Wetterich equation in the schematic form 
\begin{eqnarray}
\label{sfrg6}
k\dd_k \Gamma _k[\vp]\is \frac{\sqrt{\epsilon_g}\hbar }{2}
\Tr\big\{k\dd_k \mathcal{R}_k\cdot G_k[\vp]\big\}\,,\quad 
\bigg[\frac{\delta^{2}\Gamma_k}{\delta \vp \delta \vp } 
+\epsilon_g \mcR_k\bigg]\cdot G_k[\vp]=\eps_g \1\,.
\end{eqnarray}
The last relation highlights the key difference between both 
signatures. For the sake of argument, consider an 
$\hbar$ expansion of $\Gamma_k$ starting with $S[\vp]$
and a corresponding $\hbar$ expansion of $G_k[\vp]$.  
This gives a recursion relation for the coefficients of 
$G_k[\vp]$ detailed in the next subsection. A each order
the solution of the recursion will invoke a Green's function 
of the modified Hessian $S^{(2)}(\vp) + \eps_g \mathcal{R}_k = 
\eps_g [ - \nabla^2 + {\cal U}''(\vp,R) + \mathcal{R}_k]$.      
For Euclidean signature this will typically be a positive
(elliptic) second order operator, which has a unique inverse
on general grounds. In contrast, for Lorentzian signature $S^{(2)}(\vp)$
is a wave operator (hyperbolic) and although a unique 
Green's function is determined by the Cauchy data \cite{wavebook} 
the data themselves are insufficiently constrained by 
physically reasonable requirements (like being induced by 
a vacuum-like two-point function). The resulting ambiguity can 
be attributed to the Green's function and is the mathematical origin 
of the state selection problem in perturbative QFT on curved 
backgrounds \cite{Moretti,Hackbook}. We stress that analogous 
choices enters the FRG's 
$\hbar$ expansion via the need to invert $S^{(2)} - \mathcal{R}_k$. 

As described above, the proposal here is to fix a foliation 
and to use a purely spatial modulator. For $\eps_g =-1$ a 
foliated geometry $ds^2 = -N^2 dt^2 + \wg_{ij} (N^idt + dx^i) 
(N^j dt + dx^j)$ will be assumed to be associated with 
a globally hyperbolic (connected, time oriented, Lorentzian) 
manifold $M$. A useful characterization (see e.g.~\cite{wavebook}, 
Thm. 1.3.10) of a globally hyperbolic manifold is: $M$ is 
isometric to $\R \times \Sigma$ with metric $-\bar{N}^2 dt^2 + 
\bar{\wg}_{ij}(\bar{t},\bar{x})d\bar{x}^i d\bar{x}^j$, where 
$\bar{N}$ is real and the second term is a Riemannian metric 
on $\Sigma$ that depends smoothly on $\bar{t}$. Further, each 
$\{\bar{t}\} \times \Sigma$ is a smooth Cauchy surface in $M$. 
In physicist's terminology this means in particular that the 
`shift zero' gauge $\bar{N}^i \equiv 0$ is always attainable
and that a preferred foliation of $M$ in terms of Cauchy surfaces 
exists. The residual covariance group is then the direct product 
of purely temporal reparameterizations and the group  
${\rm Diff}(\Sigma)$ of time independent spatial diffeomorphisms.   
Occasionally we shall restrict the foliation time to an interval 
$[t_i, t_f]$; the allowed temporal reparameterizations must the 
preserve the end points and we write ${\rm Diff}[t_i, t_f]$ 
for the resulting group. From now on we shall always work 
in the `shift zero' gauge and drop the overbars from the notation,
i.e.~take  $ds^2 = -N^2 dt^2 + \wg_{ij} dx^i dx^j$ as the metric. 
After a conformal rescaling to obtain $-dt^2 + N^{-2} \wg_{ij} dx^i dx^j$ 
the global hyperbolicity of the spacetime is closely related to the 
geodesic completeness of the spatial metric $N^{-2} \wg_{ij} dx^i dx^j$ 
\cite{GlobalHyp}.   

For clarity's sake we summarize the key relations of the 
spatial FRG in the $N^i =0$ foliation:   

\underline{Modulator term:} 
\ba
\label{sfrg7}
\Delta S_{k}[\chi] \is -\frac{1}{2}  \int \! dt 
dxd x'\sqrt{N\wg}(t,x) \sqrt{N\wg}(t,x') \chi(t,x) 
R_k(t,x,x') \chi(t,x')\,.
\ea
Here $R_k(t,x,x')$ is a bi-scalar under spatial diffeomorphisms and 
a scalar under temporal reparameterizations, so as to render 
(\ref{sfrg3}) invariant. As noted before, we assume it
to be independent of the background field and  
to depend on the spatial metric $\wg_{ij}(t,x)$ only through the 
spectral values of the spatial part $\nabla^2_{\!s}$ of the Laplacian. 
Explicitly,
\ba 
\label{Lapl1} 
\nabla^2 \is - \sqrt{\wg}^{-1} N^{-1} \dd_t \big( \sqrt{\wg} N^{-1} \dd_t \big) 
+ \nabla_{\!s}^2 \,,
\nonum
\nabla_{\!s}^2 &:=& N^{-1} \sqrt{\wg}^{-1} \dd_i 
\big( N \sqrt{\wg} \wg^{ij} \dd_j \big) \,. 
\ea
For closed spatial sections $\{t\} \times \Sigma$  the 
operator $- \nabla_{\!s}^2$ is symmetric and positive definite 
with respect to the inner product $(f_1, f_2)_s := 
\int\! dt dx N \sqrt{\wg} f_1^* f_2$. In order to render the temporal 
part symmetric the domain has to be suitably restricted, either
by imposing fall-off conditions for $t \in\R$ or by imposing 
Sturm-Liouville boundary conditions for $t \in [t_i, t_f]$. 

\underline{Flow equation:} 
\be
\label{sfrg8}
\dd_k\Gm_k[\vp] 
= \frac{i\hbar}{2}\int \! dt dx dx' \sqrt{N\wg}(t,x) \sqrt{N\wg}(t,x') 
\,\dd_kR_k(t,x,x') G_k[\vp](t,x'; t,x)\,,
\ee
where we assume the temporal coincidence limit of the Green's function 
$G_k$  to be well-defined and non-zero. Then $G_k(t,x;t,x')$ is a 
temporal scalar and a spatial bi-scalar which carries a functional 
dependence on $\vp(t,x)$.  For generic $\vp(t,x)$ it is not spatially 
translation invariant, so no overcounting of the volume occurs.  

\underline{Green's function relation:} 
\be
\label{sfrg9}
-\int \!dt' dx' N(t',x')\sqrt{\wg}(t',x') 
\frac{\delta^2 \big(\Gm_k[\vp]+ \Delta S_k[\vp]\big)}%
{\delta \vp(t,x)\delta \vp (t',x')}\,
G_k[\vp](t',x'; t'',x'') = \delta(t,x;t'',x'')\,,
\ee
where $ \delta(t,x;t'',x'')$ is the covariantly normalized 
$\delta$-function.


\newsubsection{Perturbative expansion and correspondence to 
Hadamard states}  

The spatial FRG is defined by the coupled system (\ref{sfrg8}), 
(\ref{sfrg9}). An analogous coupling would hold for any other 
Lorentzian signature FRG, but the spatially modulated version 
is more amenable to analysis. To start, a perturbative analysis 
turns out to be instructive. We consider Ans\"{a}tze for $\Gamma_k$ 
and $G_k$ of the form 
\be 
\label{sfrgPT1} 
\Gamma_k[\vp] = S[\vp] + \sum_{n \geq 1} \hbar^n \Gamma_{k,n}[\vp]\,,
\quad \quad 
G_k[\vp] = G_{k,0}[\vp] + \sum_{n \geq 1} \hbar^n G_{k,n}[\vp]\,,
\ee 
where $S[\vp]$ is the renormalized action. Inserted into 
(\ref{sfrg8}), (\ref{sfrg9}) this gives rise to a closed recursive 
system for the coefficients. We first present the condensed 
version arising from the $\eps_g =-1$ version of (\ref{sfrg6}):
\ba 
\label{sfrgPT2} 
&& [-S^{(2)}(\vp) + \mcR_k]\cdot G_{k,0}[\vp] = \1\,,
\nonum
&& [-S^{(2)}(\vp) + \mcR_k]\cdot G_{k,n}[\vp] = 
\sum_{l =1}^n \Gamma_{k,l}^{(2)}(\vp)\cdot G_{k,n-l}[\vp]\,,\quad n \geq 1\,,
\\[2mm] 
&& k \dd_k \Gamma_{k,n}[\vp] =  \frac{i}{2} {\rm Tr} 
\big\{ k \dd_k \mcR_k \cdot G_{k,n-1}[\vp] \big\} \,, \quad 
n \geq 1\,. 
\nonumber
\ea 
In principle, this determines iteratively the pairs $(G_{k,n}, 
\Gamma_{k, n+1})$, $n \geq 0$, viz
\be 
\label{sfrgPT3} 
G_{k,0} \;\;\longrightarrow\;\; \Gamma_{k,1} 
\;\;\longrightarrow\;\; G_{k,1} \;\;\longrightarrow\;\; 
\Gamma_{k,2} \;\;\longrightarrow\;\; \ldots \,.
\ee  
The only - though major - complication is to address the ambiguities 
that arise at each step and how they accumulate. The flow equation is 
integrated between a reference scale $\mu$ and an ultraviolet 
cutoff $\Lambda$ leading to
\be 
\label{sfrgPT4} 
\Gamma_{\mu,n}[\vp] = \Gamma_{\Lambda, n}[\vp] 
- \frac{i}{2} \int_{\mu}^{\Lambda} \! dk {\rm Tr} \big\{ 
\dd_k \mcR_k \cdot G_{k,n-1}[\vp] \big\}\,.  
\ee
We seek to interpret $\Gamma_{\mu,n}$ as the $n$-th order renormalized 
effective action at scale $\mu$.  Field and coupling renormalizations 
are introduced with the goal of {\itshape strictly removing} the UV 
cutoff $\Lambda$, although  this strict removal may or may not be 
possible depending on the self-interaction under consideration.
The successful removal of the UV 
cutoff by tractable coupling and field renormalizations 
amounts to establishing the perturbative renormalizability 
of the QFT. In Section 3 we shall detail this process 
at one-loop order for a generic (beyond quartic) scalar potential on 
\FL backgrounds, with the result that the Ricci couplings 
(\ref{i0}) need to be turned on to achieve strict renormalizability.    

Here we focus on the, in the context of the FRG, novel ambiguities 
arising from the inversion of $S^{(2)}(\vp) - \mcR_k$, which 
drives the other part of the recursion. We write $S^{(2)}(\vp) = 
-\cD \,\delta$ with 
\ba
\label{sfrgPT5}  
\cD &:=& - \nabla^2 + \sum_{n \geq 0} \nnU{n}^{(2)}(\vp) R(g)^n\,.
\nonum
&=:& \sqrt{\wg}^{-1} N^{-1} \dd_t (\sqrt{\wg} N^{-1} \dd_t) 
- \nabla_{\!s}^2 + {\cal U}'' \,.
\ea    
In the second line the decomposition (\ref{Lapl1}) was 
inserted. Despite the complicated form of the potential ${\cal U}''(\vp,R)$ 
it can in the context of (\ref{sfrgPT2}) be viewed as a given (smooth) 
function on $M$ transforming as a scalar. The operator 
$-\nabla_{\!s}^2 + {\cal U}''$ can be rendered selfadjoint subject to 
suitable positivity conditions on ${\cal U}''$ \cite{GlobalHyp}.

For the Green's function of such wave operators powerful general 
results are available, including the unique solubility of the 
Cauchy problem and 
the existence of asymptotic expansions of the Hadamard form 
\cite{wavebook}. Unfortunately, the defining relation for 
$G_{k,0}$ is not quite of this form. In terms of $\cD$ and 
the spatial kernel $R_k$ it reads 
\be 
\label{sfrgPT6}
\cD_{t,x} G_{k,0}[\vp](t,x,t',x') + \int\! dx'' (N \sqrt{\wg})(t,x'') 
R_k(t,x,x'') G_{k,0}[\vp](t,x'';t',x') = \delta(t,x;t',x') \,.
\ee   
Although considerably more complicated than the standard case, 
locality in $t$ is preserved in that a spatial integro-differential 
operator acts at fixed $t$ on $G_{k,0}[\vp](t,x;t',x')$. 
No ready-made theory seems to be available for this situation. 
To proceed, we note that for $k \ra 0$ the relation (\ref{sfrgPT6}) 
reduces to the standard one, $\cD_{t,x} G_{0,0}[\vp](t,x,t',x') = 
\delta(t,x;t',x')$, and hence is tractable. For the moment we assume 
that a solution has been chosen according to some criterion and 
for simplicity we take it to be a Feynman Green's function.  
The relation (\ref{sfrgPT6}) can then be converted into an integral 
equation 
\ba 
\label{sfrgPT7}
&& G_{k,0}[\vp](t,x;t',x') + \int\!\! dt'' dx'' (N \sqrt{\wg})(t'',x'') \,  
{\bf R}_k(t,x;t'',x'') G_{k,0}(t'',x'';t',x') 
\nonum
&& \sspace =  G_{0,0}[\vp](t,x;t',x') \,,
\nonum
&& {\bf R}_k(t,x;t'',x'') = \int\! dx_1 (N \sqrt{\wg})(t'', x_1) 
G_{0,0}[\vp](t,x;t'',x_1) R_k(t'', x_1, x'') \,.   
\ea 
In order to obtain a Feynman-type Green's function this needs to 
be augmented by suitable boundary conditions for the 
(first derivative of the) coincidence limit. 
Both the inhomogeneity and the kernel ${\bf R}_k$ depend on the 
choice of $G_{0,0}$. Schematically (\ref{sfrgPT7}) is of the form 
$[\1 + {\bf R}_k] \cdot G_{k,0} = G_{0,0}$ and for `small' ${\bf R}_k$ 
should have a unique solution. We postpone an analysis of the 
conditions under which this holds true and note the important 
conclusion: 

\begin{quote} 
{\it Whenever (\ref{sfrgPT7}) has for given $R_k$ and a choice of 
Feynman Green's function $G_{0,0}$ a unique solution it defines 
a one-to-one correspondence between Hadamard states and 
the perturbative expansion of the spatial FRG.} 
\end{quote}  
 
This rests on the fact that physically viable Feynman Green's function 
solutions of the basic wave equation $\cD_{t,x} G_{0,0}[\vp](t,x,t',x') 
= \delta(t,x;t',x')$ are in one-to-one correspondence to Hadamard 
states. For completeness' sake we briefly outline the (known) rationale
underlying the last assertion. A Feynman Green's function can
always be written as $2 G_{0,0} = G^c + i \om^s$, where $G^c$ is the 
unique real valued causal Green's function and $\om^s$ is the 
symmetrized Wightman function.  On any globally hyperbolic manifold 
the unique existence of $G^c$ is ensured even globally, see 
\cite{wavebook} Section 3.4. The symmetrized Wightman function 
is said to be locally of Hadamard form if it can be written as 
\begin{equation} 
\label{Had1} 
\om^s(t,x;t',x') = N_d [ H_{\eps}(t,x;t',x') + W(t,x;t',x')]\,, 
\quad N_d = \frac{\Gamma(\frac{d-1}{2})}{2 (2\pi)^{(d+1)/2}}\,,
\end{equation}
where $H_{\eps}$ is the ``Hadamard parametrix'' and $W$ is a smooth    
symmetric function. The Hadamard parametrix formalizes the notion 
of `Minkowski-like' short distance singularities, which 
occur for small $\sigma_{\eps}(t,x;t',x') = \sigma(t,x;t',x') 
+ i \eps( t - t') + O(\eps^2)$, where $\sigma(t,x;t',x')$ is the 
Synge function (one-half of the geodesic distance squared between 
points with coordinates $(t,x)$ and $(t',x')$). The normalization 
$N_d$ is chosen such that leading singularity of $H_{\eps}$ is
$U(t,x',t',x')\sigma_{\eps}^{- \frac{d-1}{2}}$, with $U(t,x;t,x) =1$.  
For $d$ even this is the only singularity, for $d$ odd there is 
an additional $V(t,x;t',x') \ln \mu^2 \sigma_{\eps}(t,x;t',x')$ 
term, for some mass parameter $\mu$. Here $U$ and $V$ are smooth 
symmetric functions uniquely associated with the differential 
operator ${\cal D}$. They can be expanded in terms of 
the ``Hadamard coefficients'', which are recursively computable 
and related to those of the formal pseudo-heat kernel \cite{heatkoffdiag1}. 
On the other hand, $W$ in (\ref{Had1}) is {\it not} uniquely     
associated with ${\cal D}_{t,x}$ and is co-determined by the 
quantum state underlying the Wightman function. A quantum state
for which (\ref{Had1}) holds is called a ``local Hadamard state''.   
There is also a global characterization of a ``Hadamard state''
in terms of the ``wave front set'' of the differential operator. 
Nevertheless, the inherent ambiguity signaled in 
(\ref{Had1}) through the underdetermination of $W$ remains. 
In the global characterization a (Feynman) Green's function 
is uniquely determined by the Cauchy data on a spacelike 
hypersurface \cite{wavebook}. Alas, on a general globally hyperbolic 
manifold there is no physically preferred way of choosing these data 
such that a preferred $W$ is associated with them. The converse 
implication however works: given a Hadamard state there 
are entailed Cauchy data as well as a unique $W$ associated 
with it. By a physically viable Feynman Green's function we mean one 
of the form $2 G_{0,0} = G^c + i \om^s$, where $\om^s$ is associated 
with a Hadamard state. As such a choice of Feynman Green's function 
is in one-to-one correspondence to the choice of a Hadamard state. 
Via the above italicized statement this lifts to a one-to-one 
correspondence between Hadamard states and the spatial FRG. 

In practice, it has proven difficult to construct 
Hadamard states explicitly. Since the italicized statement 
conceptually rests on such a construction we shall not try 
to establish a general result here.  Among non-static globally 
hyperbolic manifolds \FL backgrounds are of prime interest. 
For them a fairly explicit construction of Hadamard states is 
possible \cite{Olbermann} and concomitant results \cite{BonusSLE} 
allow one to analyze (\ref{sfrgPT6}), (\ref{sfrgPT7}) in detail.

\newsubsection{FRG vs Hadamard correspondence for \FL backgrounds} 

We consider spatially flat Friedmann-Lema\^{i}tre line elements
of the form  
\be 
\label{lineel} 
ds^2 = - N(t) dt^2 + a(t)^2 \delta_{ij} dx^i dx^j = 
g^{\rm FL}_{\mu\nu} dy^{\mu} dy^{\nu}\,,
\ee
initially without imposing any field equations. It is globally 
hyperbolic because the flat spatial metric is 
geodesically complete, see \cite{wavebook}, Lemma A.5.14.  
Here the shift $N^i$ has been set to zero and we continue to write 
$N$ for the merely $t$ dependent lapse.  The form of 
the line element (\ref{lineel}) is preserved under ${\rm Diff}[t_i,t_f] 
\times {\rm ISO}(d)$ transformations, where the rotation group acts 
as global ${\rm Diff}(\Sigma)$ transformations connected to 
the identity. Under the temporal reparameterizations $a(t)$ and 
$\tilde{\chi}(t,x)$ transform as scalars, while $N(t)$ and 
$n(t) = N(t)/a(t)^d$ are temporal densities, 
$n'(t') = n(t)/|\dd t'/\dd t|$, etc..
The scalar field action from (\ref{action1}) specialized to $\wg_{ij} 
= a(t)^2 \delta_{ij}$, $n = n(t)$, and $N^i =0$ reads
\ba
\label{action2}
S[\tilde{\chi}]\is \int_{t_i}^{t_f} \! dt \int_{\Sigma} \! d^dx \,
\Big\{ \frac{1}{2n(t)} (\dd_t \tilde{\chi})^2
- \frac{1}{2} n(t) a(t)^{2 d -2} \dd_i \tilde{\chi} 
\delta^{ij} \dd_j \tilde{\chi}-n(t) a(t)^{2d} \,{\cal U}(\tilde{\chi}, R)\Big\}
\nonum
&+ &\mbox{boundary terms} \,,
\ea
where $n(t)$, $a(t)$ specify the geometry (\ref{lineel}) and 
\begin{equation} 
\label{RFL}
R(g^{\rm FL}) = 2 d \frac{(N^{-1} \dd_t)^2 a}{a} 
+ d (d\!-\!1) \frac{ (N^{-1} \dd_t a)^2}{a^2} = : R(t)\,.
\end{equation}
As indicated, we normally use a finite temporal interval 
but do not keep track of boundary terms. In the 
basic action, for example, the term proportional to $R(g^{\rm FL})$ 
could be rewritten -- modulo a Gibbons-Hawking boundary term -- 
to obtain a kinetic term for the scale factor $(N^{-1} \dd_t a)^2$. 
Since we shall not consider equations of motion for $a$ and 
$N$, subtraction of such boundary terms is inessential 
for the present purpose. Later on the effective action will 
similarly be evaluated only modulo temporal boundary terms.   

In a next step we transition to the background field formalism,
using the $\eps_g =-1$ versions of (\ref{sfrg1}), (\ref{sfrg2}), 
(\ref{sfrg6}). The main simplifying feature of  (\ref{lineel}) as a 
globally hyperbolic background is that the flat spatial sections 
allow for a spatial Fourier 
transform of all relevant quantities. Our conventions are that of $\R^d$, 
$\chi(t,x) = \int \! \frac{d^dp}{(2\pi)^d} e^{i p x } \chi(t,p)$, 
$\chi(t,p) = \int \! d^dx \, e^{- i p x} \chi(t,x)$, 
with a `$\;\widehat{}\;$' omitted on $\chi(t,p)$. The associated 
delta distribution is $\delta(x\!-\!x')$, so that the previous covariant 
delta distribution factorizes according to 
\begin{equation} 
\label{FT0} 
\delta(t,x;x',t') = \delta(t,t') \delta(x-x') \,, 
\quad \delta(t,t') := (N a^d)^{-1} \delta(t-t')\,. 
\end{equation}
Since $R_k(t,x,x')$ in Section 2.2 was normalized by 
$R_k(t,x;t',x') = N(t)^{-1} \delta(t-t') R_k(t,x,x')$ 
additional powers of $a^{-d}$ may occur in the specialization
of the general formulas. 

For example, the modulator term in the action reads 
\begin{equation}
\label{FT1}
\Delta S_k[\chi] = - \frac{1}{2} \int \!dt N(t) a(t)^{d} 
\int\! \frac{d^dp}{(2\pi)^d} R_k(t,p) |\chi(t,p)|^2\,,
\end{equation}
with $N a^d$ instead of $N a^{2d}$ occurring. The Fourier 
representation of the equal time kernel is defined by 
\ba 
\label{reg1}
R_k(t,x,x') \is a(t)^{-d} \int\!\frac{d^dp}{(2\pi)^d}\,e^{ip(x-x')} 
R_k(t,p)\,,
\nonum
R_k(t,p) \is k^2 r\bigg( \frac{p^2}{k^2 a(t)^2} \bigg)\,, 
\quad 
r \in {\cal S}(\R_+)\,, \; 
r(0)= 1\,, 
\ea 
where the factor of $a^{-d}$ is required to arrive at 
\eqref{FT1} from \eqref{sfrg7}. 
Similar modulators have been used in \cite{Kaya, Serreau1,Serreau2}.
As indicated, the $t,p$ dependence 
is stipulated to enter through the comoving momentum square 
$p^2/a(t)^2$ only. Further, $\cS(\R_+)$ is the class of radial, real 
valued Schwartz functions on 
$\R_+$, decaying together with all its derivatives faster than 
any power. The $r(0) =1$ normalization could be relaxed to any 
nonzero positive constant while the smooth fast decay will be important 
later on; note that it rules out the widely used `hockey stick' 
$r(u) = (1\!-\!u) \th(1\!-\!u)$. An admissible example is the 
exponential modulator, 
$r(u) = \alpha u/(e^{\alpha u} -1)$, $\alpha >0$.

Since the background field $\vp$ is a function of $t$ only the 
Green's' function likewise admits a Fourier realization 
\begin{equation}
\label{FT3}
G_k[\vp](t,x;t,x') = \int\!\frac{d^dp}{(2\pi)^d}\,e^{ip(x-x')}   
G_k[\vp](t,t',p)\,,
\end{equation}
and is defined by the relation
\be
\label{FLfrg2} 
- \int\! dt' {N}(t') a(t')^d \,\Gamma_k^{(2)}[\vp](t,t',p) 
G_k[\vp](t',t'', p) + R_k(t,p) G_k[\vp](t,t'';p) = 
\delta(t,t'')\,. 
\ee
Here we used $(\Delta S_k)^{(2)}(t,x,t',x') = - R_k(t,x,x')\delta(t,t')$ 
in the conventions (\ref{sfrgPT5}), (\ref{FT0}) and wrote 
$\Gamma_k^{(2)}[\vp](t,t',p)$
for the Fourier transform of $\Gamma_k^{(2)}[\vp](t,x;t',x')$. 

Due to the spatial homogeneity the volume in $\Gamma_k$ and 
its FRG is overcounted and will be factored out later on. With 
this understanding the spatial FRG (\ref{sfrg8}) assumes the form 
\be 
\label{FLfrg1} 
\dd_k \Gamma_k[\vp] = \frac{i \hbar}{2} 
\int\!\! dt d^dx \,N(t) a(t)^d 
\int\! \ddf{p}{d} \dd_k R_k(t,p) G_k[\vp](t,t,p)\,.
\ee
The system (\ref{FLfrg1}), (\ref{FLfrg2}) is the central object 
in the subsequent sections and will be referred to as the 
\FL spatial FRG, {\bf FL-sFRG}. A similar spatial FRG has been 
considered on a de Sitter background in \cite{Kaya,Serreau1,Serreau2}. 
However, these constructions utilize the maximal symmetry of de 
Sitter spacetime to select the Bunch-Davies vacuum as the unique 
de Sitter invariant Hadamard state, and hence the issue of state 
dependence of the flow equation does not arise. In cosmology one of 
the prime applications of the  { FL-sFRG} is to the exploration 
of a pre-inflationary, kinetic energy dominated era 
\cite{FRWkindom1,FRWkindom2}.

The perturbative expansion proceeds as before via Ans\"{a}tze 
(\ref{sfrgPT1}) transferred to Fourier space. The recursion reads 
\ba 
\label{FLfrg3} 
&& [\cD_{t,p} + R_k(t,p)] G_{k,0}[\vp](t,t';p) = \delta(t,t')\,,
\nonum
&& [\cD_{t,p} + R_k(t,p)] G_{k,n}[\vp](t,t';p) = 
\sum_{l=1}^n \int\! dt''N(t'')
a(t'')^d \, \Gamma^{(2)}_{k,l}(t,t'';p) G_{k,n-l}[\vp](t'',t';p) \,.
\nonum
&& \dd_k \Gamma_{k,n}[\vp] = \frac{i}{2} 
\int\!\! dt d^dx \,N(t) a(t)^d 
\int\! \ddf{p}{d} \dd_k R(t,p) G_{k,n-1}[\vp](t,t,p)\,.
\ea 
Here we used $S^{(2)}[\vp](t,t';p) = - \cD_{t,p} \delta(t,t')$, 
for the Fourier kernel of the leading order. The differential 
operator reads 
\ba  
\label{Diffop}
\cD_{t,p} \is  a^{-2d} (a^d N^{-1} \dd_t)^2 + a^{-2} p^2 + W(t)\,, 
\nonum
W(t) &:= & {\cal U}''(\vp(t), R(t)) = \sum_{n \geq 0} 
{}^nU^{(2)}(\vp(t)) R(t)^n\,. 
\ea
Importantly, the Green's function relations are 
pointwise in $p$ and $G_{k,0}$ is the Green's function 
of an ordinary, second order, differential operator with a
parametric dependence on $p^2$ and $k^2$. For $p^2\ll k^2$
the $R_k$ term will by (\ref{reg1}) just 
give rise to an additive term in the potential, $W \mapsto 
W + O(k^2)$. For $p^2 \gg k^2$ the influence of the $R_k$ 
term will  be negligible due to the fast decay. 
The construction of a Green's function $G_{k,0}$ for 
$\cD_{t,p} + R_k(t,p)$ should therefore be a `small variation' of 
the construction of a
Green's function $G_{0,0}$ of $\cD_{t,p}$. In technical terms, 
the relation between a given $G_{0,0}$ and the $G_{k,0}$ sought 
reads 
\be 
\label{FLfrg4} 
G_{k,0}[\vp](t,t';p) + \int\! dt'' (N a^d)(t'') 
G_{0,0}[\vp](t,t'';p) R_k(t'', p) 
G_{k,0}[\vp](t'', t';p) = G_{0,0}[\vp](t,t';p)\,,
\ee
where $[a^{-2d} (a^d N^{-1} \dd_t)^2 + a^{-2} p^2 
+ W] G_{0,0}[\vp](t,t';p) = \delta(t,t')$. 
Again, this has to be augmented by conditions on the 
(first derivative's) temporal coincidence limit to 
characterize a Feynman-type Green's function. We postpone the 
analysis of (\ref{FLfrg4}) and first describe how to obtain 
a physically viable $G_{0,0}$.

The construction of a `Hadamard type' Feynman Green's function 
$G_{0,0}$ has been discussed in detail in \cite{Olbermann,BonusSLE}. 
We summarize the aspects relevant for the FRG versus Hadamard 
correspondence here. In order to set the notation we include the 
mode modulation term in the basic relations.  Any Feynman type 
Green's function on a \FL background (\ref{lineel}) can in Fourier 
space be realized as 
\be 
\label{FLhad1} 
G_{k,0}[\vp](t,t';p) = i \th(t-t') T_k(t,p) T_k(t',p)^* 
+ i \th(t'-t) T_k(t,p)^* T_k(t',p)\,,
\ee 
where $T_k(t,p)$ is a Wronskian normalized solution of the 
homogeneous wave equation
\ba 
\label{FLhad2} 
&& \Big[ (a^d N^{-1} \dd_t)^2 + \om_k(t,p)^2 + p^2 \om_2(t)^2\Big] 
T_k(t,p)=0\,,
\nonum
&& \om_2(t)^2 = a(t)^{2 d-2}\,, \quad \om_k(t,p)^2 = 
a(t)^{2d} \Big[ W(t) + k^2 r\Big(\frac{p^2}{k^2 a^2}\Big)\Big]\,, 
\nonum
&& a^d N^{-1} \dd_t T_k(t,p)\, T_k(t,p)^* - 
a^d N^{-1} \dd_t T_k(t,p)^*\, T_k(t,p) = - i\,. 
\ea  
Here $\lim_{k \ra 0} \om_k(t,p)^2 = a(t)^{2 d} W(t) = \om_0(t)^2$ recovers the 
standard case with solution $T_0(t,p)$; the subscript ``$2$'' in 
$\om_2(t)$ signals the coefficient of $p^2$ (not $k\!=\!2$ in $\om_k(t,p)^2$, 
of course). 
Note that $G_{k,0}[\vp](t,t;p) = i |T_k(t,p)|^2$, 
$a^d N^{-1} \dd_t G_{k,0}[\vp](t,t';p)|_{t'=t} = 1/2$, 
so that (\ref{FLhad1}) is fully determined by $|T_k(t,p)|$. 
In fact $T_k(t,p)$ itself is - up to constant phase - determined 
by its modulus 
\be 
\label{FLhad3} 
T_k(t,p) = \big| T_k(t,p)\big| \exp\bigg\{ 
\!-\frac{i}{2} \int_{t_0}^{t} \! dt'\frac{N(t')}{a(t')^d}  
\frac{1}{ |T_k(t',p)|^2 } \bigg\}\,. 
\ee    
Inserting (\ref{FLhad3}) into (\ref{FLhad1}) produces the 
decomposition of the Feynman Green's function into the 
state-independent causal Green's function and the state-dependent 
symmetrized Wightman function
\ba 
\label{FLhad4} 
G_{k,0}[\vp](t,t';p) \is \frac{1}{2} \Delta^c_k(t,t';p) + 
\frac{i}{2} \om_k^s(t,t';p) \,,
\ea 
both of which are fully determined by the modulus of $T_k$. 

Remarkably, the modulus square satisfies itself a nonlinear second 
order differential equation
\be
\label{FLhad5} 
2 |T_k|^2 (a^d N^{-1} \dd_t )^2 |T_k|^2 - 
\big( a^d N^{-1} \dd_t |T_k|^2 \big)^2 + 
4 \big[ \om_2(t)^2 p^2 + \om_k(t,p)^2 \big]|T_k|^4 =1,
\ee   
as can be verified from (\ref{FLhad2}). This is a variant of the 
Gelfand-Dickey equation satisfied by the diagonal of the resolvent 
kernel of a Schr\"{o}dinger operator. Indeed, $G_{k,0}(t,t;p) = 
i |T_k(t,p)|^2$ is heuristically the diagonal of the inverse of 
the differential operator in (\ref{FLhad2}). Due to the 
$a^d {N}^{-1}$ factors the differential operator in (\ref{FLhad2}) 
is not directly a Schr\"{o}dinger operator and there is no 
immediate resolvent parameter. After a suitable transformation and 
without the $r$ term the coefficient 
of $p^2$ is $N^2/a^2$, which is in general not constant 
and $p^2$ does not quite play the role of a resolvent parameter
for the Schr\"{o}dinger operator. One can, however, develop a 
generalized resolvent expansion which allows the coefficient of the 
large parameter -- here $p^2$ -- to be non-constant. Moreover, 
the structure of the expansion characterizes a Hadamard 
state:

{\bf Result} \cite{BonusSLE}{\bf :} Let $T_0(t,p)$ be a solution of 
$[ (a^d N^{-1} \dd_t)^2 + \om_0(t)^2 + p^2 \om_2(t)^2] T_0(t,p) =0$
and let $\om_0^s(t,t';p)$ be the symmetrized Wightman function 
(\ref{FLhad4}) built from it. Then, its inverse Fourier transform 
is associated with a Hadamard state if and only if $|T_0(t,p)|^2$ admits an 
asymptotic expansion of the form 
\be 
\label{FLhad7}
- i G_{0,0}[\vp](t,t;p) = |T_0(t,p)|^2 \asymp \frac{1}{2 \om_2(t) p} \Big\{ 1 + 
\sum_{n \geq 1} (-)^n \bar{G}_n(t) p^{-2 n} \Big\}\,.  
\ee  
Here $\bar{G}_0 =1$ the coefficients $\bar{G}_n(t), n \geq 1$, are 
fixed {\it uniquely} by the recursion relation 
\begin{eqnarray}
\label{FLhad8} 
\bar{G}_n \is \!\!\sum_{j,l\geq 0, j+l =n-1} \!\bigg\{ 
\frac{1}{4} \frac{\bar{G}_j}{\omega_2} 
(a^d N^{-1} \dd_t)^2 \Big( \frac{\bar{G}_l}{\omega_2} \Big) 
- \frac{1}{8} (a^d N^{-1} \dd_t) \Big( \frac{\bar{G}_j}{\omega_2} \Big) 
(a^d N^{-1} \dd_t) \Big( \frac{\bar{G}_l}{\omega_2} \Big) 
\nonum
&+&  
\frac{1}{2} \frac{\omega_0^2}{\omega_2^2} \bar{G}_j \bar{G}_l \bigg\} 
-\frac{1}{2} \sum_{j,l\geq 1, j+l =n} \bar{G}_j \bar{G}_l. 
\end{eqnarray} 
Here the ``$\asymp$'' sign indicates that the difference between the left hand side and the $N^{\text{th}}$ partial sum of the right hand side is $O(p^{-2(N+1)})$, without control over the $N$ and $t$  dependent coefficient. The implied asymptotic expansion of $1/|T_0(t,p)|^2$ and
hence of the phase is understood. 

\underline{Remarks:}
\vspace{-3mm} 

\begin{itemize}[leftmargin=8mm, rightmargin=-0mm]
\itemsep -1mm
\item[(i)] The recursion follows by inserting an Ansatz of the form 
(\ref{FLhad7}) into (\ref{FLhad5}). It expresses $\bar{G}_n$ in terms 
of $\bar{G}_{n-1}, \ldots, \bar{G}_1$, and involves only differentiations. 
It follows that $\bar{G}_n$ equals $\om_2^{-4n}$ times 
a differential polynomial in $\om_0^2, \om_2^2$, with $a^d N^{-1} 
\dd_t$ as basic derivative. Moreover, one can show that 
the $\bar{G}_n, n \geq 2$, can be expressed as differential polynomials 
in only $\bar{G}_1$, using  $\om_2^{-1} a^d N^{-1} \dd_t$ as basic 
derivative. For (\ref{FLhad2}) one has $\om_2^2 = a^{2d-2}$, 
$\om_0^2 = a^{2d} W$, where positivity of $W$ is not needed. 
\item[(ii)] Writing $v := \omega_0^2,w:= \omega_2^2$ and denoting 
$a^d N^{-1} \dd_t$ differentiations momentarily 
by a ``$\,'\,$'' one finds:
\begin{eqnarray}
\label{FLhad9} 
\bar{G}_1 &\! =\! & \frac{v}{2 w} + \frac{5}{32} \frac{{w'}^2}{ w^3} 
- \frac{1}{8} \frac{w''}{w^2}\,,
\nonumber \\[1.5mm]
\bar{G}_2 &\! =\! & \frac{3}{8w^2} \Big( v^2 + \frac{1}{3} v'' \Big) 
- \frac{5}{16 w^3} \Big( v w'' + v'w' - v\frac{7 {w'}^2}{4w} \Big)     
\nonumber \\[1.5mm]
&+&\frac{1}{32 w^3} \Big(\! - w^{(4)} +\frac{21 {w''}^2}{4 w}
+\frac{7 w^{(3)} w'}{w} -\frac{231 {w'}^2w''}{8 w^2} 
+\frac{1155 {w'}^4}{64 w^3} \Big)\,.
\end{eqnarray}   
The recursion (\ref{FLhad8}) is easily programmed 
in {\tt Mathematica} and produces the $\bar{G}_n$ to reasonably high orders. 

\item[(iii)] The familiar adiabatic iteration \cite{ParkerTomsbook} is primarily organized according to the number of time-derivatives of $a(t)$, but is provenly  asymptotic only for large $p$ \cite{Luders}. In this form it enters the original proof \cite{Olbermann} of the Hadamard property of SLE. The above result eliminates the reorganization of the adiabatic iteration, and characterizes directly the coefficients of $p^{-2n-1}$.

\item[(iv)]  Being merely asymptotic, the expansion (\ref{FLhad7}) does 
{\it not} fully determine $G_{0,0}$. Rather, an independent construction 
principle for a specific solution $T_0(t,p)= T_0^{\rm Had}(t,p)$, of 
Hadamard type, valid for {\it all} $p$, needs to be known and implemented. 
While the large $p$ behavior of $T_0^{\rm Had}(t,p)$ will always be governed 
by (\ref{FLhad7}) the small $p$ behavior will reflect the nature 
of the construction principle used to obtain it. 
\end{itemize}

Occasionally, one will want to relate (\ref{FLhad2}) and the 
results based on it to a standard Schr\"{o}dinger operator
and its resolvent expansion. This can be done by expanding  
$(a^d N^{-1} \dd_t)^2$ $= a^{2d} N^{-2} [ \dd_t^2 - 
\dd_t \ln (N a^{-d}) \dd_t]$ and then removing the first order term 
by redefining $T_k(t,p) = (N(t) a(t)^{-d})^{1/2} \chi_k(t,p)$. Doing so, 
(\ref{FLhad2}) transcribes into  
\ba 
\label{FLhad10} 
&& \Big[\dd_t^2 + {N}^2 k^2 \Big( \frac{p^2}{ a^2 k^2} 
+ r\Big(\frac{p^2}{a^2 k^2}\Big)\Big) + N^2 W 
+ \frac{1}{2} S(t)\Big] \chi_k(t,p)=0\,,
\nonum
&& \dd_t \chi_k(t,p)\, \chi_k(t,p)^* - 
\dd_t \chi_k(t,p)^*\, \chi_k(t,p) = - i\,. 
\ea  
where $S(t) = - 2 (N a^{-d})^{1/2} \dd^2_t ( N^{-1} a^d)^{1/2}$ 
is the induced Schwarzian. The associated Gelfand-Dickey equation is 
\be 
\label{FLhad11} 
2 |\chi_k|^2 \dd_t^2 |\chi_k|^2 - 
\big(\dd_t |\chi_k|^2 \big)^2 + 
\Big[ 4 N^2 k^2 \Big( \frac{p^2}{ a^2 k^2} 
\!+\! r\Big(\frac{p^2}{a^2 k^2}\Big)\Big) + 4 N^2 W +2 S\Big]
|\chi_k|^4 =1\,.
\ee 
Without the $r$ term the coefficient of $p^2$ in (\ref{FLhad10}) 
is $N^2/a^2$, which is in general not constant 
and $p^2$ does not quite play the role of a resolvent parameter
for the Schr\"{o}dinger operator. This can be remedied in several 
ways: One can gauge fix temporal reparameterization invariance
by choosing conformal time, $N(t) = a(t)$. Without the 
$r$ term then $p^2$ literally plays the role of a resolvent 
parameter for the Schr\"{o}dinger operator with potential $a^2 W + S/2$. 
Alternatively, one can introduce a resolvent parameter for the original 
Hessian in (\ref{sfrgPT5}), $\cD \mapsto \cD + z$. Then ${\cal U}''$ is 
everywhere replaced with ${\cal U}'' + z$, and in (\ref{FLhad10}) 
it occurs with pre-factor ${N}^2$. This can now be rendered 
constant by choosing cosmological time, $N(t) \equiv 1$, 
which results in a Schr\"{o}dinger operator with potential 
$W + S/2$. 

With the above result in place we can return to the issue of how 
to lift a Feynman Green's function $G_{0,0}$ to its scale dependent 
counterpart $G_{k,0}$. Instead of analyzing (\ref{FLfrg4}) 
directly we return to (\ref{FLhad2}), (\ref{FLhad5}) and note 
\be 
\label{FLhad12}
- i G_{k,0}[\vp](t,t;p) = |T_k(t,p)|^2 \asymp \frac{1}{2 \om_2(t) p} 
\bigg\{ 1 + 
\sum_{n \geq 1} (-)^n \bar{G}_n(t)\Big|_{\om_0^2 \mapsto \om_k^2} 
p^{-2 n} \bigg\}\,. 
\ee  
This holds for $k$ bounded away from zero because the $r$ term 
in $\om_k(t,p)^2$ decays faster than 
any power in $p$, and thus does not interfere with the derivation of 
the recursion relation obtained by comparing powers of $1/p^2$.
The only difference is that in (\ref{FLhad8}) $\om_0^2$ will 
be replaced with $\om_k^2$, which results in (\ref{FLhad12}).  
For given $r$ one obtains a one-parameter deformation of the original 
Hadamard state. Note that no modified notion of a Hadamard state is 
needed, the asymptotic expansion (\ref{FLhad7}) simply allows for 
the deformation.

In Section 3.1 we shall keep the dimensionless ratio $\wp = p/k$ 
fixed and interpret the differential operator in (\ref{FLhad2}) as 
$(a^d N^{-1} \dd_t)^2 + a^{2d} W + k^2 a^{2 d-2} [\wp^2 + 
a^2 r(\wp^2/a^2)]$. The general result (\ref{FLhad7}), (\ref{FLhad8})
then applies with $p \mapsto k$, $\om_0^2 \mapsto 
a^{2d} W$, $\om_2^2 \mapsto a^{2 d-2} [\wp^2 + 
a^2 r(\wp^2/a^2)]$. The resulting series in $k^{-2n-1}$ can be 
seen as a re-organization of the one in (\ref{FLhad12}). In 
particular, it can still be viewed as a deformation of the 
underlying Hadamard state.

A similar deformation exists in the infrared regime. While this can 
be seen at the level of the integral equation \eqref{FLfrg4} for the 
Green's function, 
it is simpler to consider the lift from $T_0(t,p)$ to $T_k(t,p)$ 
via an expansion in convolution powers of 
$a(t)^{2d} k^2 r(p^2/(k^2 a(t)^2))$. 
\medskip

{\bf Result:} Let $T_0^{\rm Had}(t,p)$ be a given 
Wronskian normalized solution of Hadamard type of 
$\big[ (a^d N^{-1} \dd_t)^2 + a^{2d} W + a^{2d -2} p^2\big] 
T_0(t,p) =0$, and let $G_0^{\rm Had}(t,t';p)$ be the associated 
Feynman Green's function, formed according to (\ref{FLhad1}).  
Then, there exists a solution of the wave equation in (\ref{FLhad2}) 
given by 
\be 
\label{IRdef1} 
\tilde{T}_k(t,p) = T_0^{\rm Had}(t,p) + 
\sum_{n \geq 1} (-)^n \Big(\Big( G_0^{\rm Had} \cdot a^{2d} k^{2n} 
r\big(p^2/(a^2 k^2) \big) \Big)^{ \cdot \,n } \cdot T_0^{\rm Had}
\Big)(t,p)  \,,
\ee 
where $\cdot$ denotes a temporal integration over $[t_i,t_f]$ 
and $\cdot n$ its $n$-fold nested iteration. Moreover, the 
series converges uniformly on $[t_i,t_f]$ for $k < k_*(p)$. 
The solution $\tilde{T}_k(t,p)$ can be re-normalized 
to satisfy the Wronskian condition in (\ref{FLhad2}) as well.

We only sketch the proof, which parallels the one in Prop.~3.1 
of \cite{BonusSLE}. On an interval $[t_i,t_f]$ the differential 
equation in (\ref{FLhad2}) can be recast as an integral equation  
\be 
\label{IRdef2} 
\tilde{T}_k(t,p) = - k^2 \int_{t_i}^{t_f} 
\! dt' G_0^{\rm Had}(t,t';p) a(t')^{2d} r \big( p^2/(k^2 a(t')^2) \big) 
\tilde{T}_k(t',p) \,,
\ee
where the given Hadamard Green's function is used to invert the 
the $k=0$ part of the differential operator. The three terms 
comprising the integral kernel are all bounded in $[t_i,t_f]$. 
For $G_0^{\rm Had}$ this holds because it is built in analogy to 
(\ref{FLhad1}) from $T_0^{\rm Had}$; for $a^{2d}$ it holds trivially; 
for the $r$-term it holds by assumption on $r$ in (\ref{reg1}).  
This implies that the integral operator (depending parametrically 
on $k$ and $p$) 
\be 
\label{IRdef3}
f \mapsto   - k^2 \int_{t_i}^{t_f} 
\! dt' G_0^{\rm Had}(t,t';p) a(t')^{2d} r \big( p^2/(k^2 a(t')^2) \big) 
f(t')\,,
\ee
is well-defined as a self-map on the Banach space $\big( \cC([t_i,t_f]), 
\Vert \cdot \Vert_{\rm sup} \big)$ of continuous functions 
on $[t_i,t_j]$ equipped with the sup norm. Moreover, for sufficiently 
small $k < k_*(p)$ it is a contraction. As such, it possesses a fixed point  
$\tilde{T}_k(t,p)$ which solves (\ref{IRdef2}) exactly, and which may be 
reached from $T_0^{\rm Had}$ by the series (\ref{IRdef1}). A similar 
argument applies to the time derivative $a^d N^{-1} \dd_t \tilde{T}_k(t,p)$, 
for which the termwise time derivative of the series (\ref{IRdef1}) 
converges uniformly in $[t_i,t_f]$ for $k < k_*(p)$; c.f.~the 
proof of Prop.~3.1 in \cite{BonusSLE}.

Taken together, this implies the constancy of the 
Wronskian 
$a^d N^{-1} \dd_t \tilde{T}_k(t,p)\, \tilde{T}_k(t,p)^* - 
a^d N^{-1} \dd_t \tilde{T}_k(t,p)^*\, \tilde{T}_k(t,p) 
=: - i {\rm wr}_k(p)$, and of the induced series expansion 
within the radius of convergence $k < k_*(p)$.  
By construction, ${\rm wr}_k(p)$ is real with ${\rm wr}_0(p) =1$. 
Re-normalizing the solution $T_k(t,p) := 
{\rm wr}_k(p)^{-1/2} \tilde{T}_k(t,p)$ produces for $k < k_*(p)$ an 
exact solution of (\ref{FLhad2}), which is a deformation of the 
given $T_0^{\rm Had}(t,p)$. This shows the result.

In summary, a given Hadamard type Feynman Green's function 
$G_{0,0}^{\rm Had}[\vp](t,t',p)$ has a unique deformation to 
a Feynman Green's function $G_{k,0}[\vp](t,t',p)$, 
both in the ultraviolet (large $p$ at fixed $k$, or large $k$ 
at fixed $\wp = p/k$) and in the infrared (small $k$ at fixed $p$, 
in particular for small  $p$). The deformation is also of Hadamard 
type but modified, see 
(\ref{FLhad12}). Both Green's functions are related by (\ref{FLfrg4}) 
and since $G_{0,0}^{\rm Had}[\vp](t,t',p)$ is defined for 
all $p$, we expect that the deformation is unique 
also in the crossover region.


\newsubsection{States of Low Energy and their infrared behavior} 

States of Low Energy (SLE) are Hadamard states for a minimally 
coupled scalar field on a generic Friedmann-Lema\^{i}tre background.  
The name originates from the fact that a time averaged energy 
functional is minimized as part of the construction. The 
energy functional derives from the time-time component of the 
energy momentum tensor, which changes if the coupling to 
gravity is non-minimal. The original construction 
\cite{Olbermann} starts from an {\it arbitrary} (exact) fiducial 
solution $S_0(t,p)$ of the homogeneous $k=0$ wave equation 
(\ref{FLhad1}) and associates to it a special solution 
$T^{\rm SLE}_0[S_0](t,p)$ of the same wave equation. The construction 
proceeds by minimization of the time averaged energy functional 
and fixes the $p$-dependence of $T^{\rm SLE}_0[S_0](t,p)$ completely,
{\it for all} $0 < p = \sqrt{p_1^2 + \ldots + p_d^2}< \infty$. 
For large $p$ it is such that the expansion (\ref{FLhad7}) holds. 
The associated Feynman Green's function (\ref{FT3}), 
(\ref{FLhad1}) can then be shown to have Hadamard property in the 
global, wave front sense. Finally, it can be shown that although a choice of $S_0$ is made in the construction of the SLE $T_0^\text{SLE}[S_0]$, it is actually independent of this choice (up to a constant phase), i.e. any other $\tilde{S}_0$ related to $S_0$ by a Bogoliubov transformation gives the same SLE, see \eqref{gsle4} below.

While the direct link to the energy-momentum tensor and the minimization 
procedure are physically appealing, the mathematics of the Hadamard 
property rests more on the time averaging and the invariance under 
Bogoliubov transformations of $S_0$. For the standard non-minimal coupling, 
$\xi R \tilde{\chi}^2$, the positivity properties of the energy functional 
are obscured for generic values of $\xi$, and a complete proof of the 
Hadamard property is currently not available. For our Ricci tower potential 
(\ref{pot}) even the definition of a viable energy-momentum tensor would 
require a lengthy detour and the analysis of the minimizers may have to 
remain incomplete. Instead, we use here the minimizing functionals 
associated with the {\it minimally} coupled scalar and the proven Hadamard 
property of the associated SLE. Whenever the distinction needs to be 
highlighted we shall refer to them as the {\bf minimal SLE}.  
This route sacrifices a direct link to the energy-momentum tensor and 
its renormalizability would have to be explored separately.  In fact, 
one of the advantages of the FRG formalism is that for most applications 
the construction of composite operators is not needed.  

For a minimally coupled scalar field the dispersion relation is 
$\om_p(t)^2 = a(t)^{2d}[p^2/a^2 + {}^0U''(\vp(t))]$, see (\ref{Diffop}).    
For later use we generalize this to 
\be 
\label{IRdispersion}
\om_p(t)^2 = \om_0(t)^2 + p^2 w_{2,p}(t)\,, \quad 
p \mapsto w_{2,p}(\cdot) \;\;\mbox{bounded and smooth}\,.
\ee 
Specifically, $w_{2,p}(t) - w_{2,\infty}(t)$ is assumed 
to be of rapid decay in $p$ uniformly in $t \in [t_i,t_f]$,
with some $w_{2,\infty}(t) =: \om_2(t)^2 >0$. Similarly, 
$\lim_{p \ra 0} w_{2,p}(t)^2 = w_{2,0}(t)$, for some nonzero 
$w_{2,0}(\cdot )$. However, $\om_p(t)^2 \geq 0$, $t \in [t_i,t_f]$, 
is assumed throughout. This  may be viewed as the dispersion of 
an IR modified QFT, where the rapid decay in the ultraviolet leaves 
the Hadamard property unaffected. The $k$-modified 
dispersion $\om_p(t)^2 = p^2 \om_2(t)^2 + \om_k(t,p)^2$ from 
(\ref{FLhad2}) is a prime example, another one will arise 
in Section 4.3. For the dispersion (\ref{IRdispersion}) we consider 
Wronskian normalized fiducial solutions solving the modified wave 
equation $[(a^d N^{-1} \dd_t)^2 + \om_p(t)^2 ] S(t,p) =0$. To any some-such 
solution the associated minimal, {\bf IR modified SLE} 
solution is defined by 
\be
\label{gsle1}
T^{\rm SLE}[S](t,p) = \lb_p[S] S(t,p) + \mu_p[S] S(t,p)^*\,. 
\ee
Suppressing the subscripts momentarily, the coefficients are given by 
\ba 
\label{gsle2} 
\lb[S] \is - e^{- i {\rm Arg} \cD[S]} \sqrt{ 
\frac{\cE[S]}{2 \sqrt{ \cE[S]^2 - |\cD[S]|^2}} + \frac{1}{2}}\,,
\nonum
\mu[S] \is \sqrt{ 
\frac{\cE[S]}{2 \sqrt{ \cE[S]^2 - |\cD[S]|^2}} - \frac{1}{2}}\,.
\ea
Here $\cE[S]$ and $\cD[S]$ are functionals that arise by 
expanding the time averaged free Hamilton operator in terms
of creation and annihilation operators. The averaging is done 
with a positive smooth function $f(\cdot)$ of compact support 
in $[t_i, t_f]$. Explicitly,  
\ba 
\label{gsle3}
\cE[S] \is \frac{1}{2} \int\! dt \, {n}(t) f(t) 
\Big\{ | {n}^{-1} \dd_t S|^2 + \om_p(t)^2 |S(t)|^2 \Big\} \,,
\nonum
\cD[S] \is \frac{1}{2} \int\! dt \, {n}(t) f(t) 
\Big\{ ({n}^{-1} \dd_t S)^2 + \om_p(t)^2 S(t)^2 \Big\} \,. 
\ea 
In (\ref{gsle2}) a choice of sign has been made that renders 
$\mu[\cdot]$ real. Consistently, we require $\cE[S] > |\cD[S]|$, 
which is satisfied on account of $\om_p(t)^2 \geq 0$, 
$t \in [t_i,t_f]$. 

An obvious concern about the construction is the dependence on 
the fiducial solutions. Recall that the general Wronskian normalized 
solution of a homogeneous second order differential equation 
can always be written as the Bogoliubov transform of a fixed 
solution and its complex conjugate. Hence, one can probe the 
extent to which $T^{\rm SLE}[S](t,p)$ depends on 
the choice of fiducial solutions by subjecting them to a 
Bogoliubov transformation. Remarkably, $T^{\rm SLE}[S](t,p)$
turns out to be invariant up to a time independent phase 
\cite{BonusSLE}. Such a phase drops out in the Green's function 
(\ref{FLhad4}) and one has in particular
\be 
\label{gsle4} 
\om^s[a S + b S^*](t,t',p) = 
\om^s[S](t,t',p)\,, \quad |a|^2 - |b|^2 =1\,.
\ee 
Viewed as a quantum state underlying the Green's function, the 
IR-modified SLE therefore is {\it independent} of the choice of $S$.

The key property of the original SLE is that the inverse Fourier 
transform of the two-point function $T_0^{\rm SLE}[S_0](t,p) 
T_0^{\rm SLE}[S_0](t',p)$ 
has the Hadamard property \cite{Olbermann}. One can show that this 
remains true for the IR modified SLE modes $T^{\rm SLE}[S](t,p)$
defined above.  The proofs in \cite{Olbermann} are given for a 
constant potential (${\rm mass}^2$), the extension to the dispersion  
$\om_p(t)^2 = a(t)^{2d}[p^2/a^2 + {}^0U''(\vp(t))]$ is straightforward 
\cite{BonusSLE}. For the IR modified dispersion (\ref{IRdispersion}) 
the non-quadratic $p$ dependence needs to be taken into account. 
A major step in the proof of the Hadamard property is to 
show that $\lb[S]-1$ and $\mu[S]$ carry a $p$ dependence that 
falls off for large $p$ faster than any power. Since in (\ref{IRdispersion}) 
we take the difference $\om_p(t)^2 - p^2 \om_2(t)^2 
- \om_0(t)^2$ to decay itself faster than any power in $p$ 
(uniformly in $t \in [t_i,t_f]$), the additional $p$ dependence 
will not interfere with these estimates.

In accordance with the general result the IR modified SLE admit 
a large $p$ asymptotic expansion of the form (\ref{FLhad12}) 
entering (\ref{FLhad3}). It is significant that the dependence 
on the window function $f(\cdot)$ drops out to all orders of 
the expansion. This is one manifestation of the universality of 
a Hadamard state's ultraviolet behavior. In contrast, the 
coefficients of the small $p$ expansion depend manifestly on 
the window function. For later use we describe this expansion 
to low orders explicitly, allowing for a dispersion of the form 
(\ref{IRdispersion}).

The invariance of an SLE under Bogoliubov transformations 
can be rendered manifest by expressing the solution $T^{\rm SLE}_0(t,p)$ 
solely in terms of the commutator function $\Delta_p(t,t') = 
i [S(t,p) S(t',p)^* - S(t,p)^* S(t',p)]$ \cite{BonusSLE}. 
For the modulus square one has 
\be
\label{sle1}
\big| T^{\rm SLE}(t,p)\big|^2 = \frac{J_p(t)}{2 \cE_p^{\rm SLE}}\,, 
\ee
with 
\ba 
\label{sle2}
J_p(t) \is \frac{1}{2} \int\! dt' \,{n}(t') f(t') 
\big[ \big( {n}(t')^{-1}\dd_{t'} \Delta_p(t',t) \big)^2 + 
\om_p(t')^2 \Delta_p(t',t)^2 \big] \,,
\nonum
\big(\cE_p^{\rm SLE}\big)^2 \is \frac{1}{8} \int dt d t'\,
{n}(t) {n}(t') f(t) f(t') \Big\{ 
\big({n}(t)^{-1}{n}(t')^{-1} \dd_t \dd_{t'} 
\Delta_p(t,t') \big)^2 
\nonum
&+& 2 \om_p(t')^2 ({n}(t)^{-1} \dd_t \Delta_p(t',t) \big)^2 
+ \om_p(t)^2 \om_p(t')^2 \Delta_p(t,t')^2 \Big\}\,.
\ea 
Here $\om_p(t)^2$ is as in (\ref{IRdispersion}) and ${\cal E}_p^{\rm SLE}$ 
has the interpretation as the time averaged Hamiltonian's expectation 
value in the SLE. Since $\Delta_p(t,t')$ has mass dimension $-1$ so has 
$J_p(t)$, while $\cE^{\rm SLE}$ is dimensionless. 
Moreover, the quantities (\ref{sle1}), (\ref{sle2}) are known 
\cite{BonusSLE} to admit for $\om_p(t)^2 = \om_0(t)^2 + p^2 
\om_2(t)^2$ a {\it convergent} 
expansion in powers of $p^2$. In particular, 
\ba
\label{sle3} 
J_p(t) = \sum_{n \geq 0} J_n(t) p^{2n} \,, \quad \quad
(\cE_p^{\rm SLE})^2 =  \sum_{n \geq 0} \varepsilon_n^2 \,p^{2n} \,,
\ea  
where $J_n(t)$ has mass dimension $-(2n\!+\!1)$ and $\varepsilon_n$ 
has mass dimension $-n$. Note that the commutator function 
$\Delta_p(t,t') = \sum_{n \geq 0} \Delta_n(t,t') p^{2n}$ needs 
to be expanded as well. We extend the expansions (\ref{sle3}) to 
the generalized dispersion (\ref{IRdispersion}) simply by 
substituting $w_{2,p}(t)$ for $\om_2(t)^2$ and write 
\ba 
\label{sle4}
J_{n,p}(t) := J_n(t) \big|_{ \om_2^2 \mapsto w_{2,p}} \,,
\quad 
\varepsilon^2_{n,p} := \varepsilon_n^2 \big|_{ \om_2^2 \mapsto w_{2,p}} \,,
\quad 
\Delta_{n,p}(t,t') := \Delta_n(t,t')\big|_{ \om_2^2 \mapsto w_{2,p}} \,.
\ea   
The induced $p$ dependence is of bounded variation. Although it 
could in principle be expanded as well for small $p$, it 
turns out to be advantageous not to do so.

For nonzero $\om_0$ one has $\varepsilon_0 >0$ 
and the expansions (\ref{sle3}) starts at $O(p^0)$.  
For $\om_0 \equiv 0$ one has $\varepsilon_0 =0$ and the expansion
for the generalized dispersion (\ref{IRdispersion}) reads  
\ba 
\label{sle5} 
\cE_p^{\rm SLE} \is \varepsilon_{1,p} \,p + 
\frac{\varepsilon_{2,p}^2}{2 \varepsilon_{1,p}} p^3 
+ O(p^5) \,,
\nonum
|T^{\rm SLE}(t,p)|^2 \is \frac{J_0}{2 \varepsilon_{1,p}} \frac{1}{p} + 
\frac{ 2 J_{1,p}(t) \varepsilon_{1,p}^2 
- J_0 \varepsilon_{2,p}^2}{4 \varepsilon_{1,p}^3} \,p  + O(p^3)\,.
\ea
Remarkably, the leading term is $O(1/p)$, indicating a Minkowski-like 
infrared behavior for {\it all} scale factors, -- a characteristic 
property of `massless' SLE. This second case will be more relevant later on, 
so we present explicit expressions for the coefficients:%
\footnote{This corrects a factor $1/2$ in the second term of 
$J_1$ and an overall sign in $\Delta_1$ in Eqs (102), (103) 
of \cite{BonusSLE}.} 
\ba 
\label{sle6}
J_0 \is \frac{1}{2} \int \!dt \,{n}(t) f(t)\,,
\nonum
J_{1,p}(t) \is \frac{1}{2} \int \!dt' {n}(t') f(t') 
\big[ 2 {n}(t')^{-1} \dd_{t'} \Delta_{1,p}(t',t) + \Delta_0(t,t')^2 
w_{2,p}(t') \big]\,,
\ea
where $\Delta_0, \Delta_1$ are the terms in the expansion of the 
$\om_0 \equiv 0$ commutator function 
\be 
\label{sle7}
\Delta_0(t,t') = \int_{t'}^t \! ds \,{n}(s)\,,\quad 
\Delta_{1,p}(t,t') = \int_{t'}^t \! ds \,{n}(s) w_{2,p}(s) 
\Delta_0(t,s) \Delta_0(t',s)\,. 
\ee
Finally, 
\ba
\label{sle8}
\varepsilon_{1,p}^2 \is \frac{1}{2} J_0 \int \!dt {n}(t) f(t) w_{2,p}(t)\,,
\nonum
\varepsilon_{2,p}^2 \is \frac{1}{8} \int \!dtdt' {n}(t) {n}(t') 
f(t)  f(t')  \Big[ \big({n}(t)^{-1}{n}(t')^{-1} \dd_t \dd_{t'} 
\Delta_{1,p}(t,t') \big)^2 
\nonum
&-& 4 w_{2,p}(t') {n}(t)^{-1} \dd_t \Delta_{1,p}(t',t) + 
w_{2,p}(t) w_{2,p}(t') \Delta_0(t,t')^2 \Big]\,. 
\ea 
Note that for given ${n}$ and window function $f$ these are 
simple functionals of $w_{2,p}$. In the special case 
${n} \equiv 1$ and constant $w_{2,p} = w_2$ the expressions simplify 
drastically: $\Delta_0(t,t') = t-t'$, $\Delta_1(t,t') = - w_2(t-t')^3/6$,
giving $J_1(t) \equiv 0$, $\eps_1= w_2 J_0$, $\varepsilon_2^2 =0$.  
Expressions similar to (\ref{sle6}), (\ref{sle8}) exist for the case 
with nonzero $\om_0$, but $\Delta_0$, $\Delta_1$ can 
then in general not be determined explicitly.

\newpage 
\newsection{Wilsonian one-loop renormalization} 

On general grounds all Hadamard states share the same ultraviolet 
behavior. For the \FL backgrounds considered this manifests
itself in the universality of the asymptotic expansion (\ref{FLhad7}).
Here we use this non-covariant expansion to extract the 
divergent parts of the one-loop effective action by re-integrating 
the one-loop flow equation, $\dd_k \Gamma_{k,1} = ...$, in (\ref{FLfrg3}). 
For simplicity we focus on $d=3$ and use conformal time.  
Remarkably, the result is again covariant, i.e.~the $a$ dependence only 
enters through the measure and the Ricci scalar. This holds for 
any potential, with or without the Ricci tower (\ref{i0}).  

Next we aim at the absorption of the divergent parts through 
coupling and field renormalizations. Traditional one-loop renormalization 
includes only power counting 
renormalizable terms in the action, evaluates the divergent part 
of the Hessian's `tracelog' using dimensional regularization, and 
then removes the divergent parts using minimal subtraction, 
see \cite{ParkerTomsbook,Phi4onFRWPT3,Phi4onFRWPT1,Phi4onFRWPT2} 
in the present context. In a 
Wilsonian framework the action is meant to contain {\it all} interaction 
monomials compatible with a stipulated symmetry requirement, 
here $\vp \mapsto -\vp$ even terms in the scalar field. Regularization 
invokes a momentum-like ultraviolet (UV) cutoff $\Lambda$ and a running 
scale $\mu$. After evaluation of the divergent part of the 
effective action $L_1^{\rm div}$ one seeks coupling and field 
redefinitions entering the bare action $L_{\Lambda}$ such that the 
UV cutoff $\Lambda$ can strictly be removed     
\begin{equation}
\label{oneloopstrict} 
L = \lim_{\Lambda \ra \infty} \big[ L_{\Lambda} + \hbar L_1^{\rm div} \big]\,.   
\end{equation}
Conceptually, the renormalized Lagrangian $L$ ought to be identified 
with the scale dependent Wilsonian one at some scale $\mu$. 
The strict removal of the UV cutoff amounts to the successful 
identification of the unstable manifold of a UV fixed point 
of the $\mu$-flow. As such it is an integral part of the generalized 
notion of renormalizability aimed at.  
For a counterterm renormalization to be consistent therewith 
{\it non-minimal} subtraction terms need to be included 
to ensure that the bare quantities (associated with scale $\Lambda$)
and the renormalized quantities (associated with scale $\mu$) 
coincide when $\Lambda =\mu$. Through the finite terms 
this innocuous ``matching principle'' \cite{FPPT} affects the 
parts of the beta functions driven by powerlike divergences. 
We shall aim at this 
 ``strict Wilsonian one-loop renormalizability''.  

For a scalar QFT on \FL backgrounds we find that any beyond-quartic,
non-conformal self-interaction requires the inclusion of an 
infinite tower of potentials (\ref{pot}) 
in order to meet (\ref{oneloopstrict}). Despite the proliferation 
of couplings only $6$ turn out to be relevant or marginal (in $d\!=\!3$) 
and we compute the associated beta functions.


\newsubsection{Ultraviolet Divergences of Hadamard states}

From (\ref{FLfrg3}) one sees that the one-loop effective 
action is determined by the diagonal $G_{k,0}[\vp](t,t;p)$ of the tree 
level Green's function $G_{k,0}$. We transition to conformal time gauge 
$N(t) = a(t)$, and write $\eta =t$ for the so-defined time 
variable. The one-loop flow equation from (\ref{FLfrg3}) then reads   
\begin{equation}
\label{fl1l1}
k\dd_k \Gamma_{k,1}[\vp] = \frac{i}{2}\int\!\!d \eta d^dx\,a(\eta)^{d+1}\!  
\int \!\frac{d^dp}{(2\pi)^d} k  \dd_k R_k(\eta,p)G_{k,0}[\vp](\eta,\eta,p)\,.
\end{equation}
Here, the Green's function satisfies the first equation in (\ref{FLfrg3}) 
with $\cD_{t,x} = a^{-2d} (a^{d-1} \dd_{\eta})^2 + a^{-2} p^2 + W$,
$\delta(\eta,\eta') = a^{-d-1}\delta(\eta\!-\!\eta')$, and 
$W(\eta):=\sum_{n\geq 0}\!\nnU{n}''(\vp(\eta))R(\eta)^n$. 
As in (\ref{sfrgPT4}) one can integrate (\ref{fl1l1}) between a 
reference scale $\mu$ and an ultraviolet scale $\Lambda$
to obtain
\begin{equation}
\label{fl1l5}
\Gamma_{\mu,1}[\vp] = \Gamma_{\Lambda,1}[\vp] - 
\frac{i}{2} \int\!\!d \eta a(\eta)^{d+1} d^dx \!
\int \!\frac{d^dp}{(2\pi)^d} \int_{\mu}^{\Lambda} dk  
\dd_k R_k(\eta,p)
G_{k,0}[\vp](\eta,\eta;p)\,.
\end{equation}
Note that for a $k$-independent $\cD$ one can interpret 
$\dd_k R_k G_{k,0}$ as $\dd_k \ln G_{k,0}$, so that the $k$ integral 
leads to the expected `tracelog' structure.  
One loop renormalizability amounts to the existence of
coupling and field renormalizations such that the $\Lambda \ra \infty$ 
limit exists. We shall take up this task in Section 3.2.
A prerequisite is the isolation of the parts of the  
$k$ integral in (\ref{fl1l5}) that diverge as $\Lambda \ra \infty$.

It is convenient to transition to a genuine Schr\"{o}dinger
operator problem as in (\ref{FLhad10}), (\ref{FLhad11}). 
Defining $\mcG_{k,0}[\vp](\eta,\eta';\wp):= [a(\eta)a(\eta')]^{\frac{d-1}{2}}
G_{k,0}[\vp](\eta,\eta';k\wp)$, $\wp = p/k$,  
one has 
\ba
\label{fl1l3}
&\nspace & \big\{\dd_\eta^2 + k^2 \om_2(\eta,\wp)^2 + \om_0(\eta)^2 \big\}
\mcG_{k,0}[\vp](\eta,\eta';\wp)= \delta(\eta\!-\!\eta')\,,
\nonum
&\nspace & \omega_2 (\eta,\wp)^2:=\wp^2+a(\eta)^2 
r\Big(\frac{\wp^2}{a(\eta)^2}\Big)\,,\quad
\omega_0 (\eta)^2:= a(\eta)^2\Big[W(\eta)- \frac{d\!-\!1}{4d} R(\eta)\Big]\,.
\ea
Here we used $S(\eta) = - \frac{d-1}{2d} a(\eta)^2 R(\eta)$, 
obtained by rewriting the definition from (\ref{FLhad10}) via  
(\ref{RFL}). Only the diagonal $\mcG_{k,0}[\vp](\eta,\eta;\wp)$ 
$= i|\chi_k(\eta,k\wp)|^2$ enters the flow equation.   
Moreover, the diagonal is governed by the Gelfand-Dickey 
equation (\ref{FLhad11}) with $t = \eta, N= a$, and 
$S = - \frac{d-1}{2d} a^2R$. Its solution $\mcG_{k,0}(\eta,\wp):= 
\mcG_{k,0}[\vp](\eta,\eta;\wp)$ enters the flow equation 
\be
\label{fl1l7}
\dd_k \Gamma_{k,1}[\vp] = \frac{ 2 i k^{d+1}}{ (4\pi)^{d/2} \Gamma(d/2)} 
\int\!\!d \eta \,a(\eta)^2 d^3x 
\int_0^\infty\!\!d\wp\, \wp^{d-1}
\big[r(\wp^2/a^2)- (\wp^2/a^2)  r'(\wp^2/a^2)] \mcG_{k,0}(\eta,\wp)\,.
\ee
Since $\mcG_{k,0}(\eta,\wp) = i |\chi_k(\eta,k \wp)|^2$ and 
$\chi_k(\eta,p)$ solves $[\dd_{\eta}^2 + \om_0(\eta)^2 + 
k^2 \om_2(\eta,\wp)^2] \chi_k(\eta,p)=0$, 
the standard adiabatic iteration \cite{ParkerTomsbook} applies. The resulting partial sums are, however, only provenly asymptotic to an exact solution for large $k$ \cite{Luders} (this is the same argument as in Remark (iii) following the result \eqref{FLhad7}, \eqref{FLhad8} with $k$ playing the role of $p$; 
see also the comment after (\ref{FLhad12})). Accordingly, 
it is advantageous to  apply the generalized resolvent 
expansion (\ref{FLhad7}), (\ref{FLhad8}) to directly extract the coefficients of $k^{-2n-1}$. The latter expansion applies with
$\dd_{\eta}$ as the basic derivative, the indicated 
$\om_0,\om_2$, and $k^2$ as the generalized resolvent parameter. 
Note that for $r \equiv 0$ the standard resolvent expansion would 
apply, just as it does for the $N = a$ form of (\ref{FLhad11}) 
without the $r$ term. With the $r$ term the generalized resolvent 
expansion is indispensable, however. It reads 
\be
\label{fl1l8}
\mcG_{k,0}(\eta,\wp)\asymp \frac{i}{2\omega_2 (\eta,\wp)k}
\Big\{1+\sum_{n\geq 1}(-)^n\bar{G}_n(\eta,\wp)k^{-2n}\Big\}\,,
\ee
where the coefficients are uniquely defined by   
\be
\label{fl1l9} 
\bar{G}_n = \!\!\!\sum_{j,l\geq 0, j+l =n-1} \!\bigg\{ 
\frac{1}{4} \frac{\bar{G}_j}{\omega_2} 
\partial_{\eta}^2 \Big( \frac{\bar{G}_l}{\omega_2} \Big) - \frac{1}{8} 
\partial_{\eta} \Big( \frac{\bar{G}_j}{\omega_2} \Big) 
\partial_{\eta} \Big( \frac{\bar{G}_l}{\omega_2} \Big) + 
\frac{1}{2} \frac{\omega_0^2}{\omega_2^2} \bar{G}_j \bar{G}_l \bigg\} 
-\frac{1}{2} \sum_{j,l\geq 1, j+l =n} \bar{G}_j \bar{G}_l,
\ee
with $\bar{G}_0 =1$ and $\om_0, \om_2$ from (\ref{fl1l3}). 
The explicit expressions (\ref{FLhad9}) carry over with $'$ now
simply $\dd_{\eta}.$ Ultimately, it is the uniqueness ensured by the 
explicit recursion (\ref{fl1l9}) that underlies the universality 
of the {FL-sFRG}'s large $k$ behavior. Note also that 
(\ref{fl1l9}) is dimension independent; it is only upon insertion of 
(\ref{fl1l8}) into (\ref{fl1l7}) that the divergence structure 
depends on $d$.

For definiteness we now specialize to $d\!=\!3$. Then only the $n=0,1,2$ 
terms in \eqref{fl1l8} give rise to divergences. Evaluating the integral 
$-\int_{\mu }^\Lambda dk \,\dd_k \Gamma_{k,1}^{\text{div}}[\vp] $ 
yields the UV-divergent contribution to the one-loop effective action 
\be
\label{fl1l10b}
\Gamma_1^\text{div}[\vp] = \frac{1}{(4\pi)^2} \int\! d\eta dx\, 
a(\eta)^4\Big\{q_0(\Lambda^4-\mu^4)+g_1(\eta)(\Lambda^2-\mu^2) 
+g_2(\eta)\ln(\Lambda/\mu)\Big\}\,.
\ee
Here we keep the terms from the lower integration boundary, 
see remark (i) below. For the coefficients it is convenient to 
redefine $\wp=a(\eta)\varrho$ and 
$G_j(\eta,\varrho):=\bar{G}_j(\eta,\wp)|_{\wp=a(\eta)\varrho}$%
\footnote{It is to be understood that the redefinition $\wp=a(\eta)\varrho$ 
is performed {\itshape after} all the time derivatives in \eqref{fl1l9} 
have been taken.}.  
Then
\ba
\label{fl1l10}
q_0\is \int_0^\infty\!\! d\varrho \,\varrho^2
\frac{ [r(\varrho^2)-\varrho^2 r'(\varrho^2)]}{[\varrho^2 +r(\varrho^2)]^{1/2}}\,,
\nonum
g_1(\eta)\is-2\int_0^\infty\!\! d\varrho \,\varrho^2
\frac{ [r(\varrho^2)-\varrho^2 r'(\varrho^2)]}{[\varrho^2 +r(\varrho^2)]^{1/2}}
G_1(\eta,\varrho )\,,
\nonum
g_2(\eta)\is4\int_0^\infty\!\! d\varrho \,\varrho^2
\frac{ [r(\varrho^2)-\varrho^2 r'(\varrho^2)]}{[\varrho^2 +r(\varrho^2)]^{1/2}}
G_2(\eta,\varrho )\,.
\ea
Inserting $G_1,G_2$ from (\ref{FLhad9}) one finds  
\ba
\label{fl1l11}
g_1\is \breve{q}R-q_1(W-R/6)
\nonum
g_2 \is\frac{1}{2}(W-R/6)^2-\frac{1}{6}\nabla^2(W-R/6) 
\nonum 
&+& B_1\frac{a^{(4)}}{a^5}+B_2\frac{a^{(1)}a^{(3)}}{a^6}+
B_3\frac{a^{(2)\,2}}{a^6}+B_4\frac{a^{(1)\,2}a^{(2)}}{a^7}+B_5\frac{a^{(1)\,4}}{a^8}
\nonum
&+&\Big(B_6\frac{R}{6}+B_7\frac{a^{(1)\,2}}{a^4}\Big)(W-R/6)
+B_8\frac{a^{(1)}}{a^3}\dd_\eta(W-R/6)\,.
\ea
Here $\breve{q},q_1,B_1,\ldots,B_8$ are real constants that depend  
on the regulator $r$ via the $\varrho$-integrals in 
\eqref{fl1l10}. The $\eta$-dependence is carried by $W,R,a$, where 
we wrote $a^{(n)}:=\dd_\eta^na$ and $\nabla^2 = -a^{-4} \dd_{\eta} (a^2 \dd_{\eta})$,
see (\ref{sfrgPT5}).   

As may be expected from our use of a purely spatial 
regulator, the structure of the logarithmic divergence from the 
second and third lines of $g_2$ above has a non-covariant form. 
Remarkably, it follows from the properties of the regulator function 
(\ref{reg1}) that  $g_2$ can be ``covariantized'' and expressed 
solely in terms of 
\ba
\label{fl1l12}
&&R= \frac{6a^{(2)}}{a^3}\,,\quad	
\nabla^2 R= -6\frac{a^{(4)}}{a^5}+24 \frac{a^{(1)}a^{(3)}}{a^6}
+18\frac{a^{(2)\,2}}{a^6}-36\frac{a^{(1)\,2}a^{(2)}}{a^7 }
\nonum
&&-R_{\mu\nu  }R^{\mu\nu}+\frac{1}{3}R^2
= -12\frac{a^{(1)\,4}}{a^8}+12 \frac{a^{(1)\,2}a^{(2)}}{a^7}
=\frac{4}{a^4}\dd_\eta \Big(\frac{a^{(1)\,3}}{a^3}\Big)\,,
\ea
for generic regulator functions $r(x)$. This is the content of the 
Lemma A in Appendix A.

Specifically, it follows from the explicit expressions \eqref{fl1l12} 
for the curvature invariants together with Lemma A$(a),(b) $
that
\ba
\label{Bcoeff1} 
&&B_1\frac{a^{(4)}}{a^5}+B_2\frac{a^{(1)}a^{(3)}}{a^6}
+B_3\frac{a^{(2)\,2}}{a^6}+B_4\frac{a^{(1)\,2}a^{(2)}}{a^7}+B_5\frac{a^{(1)\,4}}{a^8}
\nonum
\is  b_1 \nabla^2R+b_2\Big[-R_{\mu\nu  }R^{\mu\nu}+\frac{1}{3}R^2\Big]\,.
\ea
Moreover, Lemma A($c$) entails that the final line of 
$g_2(\eta)$ in \eqref{fl1l11} may be expressed in terms of a single 
coefficient $b_3:=B_8$,
\be
\label{Bcoeff2} 
\Big(B_6\frac{R}{6}+B_7\frac{a^{(1)\,2}}{a^4}\Big)(W-R/6)
+B_8\frac{a^{(1)}}{a^3}\dd_\eta(W-R/6)= b_3a^{-4}\dd_\eta 
\big[aa^{(1)}(W-R/6)\big]\,.
\ee
Explicit expressions for $\breve{q}, q_1$ and $ b_1,b_2,b_3$ 
are relegated to Appendix A. All but the first term of $g_2(\eta)$ 
are thus total derivatives. Upon $\int d\eta d^3x a(\eta)^4$ 
integration in \eqref{fl1l10b} these evaluate to boundary terms, which 
we omit in accordance with (\ref{action2}). In summary, up to boundary 
terms, the UV-divergent correction to the one-loop effective action is 
$\hbar\Gamma_1^\text{div}[\vp]= \int \!d\eta dx \,a^4L_1^{\text{div}}$, with 
\begin{equation}
\label{fl1l15}
L_1^{\text{div}} = \frac{\hbar}{(4\pi)^2}\Big\{q_0(\Lambda^4-\mu^4)
+\big[ \breve{q}R-q_1(W\!-\!R/6)\big](\Lambda^2-\mu^2)
+\frac{1}{2}(W\!-\!R/6)^2\ln (\Lambda/\mu)\bigg\},
\end{equation}
and $W(\eta)=\sum_{n\geq 0}\!\nnU{n}''(\vp(\eta))R(\eta)^n$. 

\underline{Remarks:}
\vspace{-3mm}

\begin{itemize}[leftmargin=8mm, rightmargin=-0mm]
\itemsep -1mm
\item[(i)] In (\ref{fl1l15}) we kept the finite terms from the lower 
integration boundary. This ensures $L_1^{\rm div}|_{\mu = \Lambda} =0$ 
and the bare action $L_{\Lambda}$ can be identified with the renormalized 
one $L_{\mu}$ at the UV scale $\mu = \Lambda$. The non-minimal 
subtraction implemented in the next subsection preserves this 
structure and after strict removal of the UV cutoff (\ref{oneloopstrict}) 
allows one to literally identify the renormalized quantities with 
those at scale $\mu$.  
\item[(ii)] Although the heat kernel is not defined for non-elliptic 
Hessians, the heat kernel coefficients remain well-defined. 
Often a `pseudo-heat kernel' is used as generating 
functional for the coefficients, obtained by formally replacing 
the heat kernel time $s$ with $s= i\tilde{s}$, treating $\tilde{s}$ 
as real; see e.g.~\cite{ParkerTomsbook}. The divergent part 
of the effective action can then be computed along the usual  
lines for any pseudo-Riemannian background metric 
$g_{\mu\nu}(y) dy^{\mu}dy^{\nu}$. In the notation of Section 2.1
one has $S^{(2)}(\vp) = \eps_g \cD$, with $\cD$ from (\ref{sfrgPT5}),  
and we normalize $L_1^{\rm div}$ by $\sqrt{\eps_g} \hbar \frac{1}{2} 
({\rm Tr} \ln \cD)_{\mu,\Lambda} 
\asymp \int\! d^Dy \sqrt{g(y)} \,L_1^{\rm div}$. 
Here the regularized `tracelog' enters, which for momentum 
type cutoffs $\mu \leq \Lambda$ can be defined  
in several equivalent ways \cite{FPPT,Codelloetal}.  
In $D=4$ only the diagonal heat kernel coefficients 
$E_0,E_2,E_4$ enter. 
The result can be expressed 
in terms of the Ricci scalar $R(g)$ and the square of the
Weyl tensor $\cC^2(g)$ only. Explicitly,
\ba 
\label{covdiv} 
-\eps_g L_1^{\rm div} \is \frac{\hbar}{2(4 \pi)^2} \Big\{ q_0^{\rm cov}
(\Lambda^4 -\mu^4) + 
q_1^{\rm cov} (\Lambda^2-\mu^2) E_2 + E_4 \ln (\Lambda^2/\mu^2) \Big\}\,,
\nonum
E_2 &\simeq & -{\cal U}''(\vp,R) + \frac{1}{6} R\,, 
\\[2mm]
E_4 & \simeq & \frac{1}{2} E_2^2 + \frac{1}{120} \cC^2 
 \simeq  
\frac{1}{2} \Big( {\cal U}''(\vp,R) - \frac{1}{6} R\Big)^2  
+ \frac{1}{120} \cC^2\,. 
\nonumber
\ea  
Here `$\simeq$' denotes equality modulo total derivative terms
and the constants $q_0^{\rm cov},q_1^{\rm cov}$ depend on the choice 
of profile function entering the covariant regularization  
\cite{FPPT,Codelloetal}. The
$\mu$-dependent terms are kept with same rationale as in (i).  
For a conformally flat metric like $g_{\mu\nu}^{\rm FL}$ 
the $\cC^2$ term vanishes and (\ref{covdiv}) has the same 
structure as (\ref{fl1l15}). The coefficient of the $\ln \Lambda/\mu$ 
term is universal as expected, while the regulator dependent constants 
$q_0^{\rm cov}/2$ vs $q_0$ and $q_1^{\rm cov}/2$ vs $q_1$ will 
in general differ.  
Notably, (\ref{fl1l15}) contains an extra  term $\breve{q}R$
which is induced by the $a(t)$ dependence of the spatial modulator 
in (\ref{reg1}). It will lead 
to an additional renormalization of Newton's constant later on.
\item[(iii)] One can check that the $\breve{q}R$ term in (\ref{fl1l15}) 
is indeed induced by the $a$ dependence of the spatial modulator 
and not by the use of the generalized resolvent expansion (\ref{FLhad7}),
(\ref{FLhad8}) by pursuing a hybrid approach.  In the situation 
at hand the pseudo-heat kernel $K_{\tilde{s}}(t,x;t',x')$ can also be 
spatially Fourier transformed and its Fourier kernel 
$K_{\tilde{s}}(t,t';p)$ then satisfies the defining relations    
\begin{equation}
\label{FLphk2} 
\Big( i \frac{\dd}{\dd \tilde{s}} - \cD_p \Big) 
K_{\tilde{s}}(t,t';p) =0\,, \quad 
\lim_{\tilde{s} \ra 0} K_{\tilde{s}}(t,t';p) = a(t)^{-d} \delta(t,t')\,, 
\end{equation}  
with $\cD_{t,p}$ from (\ref{FLfrg3}). In cosmological time gauge 
$N(t) =1$ one can use the Gelfand-Dickey coefficients 
$\bar{G}_n$ to compute the coefficients $E_n^{\rm GD}$ induced by an 
asymptotic expansion of $K_{\tilde{s}}(t,t;p)$. With matched 
normalizations one finds $E_0^{\rm GD}(t) = 1$ and 
\begin{eqnarray}
\label{FLphk3}
E_2^{\rm GD}(t) \is  - W(t) + 
\frac{\ddot{a}}{a} + \frac{\dot{a}^2}{a^2}\,, 
\nonumber
\\
E_4^{\rm GD}(t) \is \frac{1}{2} W(t)^2 + \frac{1}{6} \ddot{W} - W
\Big(\frac{\ddot{a}}{a} + \frac{\dot{a}^2}{a^2} \Big) + 
\frac{1}{2} \dot{W} \frac{\dot{a}}{a}  
\\
&+& \frac{1}{2} \frac{\dot{a}^4}{a^4} + 
\frac{29}{15} \frac{\dot{a}^2 \ddot{a}}{a^3} + \frac{3}{10} 
\frac{\ddot{a}^2}{a^2} - \frac{3}{5} \frac{\dot{a} a^{(3)}}{a^2} 
- \frac{1}{5} \frac{a^{(4)}}{a}\,.
\nonumber
\end{eqnarray}
Discarding boundary terms in the integrals the expressions 
(\ref{FLphk3}) lead to the same divergence structure as (\ref{covdiv}).   
\end{itemize}

\newsubsection{Counterterm renormalization and one-loop beta functions} 

Next, we introduce the general structure of the counterterm  Lagrangian. 
Throughout we denote bare quantities (diverging as $\Lambda \ra \infty$) 
by a subscript $\Lambda$ while renormalized quantities carry no subscript 
but tacitly refer to a finite renormalization scale $\mu$. Both are as 
usual related by a renormalization constant where the subscript 
$\Lambda$ is omitted. Writing 
\begin{equation}
\label{flr1}
-L_\Lambda=-\frac{1}{2}\tilde{\chi}_\Lambda\nabla^2 \tilde{\chi}_\Lambda 
+ \sum_{n\geq 0} \nnU{n}_\Lambda(\tilde{\chi}_\Lambda)R^n\,,
\end{equation}
the bare field $\tilde{\chi}_\Lambda$ is related to the renormalized 
one $\tilde{\chi}$ by $\tilde{\chi}_\Lambda=Z^{1/2}\tilde{\chi}$. 
In the background field formalism the wave function renormalization 
constant of the fluctuation field $Z_{\chi}$ and that of the background
field $Z_{\vp}$ will coincide as long as the splitting symmetry 
$\chi \ra \chi + \zeta, \vp \ra \vp - \zeta$, is preserved.    
Since we assume the latter to be the case (see the discussion after 
(\ref{sfrg2})) we can identify $Z_{\chi} = Z_{\vp} = Z$, set the 
mean fluctuation field $\phi$ to zero, and renormalize the 
background field according to $\vp_{\Lambda} = Z^{1/2} \vp$. 
Using the parameterization (\ref{pot}) the bare potentials 
$\nnU{n}_\Lambda(\vp_\Lambda)$ then expand 
according to
\ba
\label{flr2}
&&\nnU{n}_\Lambda(\vp_\Lambda)=\sum_{j\geq 0}\frac{1}{(2j)!}
\nnu{n}_{2j,\Lambda}\vp_\Lambda^{2j}=\sum_{j\geq 0}\frac{1}{(2j)!}
\nnZ{n}_{2j}\nnu{n}_{2j}\vp^{2j}\,,
\nonum 
&& Z^j\nnu{n}_{2j,\Lambda}=\nnZ{n}_{2j}\nnu{n}_{2j}\,,\quad n\geq 0\,,\,j\geq 0\,,
\ea
where all the renormalization constants are $1+O(\hbar)$, and 
hence $L_\Lambda=L+\hbar L_\text{ct}$ defines the counter term action.

We now examine the general structure of the counterterm Lagrangian. 
It is clear from \eqref{fl1l15} that the one-loop divergence structure 
does not necessitate a non-trivial wave function renormalization, 
i.e.~$Z=1+O(\hbar^2)$. Thus the bare and renormalized couplings in 
\eqref{flr2} are related by 
\be
\label{flrg1}
\nnu{n}_{2n,\Lambda}=\nnZ{n}_{2j}\nnu{n}_{2j}\,,\quad n\geq 0\,,\,j\geq 0\,,
\ee
and we choose  appropriate Ans\"{a}tze for the renormalization constants:
\ba
\label{flrg2}
\nnZ{0}_0\is 1 + \frac{\hbar}{(4 \pi)^2} 
\Big\{\nnz{0}_{0,0} + \nnz{0}_{0,1} \ln \Lambda/\mu  
+ \nnz{0}_{0,2} (\Lambda/\mu)^2+ \nnz{0}_{0,4} (\Lambda/\mu)^4\Big\}\,,
\nonum
\nnZ{0}_{2j}\is 1 + \frac{\hbar}{(4 \pi)^2} 
\Big\{\nnz{0}_{2j,0} + \nnz{0}_{2j,1} \ln \Lambda/\mu  
+ \nnz{0}_{2j,2} (\Lambda/\mu)^2\Big\}\,,
\quad j \geq 1\,,
\nonum
\nnZ{n}_{2j} \is 1 + \frac{\hbar}{(4 \pi)^2} 
\Big\{\nnz{n}_{2j,0} + \nnz{n}_{2j,1} \ln \Lambda/\mu  + \nnz{n}_{2j,2} 
(\Lambda/\mu)^2\Big\}\,, \quad n \geq 1,\,j\geq 0\,.
\ea
In all cases $O(\hbar^2)$ corrections are implicit. The Ans\"{a}tze 
(\ref{flrg2}) contain non-divergent terms, which are however 
related to the coefficients of the powerlike terms by the ``matching 
principle'' of \cite{FPPT}: We seek to identify the renormalized 
couplings with the bare ones at scale $\mu$. This constrains the  
coefficients of the non-logarithmic terms as follows 
\ba
\label{flrg3} 
&& \nnz{0}_{0,0} + \nnz{0}_{0,2} + \nnz{0}_{0,4} =0\,, \quad 
\nnz{0}_{2j,0} + \nnz{0}_{2j,2} =0\,, \quad j
\geq 1\,, 
\nonum
&& \nnz{n}_{2j,0} + \nnz{n}_{2j,2} =0\,, \quad n \geq 1,\,j\geq 0\,.
\ea
The counterterm action thus reads
\ba
\label{flrg4}
L_{\Lambda} \is  L -  \frac{\hbar}{(4\pi)^2} 
\Big\{ \mu^{-4} \nnu{0}_0 \nnz{0}_{0,4} \,\Lambda^4 
+ F_2 \,\Lambda^2 + F_4 \ln(\Lambda/\mu) 
+ O(\Lambda^0)\Big\} \,.
\nonum
F_2 \is \mu^{-2} \sum_{n \geq 0}\sum_{j\geq0}\frac{1}{(2j)!} 
\nnz{n}_{2j,2}\!\nnu{n}_{2j}\vph^{2j} R^n \,,
\nonum
F_4\is \sum_{n \geq 0}\sum_{j\geq0}\frac{1}{(2j)!} 
\nnz{n}_{2j,1}\!\nnu{n}_{2j}\vph^{2j} R^n \,.
\ea
Then, the condition $L_\Lambda+L_1^\text{div}\stackrel{!}{=}\text{finite}$ 
as $\Lambda\to \infty$ amounts to three conditions: 
$\nnu{0}_0\nnz{0}_{0,4} \stackrel{!}{=} \mu^4 q_0$, which fixes 
${}^0z_{0,4}$, and 
\be
\label{flrg6}
F_2 \stackrel{\displaystyle{!}}{=}  \breve{q}R-q_1(W-R/6)\,,
\quad 
F_4 \stackrel{\displaystyle{!}}{=}  \frac{1}{2}(W-R/6)^2\,.
\ee
The occurrence of $W\!-\!R/6$ in (\ref{flrg6}) makes it convenient 
to define couplings $\nbu{1}_{2} :=\nnu{1}_{2}-\frac{1}{6}$, 
$\nbu{n}_{2j} := \nnu{n}_{2j}$, $j \neq 1$, and associated potentials  
\begin{equation}
\label{flrg7}
{}^{1}\bar{U}(\vp) := {}^{1}U(\vp)-\frac{1}{12}\vp^2\,,
\quad \,^{n}\bar{U}(\vp):=\,^{n}U(\vp)\,,\quad n\neq 1\,.
\end{equation} 
Re-expressing  (\ref{flrg6}) in terms of these yields
\ba
\label{flrg8}
F_2&\!\overset{\displaystyle{!}}{=}\!& -q_1\nbU{0}''(\vp)+\big[\breve{q}
-q_1\nbU{1}''(\vp)\big]R-q_1\sum_{n\geq 2}\nbU{n}''(\vp)R^n\,,
\nonum
2F_4&\!\overset{\displaystyle{!}}{=}\!&\nbU{0}''(\vp)^2
+2\nbU{0}''(\vp)\nbU{1}''(\vp)R + 
\sum_{n\geq 2}\Big[\sum_{n_1+n_2=n}\,^{n_1}\bar{U}''(\vp)\,
\,^{n_2}\bar{U}''(\vp)\Big]R^n\,.
\ea

We proceed by computing the renormalization constants $\,^nZ_{2j}$ 
by matching powers of $\vp^{2j}R^n$ in (\ref{flrg8}). 
From the $F_2$-relation one obtains
\ba
\label{flbt1}
\nnz{0}_{2j,2}\nnu{0}_{2j}\is -\mu^2q_1\nbu{0}_{2j+2}\,,\quad j\geq 0\,,
\nonum
\nnz{1}_{0,2}\nnu{1}_{0}\is \mu^2(\breve{q}-q_1\nbu{1}_{2})\,,
\nonum
\nnz{1}_{2j,2}\nnu{1}_{2j}\is -\mu^2q_1\nbu{1}_{2j+2}\,,\quad j\geq 1\,,
\nonum
\nnz{n}_{2j,2}\nnu{n}_{2j}\is -\mu^2q_1\nbu{n}_{2j+2}\,, 
\quad n\geq 2\,,j\geq 0\,,
\ea
while the logarithmic divergence  yields
\ba
\label{flbt2}
\nnz{n}_{2j,1}\nnu{n}_{2j}\is \frac{1}{2}	
\sum_{\scriptstyle n_1+n_2=n\atop\scriptstyle n_1,n_2\geq 0}
\sum_{\scriptstyle j_1+j_2=j\atop\scriptstyle j_1,j_2\geq 0}
\frac{(2j)!}{(2j_1)!(2j_2)!}\nbu{n_1}_{2j_1+2}\nbu{n_2}_{2j_2+2}\,.
\ea
These solutions can now be inserted into the Ans\"{a}tze (\ref{flrg2})  
and produce expressions for the ${}^n\!Z_{2j}$, $n,j \geq 0$, relating the 
bare and the renormalized couplings via (\ref{flrg1}). Instead of presenting 
the ${}^n\!Z_{2j}$ we insert them into (\ref{flr2}) and note the 
resulting relations between the bare and the renormalized potentials   
\ba 
\label{flpotren1} 
\nnU{0}_{\Lambda}(\vp_{\Lambda}) \is \nnU{0}(\vp) + 
\frac{\hbar}{(4\pi)^2} \Big\{ (\Lambda^4\!-\!\mu^4) q_0 - 
(\Lambda^2\!-\!\mu^2) q_1 \nnU{0}''(\vp) + 
\ln(\Lambda/\mu) \frac{1}{2} \nnU{0}''(\vp)^2 \Big\}\,,  
\nonum
\nnU{1}_{\Lambda}(\vp_{\Lambda}) \is \nnU{1}(\vp) + 
\frac{\hbar}{(4\pi)^2} \Big\{ 
(\Lambda^2\!-\!\mu^2) [\breve{q} - q_1 \nbU{1}''(\vp)] + 
\ln(\Lambda/\mu) \nnU{0}''(\vp) \nbU{1}''(\vp) \Big\}\,,  
\nonum
\nnU{n}_{\Lambda}(\vp_{\Lambda}) \is \nnU{n}(\vp) + 
\frac{\hbar}{(4\pi)^2} \Big\{ 
-(\Lambda^2\!-\!\mu^2) q_1 \nbU{n}''(\vp) 
\nonum
&+& \frac{1}{2} 
\ln(\Lambda/\mu) 
\sum_{\scriptstyle n_1+n_2=n\atop\scriptstyle n_1,n_2\geq 0}
\nbU{n_1}''(\vp) \nbU{n_2}''(\vp) \Big\}\,, \quad n \geq 2\,. 
\ea
Of course, $O(\hbar^2)$ corrections are implicit in all relations. 
\noindent

\underline{Remarks:}
\vspace{-3mm} 

\begin{itemize}[leftmargin=8mm, rightmargin=-0mm]
\itemsep -1mm
\item[(i)] The renormalization of ${}^0U$ is the same as in flat space
and is independent of the non-minimal potentials. In contrast, the 
renormalization of each ${}^nU,\,n\geq 1$, depends on 
${}^lU$, $l=0,\ldots,n\!-\!1$. 

\item[(ii)] This upward cascade in the order of the Ricci scalar halts 
only in two special cases. One is the standard case of a $\vp^4$-theory, 
where one may set $\!\nnu{n}_j=0$ for all $n\geq1,\,j\geq 0$ except for 
$\xi=\!\!\nnu{1}_2$, which corresponds to non-minimal coupling 
$\frac{1}{2} \xi\vp^2 R$ \cite{ParkerTomsbook}. The other is 
the ``conformal sector'', where
\begin{equation} 
\label{flrg10}
\nnU{1}(\vp)=\frac{1}{12}\vp^2\,,\quad \nnU{n}(\vp)\equiv0\,,\quad n\geq 2\,,
\end{equation} 
is consistent for {\it generic} even scalar potential $\!\nnU{0}(\vp)$. 
\item[(iii)] For a generic scalar potential $\nnU{0}(\vp)$ and non-conformal 
coupling $\!\nnu{1}_2\neq 1/6$, the need to include Ricci-couplings 
of all orders in the action \eqref{action2}, \eqref{flr1}, follows from 
the structure of the logarithmic divergence. Consider, for example, 
a  sextic potential $\!\nnU{0}(\vp)=
\frac{1}{2}m^2\vp^2+\frac{1}{4!}\lambda\vp^4+\frac{1}{6!}g\vp^6$ 
for non-conformal coupling. Then the logarithmic divergence in 
\eqref{fl1l15} contains a term proportional to $(\!\!\nnu{1}_2-1/6)g\,
\vp^4 R$, whose absorption requires $\!\nnu{1}_4\neq 0$ in the action 
\eqref{flr1}, \eqref{flr2}. This in turn leads to the logarithmic divergence
proportional to $(\!\!\nnu{1}_2-1/6)\!\nnu{1}_4\,\vp^2 R^2$, 
which necessitates $\!\nnu{2}_2\neq 0$, and so on.
\item[(iv)] In an attempt to avoid the infinite tower of Ricci couplings, 
one might try to remove some of the $R^n$ divergences in \eqref{fl1l15} 
by a non-linear field renormalization
\begin{equation}
\vp_\Lambda = \vp +  \frac{\hbar}{(4\pi)^2} \ln (\Lambda/\mu) 
\,\zeta(\vp) R+ O(\hbar^2)\,,\quad 
\zeta(\vp) =  \sum_{n\geq 0}\,^n\zeta(\vp)R^n\,.
\end{equation}
However, since
\begin{equation} 
S_\Lambda[\vp_\Lambda] = S_\Lambda[\vp]+\frac{\hbar}{(4\pi)^2}
\ln(\Lambda/\mu)\int\!\!d\eta d^3x\,a^4\,
\frac{\delta S}{\delta \vp}[\vp]\zeta(\vp)+O(\hbar^2)\,,
\end{equation} 
such a field renormalization does not produce any counterterms 
for on-shell background fields with $ \delta S_\Lambda /\delta \vp=0$. 
When keeping the background off-shell, a non-trivial $\zeta(\vp)$ 
clashes with a standard kinetic term and would require further modifications.
\end{itemize}

The flow equations for the couplings are derived by differentiating 
the defining relations: $\text{bare coupling}=
\text{(renormalization constant)}\times\text{(renormalized coupling)}$ 
with respect to the scale $\mu$. For the original (usually dimensionful) 
couplings, the response is always $O(\hbar)$ so that only the explicit 
$\mu$-dependence needs to be taken into account. One can view the 
renormalized potentials $\nnU{n}(\vp) = \sum_{j \geq 0} \nnu{n}_{2j} 
\vp^{2j}/(2j)!$ as generating functionals for the couplings and
differentiate the relations (\ref{flpotren1}) with respect to 
$\mu$. This gives  
\ba 
\label{flpotren2}
\mu \dd_{\mu}\! \nbU{0} \is \frac{\hbar}{(4\pi)^2} \Big\{ 
4 \mu^4 q_0 - 2 \mu^2 q_1 \nbU{0}'' + \frac{1}{2} 
\big(\!\!\nbU{0}''\big)^2 \Big\}\,, 
\nonum
\mu \dd_{\mu}\! \nbU{1} \is \frac{\hbar}{(4\pi)^2} \Big\{ 2\mu^2 
\big( \breve{q} - q_1 \nbU{1}'' \big) + 
\nbU{0}'' \nbU{1}'' \Big\}\,, 
\\[2mm] 
\mu \dd_{\mu}\! \nbU{n} \is \frac{\hbar}{(4\pi)^2} \Big\{ -2\mu^2\, 
q_1 \nbU{n}'' + \frac{1}{2} \sum_{n_1+n_2=n}\!^{n_1}\bar{U}''(\vp)\,
\,^{n_2}\bar{U}''(\vp) \Big\}, \;\; n \geq 2\,.
\nonumber
\ea 
By expansion, the induced flows of all couplings $\nnu{n}_{2j}$, 
$n,j \geq 0$, can be obtained. 

Next, we transition to dimensionless 
counterparts of the dimensionful couplings and rewrite the flow 
equations in terms of them. In preparation, we recall that the 
couplings $\!\!\nnu{n}_{2j}$ have  mass dimension $4-2(n+j)$.  
Only $6$ have non-negative mass dimension and for orientation we
note their standard notations    
\ba
\label{flrg15}
\nnu{0}_0\is \frac{\Lambda_{\text{cosm}}}{\kappa}\,,\quad 
\nnu{0}_2= m^2\,,\quad\nnu{0}_4=\lambda\,,
\nonum
\nnu{1}_0\is -\frac{1}{\kappa}\,,\quad\quad \nnu{1}_2=\xi\,,\quad 
\;\;\nnu{2}_0 = \omega_0\,.
\ea
Generally, we denote the dimensionless counterparts of the 
couplings ${}^nu_{2j}$ by 
\be
\label{flbt6}
\nnv{n}_{2j}:=\mu^{-4+2(n+j)} \nnu{n}_{2j}\,, \quad n,j \geq 0\,,
\ee
and $\varepsilon:=\!\!\nnv{0}_0=\mu^{-4}\nnu{0}_0$ for the  
dimensionless vacuum energy. 
The redefinitions \eqref{flbt6} ensure that the dimensionless coupling-flow equations  obtained from \eqref{flpotren2} are all autonomous, i.e. carry no explicit $\mu$-dependence.
The resulting flow equations for 
the dimensionless couplings of $\nnU{0}$ read:
\ba
\label{flbt8}
\mu \frac{d}{d\mu}\varepsilon\is-4\varepsilon
+ \frac{\hbar}{(4\pi)^2}\Big\{4q_0-2q_1\nnv{0}_2
+\frac{1}{2}\nnv{0}_2^2\Big\}\,,
\nonum 
\mu \frac{d}{d\mu}\!\!\nnv{0}_{2j}\is (2j-4)\nnv{0}_{2j}
\\
&\!+\!&\frac{\hbar}{(4\pi)^2}\Big\{-2q_1\nnv{0}_{2j+2}
+\frac{1}{2}\sum_{\scriptstyle j_1+j_2=j \atop\scriptstyle  j_1,j_2\geq 0}
\frac{(2j)!}{(2j_1)!(2j_2)!}\nnv{0}_{2j_1+2}\nnv{0}_{2j_2+2}\Big\}\,,
\quad j\geq 1\,,
\nonumber
\ea
These equations are 
precisely those obtained on a flat spacetime, and are not closed as 
the one for $\nnv{0}_{2j}$ also invokes  $\nnv{0}_{2j+2}$, $j\geq 0$. 
In order to solve them a truncation is required that sets all 
$\nnv{0}_{2j}$ to zero for all $j\geq j_0$, for some $j_0\in \N$. 
Truncation at order $j_0=4$ yields the fixed point
\begin{equation}
\label{flbt9}
\varepsilon^\ast = \frac{\hbar}{(4\pi)^2}q_0\,,\quad 
\!\!\nnv{0}_{2}^\ast =\!\!\nnv{0}_{4}^\ast=\!\!\nnv{0}_{6}^\ast=0\,,
\end{equation} 
consistent with the existence of only a Gaussian UV fixed point for a 
scalar theory in four spacetime dimensions. It is noteworthy, however, 
that $\varepsilon$ has a mass independent 
positive fixed point that depends only mildly on the choice of the 
regulator. This is also a feature of the effective potential approximation 
of the FRG, but it cannot be seen in dimensional regularization. 

For the non-minimal couplings we focus on the three which are 
power counting relevant or marginal, see (\ref{flrg15}). The 
equation for $\nnu{1}_0$ transcribes 
into a flow equation for the dimensionless Newton constant 
$\text{g}_N=\mu^2\kappa$ via $\!\nnv{1}_0=-1/(2\text{g}_N)$.
For the other two we retain the general definitions. The resulting 
flow equations are  
\ba
\label{flbt10}
\mu \frac{d}{d\mu}\text{g}_N\is 2\text{g}_N+\frac{\hbar}{(4\pi)^2}
\Big\{4\breve{q}\,\text{g}_N^2
+\big(\!\!\nnv{0}_2-2q_1\big)\big(\!\!\nnv{1}_2-1/6\big)\text{g}_N^2\Big\}\,,
\nonum
\mu \frac{d}{d\mu}\!\!\nnv{1}_2\is \frac{\hbar}{(4\pi)^2}
\Big\{\big(\!\!\nnv{1}_{2}-1/6\big)\nnv{0}_4
+\!\!\nnv{0}_2\nnv{1}_4-2q_1\nnv{1}_4\Big\}\,,
\nonum 
\mu \frac{d}{d\mu}\!\!\nnv{2}_0\is \frac{\hbar}{(4\pi)^2}
\Big\{\frac{1}{2}\big(\!\!\nnv{1}_{2}-1/6\big)^2
+\!\!\nnv{0}_2\nnv{2}_2-2q_1\nnv{2}_2\Big\}\,.
\end{eqnarray}
The $\text{g}_N$ flow equation admits a nontrivial fixed point 
that is best interpreted as one for the $1/\text{g}_N$ flow, 
\begin{equation} 
\label{flbt12}
1/\text{g}_N^\ast = - \frac{\hbar}{2(4\pi)^2}
\Big\{4\breve{q}+\big(\!\!\nnv{0}_2^\ast -2q_1\big)
\big(\!\!\nnv{1}_2^\ast-1/6\big)\Big\}\,,
\end{equation} 
where typically $\!\nnv{0}_2^\ast =0$ by \eqref{flbt9}. We note that 
compared to the covariant formulation, the beta function \eqref{flbt10} 
and the fixed point \eqref{flbt12} feature an additional contribution 
proportional to $\breve{q}$ arising from the time dependence of the spatial 
regulator  \eqref{fl1l1}. Since $\breve{q}>0$ this contribution (as is 
typical of matter) tends to drive the Newton coupling to negative values. 
However, in a full quantum gravity plus matter computation one expects 
that the quantum gravity contribution will turn $\text{g}_N^\ast$ 
positive again. Furthermore, it is noteworthy that the $\breve{q}$ 
term does not vanish for conformal coupling to matter.

This concludes our discussion of the one-loop renormalization flow.
As noted in the introduction it will be used to set boundary 
conditions for the FL-sFRG.

\newpage 
\newsection{Spatial cosmological EPA flow} 

Generally, the effective potential approximation (EPA) truncates 
the nonlocal effective action to a local functional
of the same form as the basic action but with a scale dependent 
potential Ansatz. This concept carries over to the FL-sFRG
(\ref{FLfrg1}), (\ref{FLfrg2}) straightforwardly. Since the 
(coefficient of the) kinetic term is taken to be $k$-independent  
the left hand side of (\ref{FLfrg1}) reduces to $-\int \! dt d^dx \,
{N} a^d \dd_k {\cal U}_k$, where ${\cal U}_k(\vp,R)$ is the 
scale  dependent effective potential to be determined.
As noted after (\ref{FLfrg3}) one has for the original action 
$S^{(2)}[\vp](t,t';p) = - \cD_p \delta(t,t')$. The $k$-dependent 
modification leads to $a^{-2d} (a^d {N}^{-1} \dd_t)^2 
+ a^{-2} p^2 + {\cal U}''_k$ as the differential operator entering 
the specialization of (\ref{FLfrg2}). Since we assume $\vp$ to 
be a function of $t$ only the spatial volume in (\ref{FLfrg1}) 
is overcounted on both sides and can be dropped. 
The temporal average can likewise be omitted. This is because 
the EPA flow equation is meant to be valid for arbitrary compact 
time intervals $[t_i,t_f]$, which enforces pointwise equality of 
the (continuous) integrands. Moreover, one typically seeks to 
study the flow of the potential rather than its spatio-temporal average; 
see Section 3.2.
The resulting spatial EPA ({\bf FL-sEPA}) 
flow equation is  
\ba 
\label{sepa1} 
&& \dd_k {\cal U}_k = -\frac{i\hbar}{2} \int\!\! \frac{d^dp}{(2\pi)^d} 
\,\dd_k R_k(t,p) G_k[\vp](t,t;p) \,,
\nonum
&& \big\{ a^{-2d} (a^d {N}^{-1} \dd_t)^2 
+ a^{-2} p^2 + R_k(t,p)+ {\cal U}''_k \big\} G_k[\vp](t,t';p) = 
\delta(t,t')\,.
\ea 
Here, ${\cal U}_k $ depends on $\vp(t)$ and the right hand side 
also carries an explicit time dependence through the background, which 
will be further discussed in Section 4.1.
As indicated, we continue to write $G_k$ for the Green's function 
entering, no confusion with the solution of the untruncated (\ref{FLfrg2}) 
should arise. Not incidentally (\ref{sepa1}) coincides 
with the one-loop equation in (\ref{FLfrg3}) upon  substitution of 
the scale dependent potential on the right hand side
$G_k[\vp](t,t';p) = G_{k,0}[\vp](t,t';p)|_{{\cal U''} \mapsto {\cal U}''_k}$. 
This will become relevant later on when examining large $k$
regime of the flow equation. In that case it is $G_{k,0}$'s large 
$k$ behavior that needs to be known as a functional of ${\cal U}''$,
and the limiting case of the flow equation arises upon substitution 
of the unknown ${\cal U}''_k$. A similar interplay holds in the 
dimensionless formulation introduced below for the small $k$ regime.

Since only the diagonal of the Green's function enters 
the flow equation it is advantageous to replace its defining relation 
by the associated Gelfand-Dickey equation for $G_k = G_k[\vp](t,t;p)$
\be
\label{sepa3} 
2 G_k (a^d {N}^{-1} \dd_t )^2 G_k - 
\big( a^d {N}^{-1} \dd_t G_k \big)^2 + 
4 a^{2d} \Big[ k^2 \Big( \frac{p^2}{ a^2 k^2} 
\!+\! r\Big(\frac{p^2}{a^2 k^2}\Big)\Big) + {\cal U}''_k 
\Big]G_k^2 =-1\,.
\ee   
Again, the solution is related to that in (\ref{FLhad5}) 
by $i |T_k(t,p)|^2 |_{W \mapsto {\cal U}''_k} = G_k[\vp](t,t;p)$. 
Of course $G_k[\vp]$ depends on $\vp$ only 
through ${\cal U}''_k(\vp)$ but the {\it explicit} ${\cal U''}_k(\vp)$ 
dependence is difficult to extract.
For generic scale factor $a$ this will only be possible via series 
expansions, either for large or for small $k$. Below we shall develop 
such series expansions, highlighting the structural difference between 
the universal large $k$ expansion and the inevitably state dependent 
small $k$ expansion. 

\newsubsection{Explicit time-dependence and dimensionless formulation}

A noteworthy feature of the flow equation (\ref{sepa1}) is its 
explicit time dependence not carried by $\vp(t)$. In the ultraviolet the 
explicit time dependence of the Green's function $G_{k,0}[\vp](t,t;p)$ 
will turn out to be solely carried by a local function of $R(t)$;
in particular ${\cal U}_k$ will for {\it large $k$} be of the form 
\begin{equation} 
\label{sepaUV}
{\cal U}_k(\vp,R) = \sum_{n \geq 0} \nnU{n}_k(\vp) R(t)^n\,. 
\end{equation}
For small $k$, on the other hand, ${\cal U}_k$ is expected to carry 
a nonlocal time dependence, so in general we write ${\cal U}_k(t,\vp)$
for the generalized effective potential sought.  In a Euclidean setting 
a nonlocal flow equation would be expected from heat kernel 
resummations; see e.g. \cite{Codelloetal} and the references therein. 
In the present context,
the infrared behavior is more drastically altered and lies
outside the realm of any heat-kernel methodology. 

To proceed, we shall expand the unknown time dependence in a
suitable basis of orthonormal polynomials. In order to retain 
contact to (\ref{sepaUV}) we use polynomials in $R(t)$ rather than polynomials in $t$. To this end, we assume the 
powers of $R(t)$ to be linearly independent. This is warranted, as for monotonically increasing $a(t)$ the Ricci scalar 
$R(t)$ given by (\ref{RFL}) is typically strictly monotonically 
decreasing in $t$ on bounded intervals, like 
$[t_i, t_f]$ in the present context. Next, we choose a nonnegative 
weight function $w$ with mass dimension +1, smooth with compact support 
in $[t_i,t_f]$. In particular, this renders the measure $dt w(t)$ 
dimensionless. In terms of it we define 
\be 
\label{ONB1} 
\sigma_i = \int\! dt w(t) R(t)^i\,,\quad i \geq 0\,,\quad 
D_l = \det(\sigma_{i+j})_{0 \leq i,j \leq l} \,,\;\;l\geq 0\,, 
\ee
and $\mathfrak{p}_0=\sigma_0^{-1/2} ,$ while for $l \geq 1$, 
\ba 
\label{ONB2} 
\mathfrak{p}_l(t) = \frac{1}{\sqrt{D_l D_{l-1}}} \det \begin{bmatrix}
\sigma_0 & \sigma_1 & \ldots & \sigma_{l-1} & 1\\
\sigma_1 & \sigma_2 & \ldots& \sigma_l & R(t) \\
 \vdots  &          &                 & \vdots \\
\sigma_{l-1} & \sigma_l & \ldots & \sigma_{2 l-2} & R(t)^{l-1} \\
\sigma_l & \sigma_{l+1} & \ldots & \sigma_{2 l-1} & R(t)^l 
\end{bmatrix} \,.
\ea 
Each $\mathfrak{p}_l(t)$ is a polynomial of degree $l$ in $R(t)$ 
with coefficients whose squares are rational functions of the 
{\it averages} of the Ricci powers. Hence $\mathfrak{p}_l$ is both 
a (nonlocal) functional of $R(\cdot)$ and a (local) function of
$R(t)$. Whenever needed we indicate this by writing $\pf_l[R](t)$ 
and $\sigma_i[R]$. Collectively the $\pf_l,l\geq 0$,  form a 
system of orthonormal polynomials 
\be 
\label{ONB3} 
\int \! dt w(t) \,\pf_l(t) \pf_{l'}(t) = \delta_{l,l'}\,. 
\ee
They are also invariant under constant rescalings of $R$
\be 
\label{ONB4} 
\pf_l[\lb R](t) = \pf_l[R](t)\,, \quad \lb >0\,,\;\; l \geq 0\,,
\ee
as can be seen by noting that all diagonals in the determinants 
have constant degrees. 
The proof of (\ref{ONB3}) specializes Theorem 2.1 of \cite{Szegobook} to 
powers of the Ricci scalar. Further, the set of these polynomials is 
closed in $[t_i,t_f]$ and a square integrable function 
$h \in L^2(wdt)$ can be expanded such that Parseval's 
identity  holds 
\be
\label{ONB5} 
h(t) = \sum_{l \geq 0} h_l \,\mathfrak{p}_l(t)\,,\quad 
h_l = \int\! dt w(t) \mathfrak{p}_l(t) h(t) \,,\quad 
\sum_{l \geq 0} |h_l|^2 = \int \! dt w(t) |h(t)|^2 \,.  
\ee    
Convergence holds in the $L^2$ sense, see Theorem 3.1.5 
of \cite{Szegobook}. This implies almost everywhere pointwise 
convergence of a subsequence of the partial sums. Later on, the 
functions $h$ considered will be at least continuous, so that 
pointwise convergence everywhere is ensured. Finally, since the 
measure $wdt$ is dimensionless and so are by (\ref{ONB4}) 
the $\mathfrak{p}_l$, the coefficients $h_l$ will have the same 
dimension as $h$.  

In the context of the FL-sFRG (\ref{sepa1}), we apply this expansion  
to parameterize the non-local time dependence of the generalized 
potential $\mcU_k$ expected to arise in the infrared part of the flow. 
This yields an alternative set of potentials ${}_l{\cal U}_k(\vp)$, 
$l \geq 0$, which in turn are expanded in powers of the field, 
\begin{equation}
\label{sepaIR1}
{\cal U}_k(\vp,t) = \sum_{l \geq 0} {}_l{\cal U}_k(\vp) \mathfrak{p}_l(t) \,,
\quad \quad 
{}_l{\cal U}_k(\vp) = \sum_{j \geq 0} {}_lu_{2j,k} \frac{\vp^{2j}}{(2j)!}\,. 
\end{equation}
In this parameterization ${\cal U}_k$ depends on the Ricci scalar 
also nonlocally through the $\mfp_l[R](t)$. As in (\ref{ONB2}) 
we do not display this dependence and just indicate the time 
dependence.

As seen in Section 3, in the ultraviolet the potentials ${}^nU(\vp)$, 
$n \geq 0$, entering (\ref{sepaUV}) are preferred and for large $k$
both sets are related by 
\begin{equation}
\label{sepaIR2} 
{}_l{\cal U}_k(\vp) = \sum_{n \geq l} {}^nU_k(\vp) \int\!\! dt w(t) 
\,\mathfrak{p}_l(R(t))\, R(t)^n\,,
\end{equation} 
and similarly for the couplings. 
Here we used that $\mathfrak{p}_l$ is orthogonal to any monomial 
of degree less than $l$. Generally, the ${}_l{\cal U}_k(\vp)$ may be 
viewed as potentials whose coefficients are determined nonlocally 
in time.  

We now first implement the transition to dimensionless variables for 
the ultraviolet potentials in (\ref{sepaUV}). In order to render 
$R(t)$ dimensionless, while preserving the autonomous nature of 
the coupling flow in the ultraviolet, we introduce 
a mass scale $\kappa(k)$ that interpolates monotonically between 
$k$ in the ultraviolet and some inverse background length scale 
in the infrared: 
\be
\label{kappadef}
\kappa, \kappa': \R_+ \ra \R_+\,,\quad \kappa(k)- k \in \cS(\R_+)\,,
\quad 0< \lim_{k \ra 0} \kappa(k) < \infty\,,
\ee
with $\cS(\R_+)$ defined in the paragraph following \eqref{reg1}. 
In particular, $\lim_{k \ra 0} [ k \dd_k \ln \kappa(k)] = 0$.  The property 
$\kappa(k)\sim k$ for large $k$ will ensure in Section 4.2 the autonomy 
of the dimensionless coupling flow, mirroring \eqref{flbt6} ff. 
In the infrared, this cannot expected to be true, and $\kappa(0)$ is 
needed to set an external, finite, infrared mass scale. As an external 
scale, it is natural to relate $\kappa(0)$ to a background scale, 
e.g.~the inverse Hubble radius $1/d_H(t_*)$ at some reference time $t_*$.

Using $k$ and $\kappa(k)$ as reference scales, dimensionless couplings 
and fields are introduced by  
\be 
\label{dimless0} 
{}^n v_{2j,k} := c_0^{1-j}\kappa(k)^{2n} k^{(d-1)j - (d+1)}\,\, {}^n u_{2j,k} \,, 
\quad n,j \geq 0\,, \quad 
\vp_0 = c_0^{-1/2} k^{- \frac{d-1}{2}} \vp\,.
\ee 
where $c_0$ is a fudge factor inserted for later convenience. This entails 
\be
\label{dimless1} 
{}^nU_k(\vp) = c_0 k^{d+1} \kappa(k)^{-2n} \,
{}^n V_k\big(c_0^{-1/2} k^{- \frac{d-1}{2}} \vp\big) \,, 
\quad {}^nV_k(\vp_0) := \sum_{j \geq 0} {}^nv_{2j,k} \frac{\vp_0^{2j}}{(2j)!}\,,
\ee
Note that $\kappa(k)$ only enters the non-minimal sectors, $n \geq 1$. 
In terms of $\scriptr_k(t) := R(t)/\kappa(k)^2$ this gives 
\be
\label{dimless2}
{\cal U}_k(\vp, R) = c_0 k^{d+1} \cV_k(\vp_0, \scriptr_k) \,,
\quad
\frac{\dd^2 {\cal U}_k}{\dd^2 \vp} = k^2 
\frac{\dd^2 \cV_k}{\dd^2 \vp_0} \,,
\ee
where $\cV_k(\vp_0,\scriptr_k) = \sum_{n \geq 0} {}^nV_k(\vp_0) \scriptr_k^n$
and we also write $\cV''_k = \dd^2 \cV_k/\dd \vp_0^2$.  Inevitably  
(\ref{dimless0}) imprints on $\cV_k$ an additional $k$ dependence 
beyond that of the couplings ${}^nv_{2j,k}$, namely that carried by 
$\scriptr_k$. The property $\kappa(k) - k \in \cS(\R_+)$ ensures that this 
parameterization agrees with the standard one for large $k$ and 
then leads to autonomous flow equations for all ${}^nv_{2j,k}$, $n,j \geq 0$.   

In the infrared we aim at a dimensionless version of \eqref{sepaIR1}. 
As noted earlier, the orthogonal polynomial basis $\mfp_l$ is dimensionless, 
so in $\sum_{l \geq 0} {}_l{\cal U}_k(\vp) \mathfrak{p}_l(t)$ the dimension 
enters only through the potentials  ${}_l{\cal U}_k(\vp)$. Clearly, 
these have the same dimension as the generalized potential itself. 
One may thus transition to a dimensionless  formulation by introducing
\begin{eqnarray}
\label{irdiml1}
{}_lv_{2j,k}\is c_0^{1-j}k^{(d-1)j-(d+1)}\, {}_l u_{2j,k}\,,\quad n,j \geq 0\,, \quad 
\vp_0 = c_0^{-1/2} k^{- \frac{d-1}{2}} \vp\,,
\nonum
{}_l{\cal U}_k(\vp) \is c_0 k^{d+1}\,
{}_l\cV_k\big( c_0^{-1/2} k^{- \frac{d-1}{2}} \vp\big)\,,
\quad {}_l \cV_k(\vp_0)=\sum_{j\geq 0}{}_lv_{2j,k}\frac{\vp_0^{2j}}{(2j)!}\,.
\end{eqnarray}
This entails
\be
\label{irdiml2}
\mcU_k(\vp,t) = c_0 k^{d+1}\cV_k(\vp_0,t)\,,\quad 
\cV_k(\vp_0,t)=\sum_{l\geq 0}{}_l \cV_k(\vp_0)\mfp_l(t)\,,
\quad 
\frac{\dd^2 {\cal U}_k}{\dd^2 \vp} = k^2 \frac{\dd^2 \cV_k}{\dd^2 \vp_0}\,.
\ee
Since the parameterization \eqref{sepaIR1}, \eqref{irdiml2} is valid for all 
$k$, not just in the infrared, we discuss the connection between the 
${}_l\cV_k(\vp_0)$ and ${}^nV_k(\vp_0)$ in a cross-over regime. 
The link follows by expressing the polynomials $\mfp_l$ as 
\begin{eqnarray}
\label{irdiml4}
\mfp_l(R(t))\is \sum_{n=0}^l\Xi^n_l\big(\sigma_1[R],\ldots,
\sigma_{2l}[R]\big)R(t)^{n}\,,
\end{eqnarray}
where in the coefficients $\Xi_l^n$  we indicate the dependence of 
the $\sigma_i[R]$ of \eqref{ONB1} on $R$. It follows from the scaling 
invariance \eqref{ONB4} that
\begin{eqnarray}
\label{irdiml5}
\Xi_l^n\big(\sigma_1[\lb R],\ldots,\sigma_{2l}[\lb R]\big)
=\lambda^{-n}\,\Xi_l^n\big(\sigma_1[R],\ldots,\sigma_{2l}[R]\big)\,.
\end{eqnarray}
We may thus write
\be
\label{irdiml6}
\cV_k(\vp_0,t) = \sum_{l\geq 0}{}_l \cV_k(\vp_0)\mfp_l(t) = 
\sum_{n\geq 0}\Big( \sum_{l\geq n}\Xi_l^n
\big(\sigma_1[\scriptr_k],\ldots,\sigma_{2l}[\scriptr_k]\big)\,
{}_l \cV_k(\vp_0)\Big)\scriptr_k(t)^n\,,
\ee
where we used $R(t)=\kappa(k)^2\scriptr_k(t)$ and \eqref{irdiml5} to 
absorb the resulting $\kappa(k)^{2n}$ factor 
into $\Xi_l^n$. Comparing to the parameterization 
$\cV_k=\sum_{n\geq 0}{}^nV_k(\vp_0)\scriptr_k(t)^n$ identifies
\begin{eqnarray}
\label{irdiml7}
{}^nV_k(\vp_0)\is \kappa(k)^{2n} \sum_{l\geq n}\Xi_l^n
\big(\sigma_1[\scriptr_0],\ldots,\sigma_{2l}[\scriptr_0]\big)
\,{}_l \cV_k(\vp_0)\,.
\end{eqnarray}
This last relation highlights that the distinction between `explicit' 
and `implicit' $k$ dependence will become blurred in a cross-over regime. 
While the ${}^nV_k(\vp_0)$ are designed to obey a system of autonomous 
(not explicitly $k$-dependent) flow equations for large $k$ (see Section 4.2), this will 
in general not be the case towards the infrared. As illustrated in Appendix B, 
this feature is rooted in the temporal nonlocality argued  after \eqref{sepaUV} to be 
a generic feature of the infrared dynamics. Hence, the autonomy of the 
small $k$ flow equation cannot be used as a criterion to adjust   
the transition formulas to a dimensionless formulation. Instead, the  
flow equation itself will govern the total $k$-dependence of the 
couplings ${}_lv_{2j,k}$ defined by (\ref{irdiml1}). As a mnemonic we 
shall write $kd/dk$  for the derivative terms.

With the above preparations at hand, the original form (\ref{sepa1}) 
of the FL-sEPA flow equation can be converted into a dimensionless form 
useful to explore the infrared (IR) regime. The Green's function relation 
in (\ref{sepa1}) is expressed in terms of a dimensionless momentum 
$\wp = p/k$ and assumes the form 
\be 
\label{dimless11} 
\big\{ a^{-2d} (a^d {N}^{-1} \dd_t)^2 
+ k^2 [\wp^2/a^2 + r(\wp^2/a^2) + \cV''_k ] \big\} G_k[\vp_0](t,t';\wp) = 
\delta(t,t')\,.
\ee
We regard the solution of (\ref{dimless11}) as a functional of $\vp_0$ 
entering via $\cV_k''(\vp_0,t)$, expanded as in (\ref{irdiml2}). 
After projection onto the $\mfp_l(t)$ basis one has for 
${}_l\!\cV_k = {}_l\!\cV_k(\vp_0)$   
\ba
\label{dimless12}  
&\nspace & 
k \frac{d}{d k}\, {}_l\!\cV_k + (1\!+\!d) \,{}_l\!\cV_k - \frac{d\!-\!1}{2} 
\vp_0 \frac{\dd}{\dd \vp_0} \,{}_l\!\cV_k = 
\frac{-i2 \hbar}{c_0(4\pi)^{d/2} \Gamma(d/2)} 
\int_0^{\infty} \! d\wp \wp^{d-1} 
\nonum
&\nspace & \quad \quad \times 
\int \! dt \,w(t) \mathfrak{p}_l(t) 
\big[  r(\wp^2/a(t)^2) - (\wp^2/a(t)^2) r'(\wp^2/a(t)^2) \big] 
k G_k[\vp_0](t,t;\wp)\,.
\ea
In Section 4.3 we shall use (\ref{dimless12}) to examine 
the IR fixed point regime, $k \ra 0$, of the FL-sEPA flow. 
Since this is inevitably state dependent we will choose specifically 
a State of Low Energy from Section 2.4 as the underlying Hadamard state.

\newsubsection{Universal large $k$ flow and UV fixed point} 

The large $k$ regime is best studied in the dimensionful formulation 
(\ref{sepa1}).  
As in Section 3 we now focus on $d=3$ and adopt the conformal time 
gauge ${N} =a$, renaming $t$ into $\eta$. Following 
the same steps as in (\ref{fl1l3}) -- (\ref{fl1l7}) one finds  
\be
\label{sepa4}
\dd_k {\cal U}_k = -\frac{i \hbar \,8k^4}{(4\pi)^2 a^2} 
\int_0^\infty\!d\wp\, \wp^2
\big[r(\wp^2/a^2)- (\wp^2/a^2)  r'(\wp^2/a^2)] 
\mcG_{k,0}(\eta,\wp)\big|_{W \mapsto {\cal U}''_k}\,,
\ee
where $\wp = p/k$ and $-i\mcG_{k,0}(\eta,\wp) = 
-i\mcG_{k,0}[\vp](\eta,\eta;p)|_{p=k\wp}$ is a solution of (\ref{FLhad11})
with $N=a$.  
For large $k$ the right hand side of the flow equation (\ref{sepa4}) 
can be expanded using (\ref{fl1l8}). This gives 
\be
\label{sepa5}
\dd_k {\cal U}_k = \frac{\hbar}{(4\pi)^2} 
\bigg\{ 4 k^3 q_0 + 2 k \,g_1(\eta)|_k + \frac{g_2(\eta)|_k}{k} 
+ \sum_{n \ge3 } \frac{2\!-\!n}{2} \frac{g_n(\eta)|_k}{k^{2n -3}} 
\bigg\}\,. 
\ee 
Here $q_0, g_1(\eta), g_2(\eta)$ are as in (\ref{fl1l10}) and we 
extend the definition to 
\be 
\label{sepa6} 
g_n(\eta) := (-)^n\frac{2}{2-n} \int_0^{\infty} d \varrho \,\varrho^2 
\frac{ [r(\varrho^2)-\varrho^2 r'(\varrho^2)]}{[\varrho^2 +r(\varrho^2)]^{1/2}}
G_n(\eta,\varrho )\,, \quad n \geq 3\,.
\ee 
For all $n \geq 1$ we use $g_n(\eta)|_k := g_n(\eta)_{W \mapsto {\cal U}''_k}$ 
as a shorthand. As seen in Section 3.1
the $g_1,g_2$ evaluate up boundary terms to 
\ba 
\label{sepa7} 
g_1(\eta)|_k \is \breve{q} R - q_1 \big( {\cal U}_k''- R/6\big) 
\nonum
\is -q_1\nbU{0}''_k(\vp)+\big[\breve{q}
-q_1\nbU{1}''_k(\vp)\big]R-q_1\sum_{n\geq 2}\nbU{n}''_k(\vp)R^n\,,
\nonum
g_2(\eta)|_k &\simeq& \frac{1}{2}\big( {\cal U}_k''- R/6\big)^2 = 
\frac{1}{2} \nbU{0}''_k(\vp)^2 +\nbU{0}''_k(\vp)\nbU{1}_k''(\vp)R
\nonum
&+&\sum_{n\geq 2}\Big[\frac{1}{2} \sum_{n_1+n_2=n}\!^{n_1}\bar{U}''_k(\vp)\,
\,^{n_2}\bar{U}''_k(\vp)\Big]R^n\,. 
\ea 
Here the $\nbU{n}_k(\vp)$  are defined as in (\ref{flrg7}) but 
refer to $k$-dependent couplings $\nbu{n}_{2j,k}$, $n,j \geq 0$. 
Comparing powers of $R$ one finds 
\ba 
\label{sepa8}
\dd_k\! \nnU{0}_k \is \frac{\hbar}{(4\pi)^2} \Big\{ 
4 k^3 q_0 - 2 k q_1 \nbU{0}''_k + \frac{1}{2 k} 
\big(\!\!\nbU{0}''_k\big)^2 + O(1/k^3) \Big\}\,, 
\nonum
\dd_k\! \nnU{1}_k \is \frac{\hbar}{(4\pi)^2} \Big\{ 2k 
\big( \breve{q} - q_1 \nbU{1}''_k \big) + \frac{1}{k} 
\nbU{0}''_k \nbU{1}''_k + O(1/k^3) \Big\}\, 
\\[2mm] 
\dd_k\! \nnU{n}_k \is \frac{\hbar}{(4\pi)^2} \Big\{ -2k\, 
q_1 \nbU{n}''_k + \frac{1}{2k} \sum_{n_1+n_2=n}\!^{n_1}\bar{U}''_k(\vp)\,
\,^{n_2}\bar{U}''_k(\vp) + O(1/k^3) \Big\}, \;\; n \geq 2\,.
\nonumber
\ea 
One may check that the transition formulas \eqref{dimless2} lead to 
an autonomous system of flow equations for the dimensionless potentials 
${}^nV_k(\vp_0)$, $n \geq 0$. The subleading terms are determined by 
$G_n, n \geq 3$, in the recursion (\ref{fl1l9}). We omit their analysis 
for now and 
relate the displayed terms in (\ref{sepa8}) to the perturbative 
analysis of Section 3. Comparison with (\ref{flpotren2})
exhibits an exact match to the one-loop flows of all 
potentials. Hence, the coupling flow determined from the EPA 
(\ref{sepa1}), (\ref{sepa4}) coincides -- for large enough $k$ -- 
with the one-loop flow of the renormalized couplings after 
strict removal of the UV cutoff: 
\be 
\label{PTmatch}
\nnu{n}_{2j,k} = \nnu{n}_{2j}(\mu), \;\; n,j \geq 0\,,\quad \mbox{for} 
\quad \mu = k \quad \mbox{large} 
\ee
Underlying this seemingly trivial match are some structural 
elements \cite{FPPT} that are easily overlooked: 
\begin{itemize} 
\item  The use of a non-minimal subtraction scheme, see (\ref{flrg2}),
which allows one to treat $\mu$ and $\Lambda$ as an instances of $k$
and the $k=\mu$ quantitites as the renormalized ones.   
\item  The inclusion of all interaction monomials, here (\ref{pot}), 
needed for strict removal of the UV cutoff. Only then can one 
make $k < \Lambda$ arbitrarily large and identify the unstable 
manifold so as to justify the interpretation of the couplings 
as the renormalized ones (as opposed to merely 
running Wilsonian parameters). 
\item  Use of the same regulator to set up perturbation theory 
and the EPA. 
\end{itemize} 
The previous comments are not specific to the FL-sFRG. 
With Euclidean signature the reintegration of the covariant 
$O(\hbar)$ FRG, $2 \dd_k \Gamma_{k,1}[\vp] = {\rm Tr}
[ \dd_k R_k(S^{(2)}(\vp) + R_k)^{-1}]$, is often used in 
combination with the heat kernel expansion to obtain the full
renormalized  one-loop effective action at $k=0$. Since the heat 
kernel expansion (even in variants where nonlocal terms are included) 
is an asymptotic expansion tailored to the ultraviolet, the integration 
down to $k=0$ is conceptually unwarranted. Replacing it with an 
integration from $\mu$ to $\Lambda$ for some Wilsonian scale $\mu$, 
it can be seen to produce termwise (precisely) the same results for 
the divergent parts as the direct evaluation of $L_{\Lambda} + L_1^{\rm div}$, 
provided a regularization is used that facilitates such a comparison. 
The next step, of attempting to strictly remove the UV cut-off $\Lambda$, 
is often omitted. Doing so somewhat trivializes the notion of  
one-loop renormalizability, fails to identify the unstable manifold, 
and may lead to inconsistent truncations.

In Section 3 of \cite{Codelloetal}  the removal of the UV cutoff is 
discussed for a scalar field theory in flat space. Unfortunately, in 
their Eq.~(49) several typos crept in which obscure the fact that 
the UV beta functions $(\Lambda \dd/\dd \Lambda$ response of the 
bare couplings) must in this framework coincide with the IR beta 
functions ($k \dd/\dd k$ response of the parameters in the EPA ansatz). 
In Section 7 of \cite{Codelloetal} gravity coupled to a scalar is studied. 
In their Eq.~(180) it is noted that several terms are generated not present 
in the original action, but no attempt is made to absorb them through 
coupling or field redefinitions. 

The EPA flow for scalar QFT on a generic Riemannian manifold has 
also been considered in 
\cite{CosmFRGtrad0,CosmFRGtrad1,CurvedLPA1,CurvedLPA2},
primarily using the heat kernel methodology to compute the 
right hand side of the flow equation. In \cite{CurvedLPA2}
a potential for the nonminimal coupling proportional to $R$ 
was included (our $\nnU{1}(\vp) R$ term). Taking into account 
${}^n \bar{U}_k'' = {}^nU_k'' - (1/6)\delta_{n,1}$, the structure 
of the $n=0,1$ flow equations (\ref{sepa8}) parallels those 
in Eqs (26),(27) of \cite{CurvedLPA2} and the modulator 
independent coefficients agree exactly. The restriction to linear order 
in $R$ is seen as truncation, so neither does the criterion 
(\ref{oneloopstrict}) enter nor the ensued Ricci tower.

\newsubsection{Infrared regime for SLE based EPA flow}

We return to the dimensionless form (\ref{dimless12}) of the EPA flow and 
seek to analyze its structure for small  $k$. On the right-hand side 
the dependence of the Green's function on $\cV''_k$ needs to be known. 
In the case of a de Sitter background \cite{Kaya,Serreau1,Serreau2}, 
as mentioned after (\ref{FLfrg1}), the  issue of state dependence of 
the flow equation does not arise due to the maximal symmetry of 
the spacetime. The right-hand side of \eqref{sepa1} can be then 
be determined in closed form for the `hockey stick' FRG modulator, 
and (assuming that $\cV''_k$ remains bounded as $k\to 0$) the well 
known infrared enhancement of the Bunch-Davies Green's function leads 
to a ``dimensional reduction'' in the small $k$ form of the 
dimensionless EPA flow equation \cite{Serreau2}. For a generic \FL 
background, however, the small $k$ behavior of the FL-sEPA flow will 
normally be very different.

In analyzing the small $k$ form of the EPA flow (\ref{sepa1}), we seek 
to keep the scale factor $a(t)$ and the modulator function  
\eqref{reg1} generic, and may allow for a time dependent $\vp_0(t)$. 
Then even the 
modulator independent basic wave equation can not be solved 
in terms of known special functions. As argued before, the 
inevitable state dependence of the small $k$ flow is a major 
difference to the large $k$ regime. Rather than 
restricting attention to some drastic specializations 
and truncations we aim at developing an analytically 
controllable series expansion of the Green's function for 
small $k$. Inevitably, this requires a choice of underlying 
Hadamard state, and we shall focus on the States of Low 
Energy ({\bf SLE}) described in Section 2.4. With the assumption 
that $\cV''_k$ remains bounded as $k \ra 0$, an analytically 
controllable small $k$ expansion turns out to be feasible.

The role of  $p$ as an expansion parameter in \eqref{sle5} is played by 
$k$ after transition to dimensionless variables. Using (\ref{irdiml2}) 
the dimensionless form of the basic wave equation reads 
\be
\label{Vexp2} 
\big\{ (a^d {N}^{-1} \dd_t)^2 
+ k^2a^{2d}  [\wp^2/a^2 + r(\wp^2/a^2) + \cV''_k ] \big\}  S_k(t,\wp) =0\,. 
\ee
An arbitrary Wronskian normalized solution of (\ref{Vexp2}) 
is used as input for the construction of the SLE solution
$T_k^{\rm SLE}[S_k](t,k\wp)$, see (\ref{gsle1}).  
Comparing with (\ref{IRdispersion}) and 
the wave equation based on it, one is lead to the identifications
$\om_0(t) \mapsto 0$ and 
\ba
\label{sleepa0} 
&& p \mapsto k\,, \quad 
w_{2,p}(t) \mapsto a(t)^{2d} \big[ \cV''_k  
+ \wp^2/a^2 + r(\wp^2/a^2) \big] =: w_2(\cV_k'',\wp)\,,\quad 
\ea
We relegate the justification of this substitution rule to Appendix C 
and note here its important consequence: the coincidence limit 
$kG_k[\vp_0](t,t;\wp)| = ik|T_k^{\rm SLE}(t,k\wp)|^2$ of the intractable 
(dimensionless) Green's function admits a convergent series expansion 
in powers of $k^2$. The coefficients can be computed analytically from 
(\ref{sle3}) -- (\ref{sle8}) in combination with (\ref{sleepa0}).  
We thus define 
\ba 
\label{sleepa1} 
\varepsilon_{n,\wp}[\cV''_k] &:= & \varepsilon_n\big|_{ \om_2^2 = w_{2}(\cV''_k,\wp)} 
\nonum
J_{n,\wp}[\cV''_k](t) &:=& J_n(t) \big|_{\om_2^2 = w_{2}(\cV''_k,\wp)}\,,
\quad n \geq 0\,,
\nonum
\bar{a}_{\wp}[\cV''_k] &:=& \frac{J_{0,\wp}}{\varepsilon_{1,\wp}} = 
\left( \dfrac{\int\! dt {n}(t) f(t)}%
{\int \! dt {n}(t) a(t)^{2d} f(t)  
\big[ \cV''_k + \wp^2/a^2 + r(\wp^2/a^2) \big]} \right)^{1/2}
\!\!.
\ea 
The trivial 
$J_0 = J_{0,\wp} = \frac{1}{2} \int\! dt \,{n}(t) f(t)$
could be normalized to some value, but is kept for dimensionality reasons.  
Then, $\bar{a}_{\wp}$ is dimensionless while $J_{n,\wp}$ and $\varepsilon_{n,\wp}$ 
have mass dimension $-(2n\!+\!1)$ and $-n$, respectively. 
In this notation
\be
\label{sleepa2} 
|T_k^{\rm SLE}(t, k \wp) |^2 
= \frac{\bar{a}_{\wp}[\cV''_k]}{2k} 
\bigg\{1 + \bigg( \frac{J_{1,\wp}[\cV''_k](t)}{J_0} -
\frac{\bar{a}_{\wp}[\cV''_k]^2 \,\varepsilon_{2,\wp}^2[\cV''_k]}{2 J_0^2} \bigg)k^2 
+ O(k^4) \bigg\} \,. 
\ee 
This has to be inserted into the projected FL-sEPA equation 
(\ref{dimless12}). From now on we choose the weight in the orthonormal 
basis proportional to the window function of the SLE, specifically 
\be
\label{sleweight}  
w(t) = \kappa(0) n(t) a(t)^{2d} f(t) \,,
\ee 
where $n(t) = N(t) a(t)^{-d}$ and the mass scale $\kappa(0)$ 
simply accounts for dimensions, see the paragraph preceding \eqref{ONB1}.
Further, we take $\vp_0$ to be time independent.  
Due to the orthonormality \eqref{ONB3}, in the denominator 
of $\bar{a}_{\wp}[\cV''_k]$ then only the $l=0$ term in the 
decomposition $\cV_k''(\vp_0,t) = 
\sum_{l \geq 0} {}_l\cV''_k(\vp_0) \pf_l(t)$ contributes:
\ba 
\bar{a}_{\wp}[\cV''_k] \is \bigg( 
\frac{2 J_0 \kappa(0) \pf_0}{ {}_0\cV_k''(\vp_0) 
+ {}_0\rho(\wp)} \bigg)^{1/2}\,,
\nonum
{}_l\rho(\wp) &:=&\int\! dt w(t) \pf_l(t)
\big[\wp^2/a(t)^2+r(\wp^2/a(t)^2) \big]\,. 
\ea  
In turn, this suggests to define for a function $h(t,\wp)$  
\ba 
\label{sleepa4} 
&\nspace & {}_lQ_m(h|V) := 
\frac{\hbar}{c_0(4\pi)^{d/2} \Gamma(d/2)} 
\int_0^{\infty} \! d\wp \wp^{d-1} 
\bigg( \frac{2 J_0 \kappa(0) \pf_0}{ V + {}_0\rho(\wp)} \bigg)^{1/2+m}
\\[2mm] 
&\nspace & \quad \quad \times 
\int \! dt \,w(t) \mathfrak{p}_l(t) 
\big[  r(\wp^2/a(t)^2) - (\wp^2/a(t)^2) r'(\wp^2/a(t)^2) \big] 
h(t,\wp) \,, \quad l,m\geq 0\,.
\nonumber
\ea
All terms on the right hand side of (\ref{dimless12}) can be 
expressed in terms of these averages.  We arrive at the 
{\bf SLE based FL-sEPA} flow equation: 
\ba
\label{sleepa3}  
&\nspace & 
k \frac{d}{dk}\, {}_l\!\cV_k(\vp_0) + (d\!+\!1) \,{}_l\!\cV_k(\vp_0) 
- \frac{d\!-\!1}{2} \vp_0 \frac{\dd}{\dd \vp_0} \,{}_l\!\cV_k(\vp_0)
\nonum
&\nspace& = {}_lQ_0\big(1|\,{}_0\cV_k''\big)  
- \frac{k^2}{2 J_0^2}\;{}_lQ_1\big(1|{}_0\cV_k''\big) 
\sum_{l_1,l_2\geq 0} {}_{l_1}\!\cV_k''(\vp_0) {}_{l_2}\!\cV_k''(\vp_0)
E_{l_1,l_2}  
\\[2mm]
&\nspace & \quad + \frac{k^2}{J_0} \sum_{l_1 \geq 0} {}_{l_1}\!\cV_k''(\vp_0) 
\bigg( {}_lQ_0\big(\jmath_{l_1}| {}_0\cV_k'' \big) 
-\frac{1}{2J_0} \,{}_lQ_1\big(E_{l_1}| {}_0\cV_k'' \big) \bigg)
\nonum
&\nspace & \quad + \frac{k^2}{J_0} \bigg({}_lQ_0\big(\jmath|{}_0\cV_k''\big) 
- \frac{1}{2J_0} {}_lQ_1\big(E|{}_0\cV_k''\big) \bigg)
+ O(k^4)\,.
\nonumber
\ea 
The functions $E_{l,l'},E_l(\wp),E(\wp)$ 
and $\jmath_l(t),\jmath(t,\wp)$ are collected in Appendix C.    
Equation (\ref{sleepa3}) is the main result of this section. 

\underline{Remarks:}
\vspace{-3mm}

\begin{itemize}[leftmargin=8mm, rightmargin=-0mm]
\itemsep -1mm
\item[(i)] Both sides of (\ref{sleepa3})
depend on the choice of window function $f$. The support of 
$f$ will be chosen in accordance with the cosmological period 
where one seeks to inject information about the primordial vacuum 
state. Since general relativity demands a pre-inflationary period, 
locating $f$ immediately after a quantum gravity epoch 
is a natural choice. 
\item[(ii)] The scale factor $a$ can be chosen to model a 
realistic expansion history. For example, one can 
have a pre-inflationary period of non-accelerated 
expansion followed by a de Sitter like acceleration, etc. 
This avoids the technically clumsy matching of piecewise 
powerlike $a$'s. 
\item[(iii)] The flow equation (\ref{sleepa3}) is non-autonomous, 
i.e.~$k$ occurs explicitly in the dimensionless formulation. 
The $O(k^4)$ and higher order corrections refer to (\ref{sle3}), 
(\ref{sle4}) and can be computed systematically.  
This feature differs from the UV flow (\ref{sepa8}) where the same 
substitution (\ref{dimless0}), (\ref{dimless1}) leads to an 
autonomous system of flow equations for the dimensionless 
potentials ${}^nV_k(\vp_0)$, $n \geq 0$. In Appendix B we  
analyze the instantaneous limit of (\ref{sleepa3}) where     
again an autonomous system of flow equations for the  
${}^nV_k(\vp_0)$ arises. The non-autonomous character of 
the (\ref{sleepa3}) therefore originates in the temporal averaging 
(and the ensued temporal nonlocality) needed to render the underlying 
state (of low energy) mathematically and physically well-defined
\cite{FRWHamdiag0,Olbermann}.   
\item[(iv)]  The small $k$ expansion \eqref{sleepa2} entails that the flow equation specializes correctly to the spatial EPA flow in Minkowski space. 
Since ${n} \equiv 1 = a$ one has $\Delta_p(t,t') = p^{-1} \sin p (t-t')$. 
Then (\ref{sle2}) reduces to $J_p(t) = J_0$, $\cE_p^{\rm SLE} = p 
\varepsilon_1$ and only the leading term on the right hand side 
of (\ref{sleepa2}) remains. For a constant 
$\vp_0$ the window function drops out in $\bar{a}_{\wp}[\cV''_k]$ 
and the corresponding flow equation reduces to (\ref{Mink5}), 
as required. 
\end{itemize} 

Importantly, (\ref{sleepa3}) has a well-defined IR fixed point equation
\be 
\label{fp1}
(d\!+\!1) {}_l\cV_* - \frac{d\!-\!1}{2} \vp_0 \frac{\dd}{\dd \vp_0} 
{}_l\cV_* = {}_lQ_0\big(1|\,{}_0\cV''_*\big)\,, 
\ee 
which resembles that in Minkowski space but refers to a curvature 
dependent generalized potential, $\cV_* = \cV_*(\vp_0, t)= 
\sum_{l \geq 0} {}_l\cV_*(\vp_0) \mathfrak{p}_l(t)$. To proceed, we insert 
power series expansions 
\begin{eqnarray}
{}_l\cV_{\ast}(\vp_0)=
\sum_{j \geq 0} {}_lv_{*,2j} \frac{\vp_0^{2j}}{(2j)!}\,,\quad 
{}_0\cV''_{\ast }(\vp_0) = {}_0v_{*,2} + 
\sum_{j \geq 1} {}_lv_{*,2j+2}\frac{\vp_0^{2j}}{(2j)!}\,,
\end{eqnarray}
 and expand ${}_lQ_0(1|\,{}_0\!\cV_*'')$ in powers of the field. This gives 
\ba 
\label{fp5} 
\!\!\! (d\!+\!1){}_lv_{*,0} \is 
\,{}_lQ_0\big(1|{}_0v_{*,2}\big)\,,
\\[2mm]
\!\!\! [d\!+\!1 - j(d\!-\!1)] {}_lv_{*,2j} \is 
\sum_{m=0}^j { -1/2 \choose m} 
\frac{ {}_lQ_m\big(1|{}_0v_{*,2}\big)}{(2 J_0 \pf_0)^m} 
\nonum
&\times & 
\sum_{\scriptstyle j_1+\ldots + j_m=j\atop\scriptstyle j_1,\ldots, j_m\geq 1}
{ 2j \choose 2j_1, \ldots, 2j_m} {}_0v_{*,2j_1+2} \ldots  {}_0v_{*,2j_1+2} \,.
\nonumber
\ea  
The $l=0$ instance of the second equation is a coupled system 
for the ${}_0v_{*,2j}$, $j \geq 0$. Assuming that a solution has been 
found all other ${}_lv_{*,2j}, l\geq 1, j\neq 1$ are determined by them.   
The same holds for $d >3$ for $j=1$, while for $d=3$ and $j=1$ 
the coefficient of ${}_l v_{*,4}$ vanishes and the ${}_0v_{,2j}$ are 
further constrained. With the possible exception of ${}_lv_{*,4}$ 
the couplings associated with the temporally averaged potential 
${}_0\cV_*(\vp_0)$ therefore determine all other ${}_l \cV_*(\vp_0)$,
$l \geq 0$. Although time dependent, the fixed point solution 
$\cV_*(\vp, t)$ is therefore essentially 
determined by its temporal average.     

The derivation of the $O(k^2)$ correction terms in (\ref{sleepa3}) 
is relegated to Appendix C. The further analysis of (\ref{sleepa3}) 
will begin with a linearization around the fixed point solutions(s)
and then use a shooting technique to numerically relate the 
infrared couplings ${}_lv_{2j,k}$, $l \geq 0$, to the ultraviolet ones 
${}^nv_{2j,k}$, $n\geq 0$, using the respective flow equations.     
For length reasons a detailed investigation will have to be relegated 
to a separate paper. To the best of our knowledge the flow 
(\ref{sleepa3}) is the first result that allows one to
explore the consequences of the Hadamard property for the
non-perturbative {\it infrared dynamics} of quantum fields 
on a \FL background with a realistic expansion history.

\newpage 
\newsection{Conclusions} 

The main take away from the results obtained here is the inevitable 
state dependence of a Lorentzian signature FRG on a generic 
foliated background. At present, no selection criteria 
for physically viable vacuum like states are known beyond
perturbation theory. The advocated correspondence between 
vacuum-like states and a spatial FRG may in fact help to 
develop such criteria. The resulting interplay between 
state selection, Wilsonian perturbation theory, and a 
non-perturbative spatial FRG flow has been summarized in Figure 1. 
Details were elaborated for \FL backgrounds with States of 
Low Energy as the Hadamard states of choice. In particular, 
systematic expansions for the infrared and ultraviolet regime 
of the associated effective potential flow equations have been
presented. For length reasons, a detailed quantitative investigation 
of the ensued flows has to be relegated to a separate study. 

There are many threads of research to be taken up from here. 
Most immediate is the non-perturbative flow of the scalar power 
spectrum. A case study in \cite{BonusSLE} suggests that the States 
of Low Energy are viable candidates for pre-inflationary vacua. 
In particular, the low multipole moments in the CMB spectrum 
are suppressed in a natural way. The approximately scale 
invariant power spectrum arises in a cross over region,
which in a spatial FRG setting has to be explored numerically. 
On a de Sitter background Stochastic Inflation provides an 
alternative framework to capture the non-perturbative inflaton 
dynamics and has some interface with the FRG flow 
\cite{Kaya,Serreau1,Serreau2}. Since general relativity 
demands a pre-inflationary (kinetic energy dominated) period, 
the generalization of this interplay to \FL 
backgrounds other than de Sitter is an important desideratum. 
A third thread is the extension to other types of cosmological 
backgrounds. Since the framework of Figure 1 requires some degree 
of computational control over the Hadamard states in question, 
the choices are presently limited. The States of Low Energy can 
however be extended to spatially inhomogeneous generalized FRW 
geometries \cite{SLE2}. Another option are temporally averaged 
Sorkin-Johnston states \cite{SJstates,SJHadamard}. Eventually, one 
would like to go beyond the Effective Potential Approximation, 
for which the spatial hopping expansion \cite{Graphpaper,Mainzproc} 
provides a starting point.  

A spatial FRG should also be the appropriate tool to investigate 
Quantum Gravity at strong coupling. This takes foliated 
geometries as basic as in \cite{FRGfoliated}, though with a 
different dynamics that underlies an associated Anti-Newtonian 
expansion \cite{Mainzproc}, antipodal to perturbation theory.  
\bigskip

{\itshape Acknowledgements:} This research was partially supported 
by PittPACC. R.B. also acknowledges support through a 
Andrew P. Mellon fellowship from the University of Pittsburgh.
We also wish to thank the referee for an unusually thorough 
reading of the manuscript, which lead to a variety of 
improvements.

\newpage 

\appendix

\newappendix{Regulator dependence of $\Gamma_{k,1}^{\rm div}$ coefficients} 

In this appendix we present a proof of the Lemma entering (\ref{Bcoeff1}), 
(\ref{Bcoeff2}) together with explicit formulas for the regulator 
dependent coefficients $\breve{q},q_1,b_1,b_2,b_3$. 
We begin by formulating 

\noindent
{\bfseries Lemma A:}
{\itshape 
Let $r(x)$ be a generic FRG regulator  function as defined in \eqref{reg1}.
\begin{enumerate}
\item[(a)] The regulator dependent coefficients $B_1,B_2,B_3$ always satisfy 
the ratio
\begin{eqnarray}\label{frgA1}
B_1:B_2:B_3\is -1:4:3\,,
\end{eqnarray}
and hence can be expressed in terms of a single regulator dependent 
constant $b_1$,
\begin{eqnarray}\label{frgA2}
B_1=-6b_1\,,\quad B_2=24 b_1\,,\quad B_3=18b_1\,.
\end{eqnarray}
\item[(b)]  $B_4+36 b_1+B_5=0$.
\item[(c)] $B_6=B_7=B_8$.
\end{enumerate}}

Before proceeding to the proof, we note the explicit expressions for 
$B_1,\ldots,B_8$ (obtained with {\ttfamily Mathematica}):
\ba
\label{frgA3}
B_1\is - \int_0^\infty\!d\varrho\,\varrho^2 
\frac{\big[r(\varrho^2)-\varrho^2\reg{1}\big]^2}%
{4\big[\varrho^2+ r(\varrho^2)\big]^{7/2}}\,,
\nonum
B_2\is -\int_0^\infty\!d\varrho\,\varrho^2 
\frac{\big[r(\varrho^2)-\varrho^2\reg{1}\big]^2}%
{2\big[\varrho^2+ r(\varrho^2)\big]^{9/2}}
\Big\{\varrho^4\reg{2}\big(\varrho^2+ r(\varrho^2)\big)
\nonum
&-&4\big(r(\varrho^2)-\varrho^2\reg{1}\big)\big(5r(\varrho^2)
-\varrho^2(2+7\reg{1})\big)\Big\}\,,
\nonum 
4B_3\is 3B_2
\nonum
B_4\is \frac{3}{8} \int_0^\infty\!d\varrho\,\varrho^2 
\frac{\big[r(\varrho^2)-\varrho^2r^{(1)}(\varrho^2)\big]^2}%
{\big[\varrho^2+ r(\varrho^2)\big]^{9/2}}\Big\{-35r(\varrho^2)^3
\nonum
&\!+\!&\!\!\varrho^2r(\varrho^2)^2\Big[42+147 r^{(1)}(\varrho^2)
+108\varrho^2 r^{(2)}(\varrho^2)+16\varrho^4r^{(3)}(\varrho^2)\Big]
\nonum 
&\!+\!&\!\!\varrho^6\Big[42 r^{(1)}(\varrho^2)^2+77 r^{(1)}(\varrho)^3
+24 \varrho^2r^{(2)}(\varrho^2)-84\varrho^2r^{(1)}(\varrho^2)r^{(2)}(\varrho^2)
+ 16 \varrho^4 r^{(3)}(\varrho^2)\Big]
\nonum 
&\!+\!&\!\!\varrho^4 r(\varrho^2)\Big[132 \varrho^2 r^{(2)}(\varrho^2)
-21r^{(1)}(\varrho^2)\big(4+9r^{(1)}(\varrho^2)+4\varrho^2 r^{(2)}(\varrho^2)\big)
+ 32\varrho^2r^{(3)}(\varrho^2)\Big]\Big\}\,.
\nonumber
\end{eqnarray}
\begin{eqnarray}
B_5\is -\int_0^\infty\!d\varrho\,\varrho^2 
\frac{r(\varrho^2)-\varrho^2r^{(1)}(\varrho^2)}%
{32\big[\varrho^2+ r(\varrho^2)\big]^{9/2}}\Big\{-315r(\varrho^2)^4
\nonum
&\!+\!&\!\!4\varrho^2r(\varrho^2)^3\Big[189+504r^{(1)}(\varrho^2)
+606\varrho^2r^{(2)}(\varrho^2)
+192 \varrho^4 r^{(3)}(\varrho^2)+16 \varrho^6r^{(4)}(\varrho^2)\Big]
\nonum 
&\!+\!&\!\!\varrho^8\Big[-924 r^{(1)}(\varrho^2)^3-1155
\reg{1}^4+240\varrho^2 \reg{2}
+ 84\reg{1}^2\big(22\varrho\reg{2}-1\big)
\nonum 
&\!-\!&\!\!112\reg{1}\big(3\varrho^2\reg{2}+4\varrho^4\reg{3}\big)
\nonum 
&\!+\!& \!\!
16 \varrho^4\big(4\varrho^2\reg{4}+20\reg{3}-21\reg{2}^2\big)\Big]
\nonum 
&\!+\!& \!\!4\varrho^6r(\varrho^2)\Big[924\reg{1}^3+264\varrho^2\reg{2}
+21 \reg{1}^2\big(31+22 \varrho^2\reg{2}\big)
\nonum 
&\!-\!&\!\!14\reg{1}\big(16\varrho^4\reg{3}+78\varrho^2\reg{2}-3\big)
\nonum 
&\!+\!& \!\!
8\varrho^2\big(6\varrho \reg{4}+44 \reg{3}-21 \reg{2}\big)\Big]
\nonum 
&\!+\!& 2 \varrho^2 r(\varrho^2)^2\Big[-2121\reg{1}^2
+6\big(270 \varrho^2\reg{2}-7\big)
\nonum 
&\!-\!& \!\!14 \reg{1}\big(16 \varrho^4 \reg{3}+144 \varrho^2 \reg{2}+87\big)
\nonum 
&\!+\!& \!\!
8 \varrho^2\big(12\varrho^2 \reg{4}
+116 \reg{3}-21 \reg{2}^2\big)\Big]\Big\}\,,
\nonum
B_6\is \int_0^\infty\!d\varrho\,\varrho^2 
\frac{r(\varrho^2)-\varrho^2r^{(1)}(\varrho^2)}%
{2\big[\varrho^2+ r(\varrho^2)\big]^{7/2}}
\Big\{-3r(\varrho^2)+\varrho^2\big(2+5\reg{1}(\varrho^2)\big)\Big\}\,,
\nonum
B_7\is \int_0^\infty\!d\varrho\,\varrho^2 \frac{r(\varrho^2)
-\varrho^2r^{(1)}(\varrho^2)}{4\big[\varrho^2+ r(\varrho^2)\big]^{9/2}} 
\Big\{-9r(\varrho^2)^2
+ 2\varrho^2 r(\varrho^2)\Big[11+20\reg{1}+10 \varrho^2 \reg{2}\Big]
\quad 
\nonum
&\!+\!&\varrho^4\Big[20 \varrho^2 \reg{2}-5 \reg{1}\big(6+7\reg{1}\big)
-4\Big]\Big\}\,,
\nonum
B_8\is B_6\,.
\end{eqnarray}
Note that we use the notation $r^{(n)}(x):=\frac{d^n}{dx^n}r(x)$, $n\geq 0$, throughout.

{\it Proof of Lemma A.} \
\begin{enumerate}
\item[(a)]  The ratio $B_2:B_3=4:3$ is clear from \eqref{frgA3}, 
so it remains to show 
\begin{eqnarray}
4B_1+B_2=0\,.
\end{eqnarray}
Indeed, it follows by direct computation that
\begin{eqnarray}
\label{frgA5}
4B_1+B_2\is \frac{1}{2}  \int_0^\infty\!d\varrho\,
\frac{\dd}{\dd \varrho}\bigg\{
\frac{\varrho^3r(\varrho^2)^2-2\varrho^5 r(\varrho^2)\reg{1}
+\varrho^7 \reg{1}^2}{\big[\varrho^2+r(\varrho^2)\big]^{7/2}}\bigg\}
\nonum
\is \lim_{\varrho\to \infty}\frac{\varrho^3r(\varrho^2)^2-2\varrho^5 
r(\varrho^2)\reg{1}+\varrho^7 \reg{1}^2}%
{2\big[\varrho^2+r(\varrho^2)\big]^{7/2}}\,.
\end{eqnarray}
By (\ref{reg1}) $r(\varrho^2)$ and its derivatives approach zero 
as $\varrho\to \infty$ faster than any polynomial in $\varrho$. 
Thus \eqref{frgA5} vanishes,  i.e.~$4B_1+B_2=0$, establishing 
Lemma A($a$).

\item[(b)] Similarly, one finds
\begin{eqnarray}
&& B_4+36b_1+B_5 = \int_0^\infty\!d\varrho\,\,
\frac{\dd}{\dd \varrho}\bigg\{\frac{\varrho^3}%
{32 \big[\varrho^2+r(\varrho^2)\big]^{11/2}}\Big[5r(\varrho^2)^4
\nonum 
&& \sspace -4\varrho^2 r(\varrho^2)^3\big(22+27 \reg{1}
+36\varrho^2 \reg{2}+8\varrho^4 \reg{3}\big)
\nonum 
&& \sspace + \varrho^8\Big(112\reg{1}^3+105\reg{1}^4-16 \varrho^4 \reg{2}^2
\nonum 
&& \sspace + 4\reg{1}^2[3-28\varrho^2 \reg{2}]
+32 \varrho^2 \reg{1}[\reg{2}+\varrho^2 \reg{3}]\Big)
\nonum 
&& \sspace + 2\varrho^4r(\varrho^2)^2\Big(6+153\reg{1}-88\varrho^2 \reg{2}
\nonum 
&& \sspace - 8\varrho^2[\reg{2}^2+4\reg{3}]+16\reg{1}[9+8\varrho^2 \reg{2}
+\varrho^4\reg{3}]\Big)
\nonum 
&& \sspace - 4 \varrho^6 r(\varrho^2)\Big(77\reg{1}^3
+2\reg{1}^2[39+14\varrho^2 \reg{2}]
\nonum 
&& \sspace + 8\varrho^2 \reg{2}(\varrho^2)+8\varrho^2[\reg{2}^2+\reg{3}]
\nonum 
&& \sspace - 
2\reg{1}[8\varrho^2\reg{3}+36 \varrho^2\reg{2}-3]\Big)\Big]\bigg\}\,,
\end{eqnarray}
which vanishes for a generic regulator function $r$ along the same lines 
as \eqref{frgA5}. This establishes Lemma A($b$).
\item[(c)] As above, direct computation yields
\begin{eqnarray}
B_6-B_7 \is -\int_0^\infty\!d\varrho \,\frac{\dd }{\dd \varrho}
\bigg\{\frac{5 \varrho^3 \big[r(\varrho^2)-\varrho^2 \reg{1}\big]^2}%
{4 \big[\varrho^2+r(\varrho^2)\big]^{7/2}}\bigg\}=0\,.
\end{eqnarray}
Together with \eqref{frgA3} this completes our proof of Lemma A. 
\begin{flushright}
	\qed
\end{flushright}
\end{enumerate}

We end this appendix with a summary of the expressions for the 
regulator dependent coefficients in \eqref{Bcoeff1}, \eqref{Bcoeff2}. 
\begin{eqnarray}
\label{frgA6}
\breve{q}\is \frac{1}{12}\int_0^\infty\!d\varrho\,\varrho^2 
\frac{\big[r(\varrho^2)-\varrho^2 \reg{1}\big]^2 }%
{\big[\varrho^2+r(\varrho^2)\big]^{5/2}}\,,
\quad
q_1  =\int_0^\infty\!d\varrho\,\varrho^2 
\frac{r(\varrho^2)-\varrho^2 \reg{1} }{\big[\varrho^2+r(\varrho^2)\big]^{3/2}}\,,
\nonum
b_1	\is \frac{3}{2}\int_0^\infty\!d\varrho\,\varrho^2 
\frac{\big[r(\varrho^2)-\varrho^2r^{(1)}(\varrho^2)\big]^2}%
{\big[\varrho^2+ r(\varrho^2)\big]^{7/2}}\,,\quad 
\quad \;b_2=-\frac{B_5}{12}\,,
\nonum
b_3	\is \frac{1}{2} \int_0^\infty\!d\varrho\,\varrho^2 
\frac{r(\varrho^2)-\varrho^2r^{(1)}(\varrho^2)}%
{\big[\varrho^2+ r(\varrho^2)\big]^{7/2}}
\Big\{\!-3r(\varrho^2)+\varrho^2\big(2+5\reg{1}(\varrho^2)\big)\Big\}\,.
\end{eqnarray}

\newappendix{Spatial EPA in Minkowski space and SLE instantaneous limit } 

The analysis of the spatial EPA (\ref{sepa1}) is normally 
hampered by the difficulty of gaining analytical control over 
the Green's function $G_k[\vp](t,t;p)$. There are two instructive 
exceptions which we present in this appendix: Minkowski space and 
the ``instantaneous limit'' of the version of (\ref{sepa1}) based 
on a State of Low Energy (SLE), see Section 2.4.  

In Minkowski space $a = {N} =1$, the generalized potential reduces 
to the standard one ${\cal U}_k(\vp,R) = \nnU{0}_k(\vp) =: U_k(\vp)$, 
and we may take the field $\vp$ to be constant. In line with
the general  discussion we choose the Feynman type solution 
of the Green's function relation in (\ref{sepa1}), viz. 
\be 
\label{Mink1} 
G_{k}[\varphi](t,t';p) = i 
\dfrac{e^{ -i |t-t'| \sqrt{ p^2 + R_k(p) + U_k''(\vp)}}}%
{ 2 \sqrt{ p^2 + R_k(p) + U''_k(\vp)}}\,. 
\ee
This corresponds to (\ref{FLhad1}) with a Wronskian normalized 
positive frequency solution of the homogeneous wave equation. 
The spatial EPA flow equation then reads
\be 
\label{Mink2} 
\dd_k U_k = \frac{\hbar}{4} \int\! \frac{d^dp}{(2\pi)^d} 
\dfrac{ \dd_k R_k(p)}{\sqrt{p^2 + R_k(p) + U''_k}}\,.
\ee
We note that no contour deformation or other indirect 
reference to a Euclidean time or momentum regime enters in 
arriving at (\ref{Mink1}), (\ref{Mink2}). Other Minkowski space 
EPA flow equations can be found in \cite{FRGWick1,FRGWick2}.
After transition to radial momenta $|p|$ and $\wp = |p|/k$ 
one has 
\be
\label{Mink3} 
\dd_k U_k =  \hbar \frac{k^d}{(4\pi)^{d/2} \Gamma(d/2)} 
\int_0^{\infty} \! d\wp \, \wp^{d-1} \dfrac{ r(\wp^2) - \wp^2 r'(\wp^2)}%
{ \sqrt{ \wp^2+ r(\wp^2) + U_k''/k^2}} \,,
\ee
using ${\rm vol}(S^{d-1}) = 2 \pi^{d/2}/\Gamma(d/2)$ for the volume 
of the $S^{d-1}$ sphere. The transition to a dimensionless formulation is  
effected by 
\be 
\label{Mink4} 
U_k(\vp) = c_0 k^{1+d} V_k( \vp_0) \Big|_{ \vp_0 = c_0^{-1/2} 
k^{- \frac{d-1}{2}} \vp}\,,
\ee 
where $c_0>0$ a fudge factor to be adjusted later on. In terms 
of $V_k(\vp_0)$ the spatial EPA flow equation reads
\be
\label{Mink5}
k \dd_k V_k(\vp_0) + (1\!+\!d) V_k(\vp_0) - \frac{d\!-\!1}{2}
\vp_0 V'_k(\vp_0) = \hbar q_0\big(V_k''(\vp_0)\big) \,.      
\ee
For re-use later on we defined here 
\be 
\label{Mink7} 
q_{\nu}(v) :=  \frac{1}{c_0(4\pi)^{d/2} \Gamma(d/2)}
\int_0^{\infty} \! d\wp \, \wp^{d-1} \frac{r(\wp^2) - \wp^2 r'(\wp^2)}%
{(\wp^2 + r(\wp^2) + v)^{1/2 + \nu}}\,, 
\ee
with $\nu \in \R$, $v\geq 0$. Following the general discussion in 
Section 4 we use (\ref{Mink3}) for the large $k$ expansion and 
(\ref{Mink5}) to study the fixed point regime. 

For the large $k$ expansion the dimensionful potential should be 
treated as $k$ independent, with $U'' \mapsto U_k''$ re-substituted 
after expansion. This gives  
\begin{equation} 
\label{Mink6}
\dd_k U_k = \hbar c_0 \sum_{n \geq 0} { -1/2 \choose n} q_n(0) 
\Big( \frac{U_k''}{k^2} \Big)^n\,.
\end{equation}
The coefficients are just the binomial ones for which we note 
${ -1/2 \choose n} = (-)^n \frac{(2n\!-\!1)!!}{2^n n!}$. 
Comparing with the general large $k$ 
expansion (\ref{sepa5}), (\ref{sepa6}) one can read off 
\be 
\label{Mink9}
G_n = \frac{(2n\!-\!1)!!}{2^n  ( \wp^2 + r(\wp^2))^n} 
\frac{(U'')^n}{n!} \,,
\quad n \geq 0\,.
\ee
For constant $\vp$ the $(U'')^n/n!$ are in fact the 
conventionally normalized heat kernel coefficients
of the Schr\"{o}dinger operator $\dd_t^2 + U''(\vp)$. 
For $d\!=\!3$ the $n=0,1,2$ terms match those for $\nnU{0}_k$ 
in (\ref{sepa8}), using $4\pi^2 c_0 q_0(0) = q_0$, $4\pi^2 c_0 q_1(0) 
=  q_1$, $4 \pi^2 c_0q_2(0) = 1/3$. 

The fixed point regime of the flow is explored via the dimensionless 
flow equation (\ref{Mink5}). Its right hand side can be expressed 
as $q_0(V''_k)$. Typically, $V''$ has a field independent part $v_2$ 
which one will split off when comparing powers of $\vp_0^2$ later on.
The relevant expansion then is 
\be 
\label{Mink10}
q_0( v_2 + \Delta V'') = 
\sum_{n \geq 0} { -1/2 \choose n} q_n(v_2) (\Delta V'')^n,
\ee 
where $\Delta V'' = \sum_{j \geq 1} (\vp_0^{2j}/(2j)!) v_{2j+2}$. 
Expanding around a non-zero $v_2$ gives a better approximation 
to the exact $q_0(V'')$ than the $q_n(0)$'s entering (\ref{Mink6}). 
Note that neither (\ref{Mink6}) nor (\ref{Mink10}) explore the 
large field regime of the nonlinear flow (\ref{Mink3}) or 
(\ref{Mink5}), which would amount to an expansion in inverse 
powers of $U''_k$ or $V''_k$. This reflects the choice of 
a potential Ansatz in positive (even) powers of the fields.

Comparing in (\ref{Mink5}) powers of $\vp_0^2$ results in a `upward' 
coupled system of flow equations for the dimensionless couplings.
Setting $\hbar=1$  the first few read
\ba 
\label{Mink11} 
k \dd_k v_{k,0} \is -(d+1) v_{k,0} + q_0(v_{k,2})\,, 
\nonum
k \dd_k v_{k,2} \is - 2 v_{k,2} - \frac{1}{2} q_1(v_{k,2}) v_{k,4} \,, 
\nonum
k \dd_k v_{k,4} \is  (d-3) v_{k,4} - \frac{1}{2} q_1(v_{k,2}) v_{k,6}  + 
\frac{9}{4} q_2(v_{k,2}) v_{k,4}^4 \,. 
\ea 
The further analysis now proceeds along the familiar lines: 
Truncating the system (\ref{Mink1}) at some order $N$ (i.e. 
$v_{k,2j} \equiv 0$, $j \geq N$) results in a closed coupled 
system of ordinary differential equations which can be
integrated numerically. The fixed point equations are 
coupled algebraic equations for which one finds many solutions. 
In $d=2$ only two of them have a stability matrix with a single 
negative eigenvalue, however. These are the Gaussian 
fixed point and the Minkowski space counterpart of the Wilson-Fisher 
fixed point. For $d \geq 3$ only the Gaussian fixed point has a stability 
matrix with a single negative eigenvalue. Linearizing 
the flow around the chosen fixed point sets the target for the numerical 
shooting technique. In the ultraviolet boundary conditions 
are set by the expansion of (\ref{Mink6}). The quantitative 
results so obtained do not differ significantly from those 
obtained with Euclidean signature and the standard EPA flow,
see e.g.~\cite{FRGreview} and references therein.  
This validates (\ref{Mink2}) as a viable Minkowski space EPA 
flow equation. 

A flow equation of the form (\ref{Mink2}) also arises in the 
instantaneous limit of the SLE based version of (\ref{sepa1}). 
Recall that the equal time limit of  the Green's function 
$G_k[\vp](t,t;p)$ entering (\ref{sepa1}) is then given by 
$i |T^{\rm SLE}[S_k](t,p)|^2$, where the fiducial solution 
$S_k(t,p)$ is an arbitrary solution of (\ref{FLhad2}).   
The modulus square can be expressed as in (\ref{sle1}) and 
involves a window function $f(t)$ of compact support in $[t_i,t_f]$. 
Although this quantity is amenable to analytical control, it 
is fairly complicated. A remarkable simplification occurs when 
the window function $f$ of the SLE is increasingly centered 
around an instant $t_0 \in (t_i, t_f)$, giving 
$f(t)^2 \ra {n}(t)^{-1} \delta(t\!-\!t_0)$ in a formal limit. 
If the underlying second order operator is of Schr\"{o}dinger type 
(i.e.~without first order term) the limit of the modulus square of 
the SLE solution equals $1/(2\sqrt{\mbox{full potential}})$, 
see \cite{BonusSLE}. In the situation at hand this gives 
\be 
\label{sleinst0}
\big|T^{\rm SLE}[S_k](t,k \wp)\big|^2 \rra 
\frac{1}{2 k a(t_0)^d \sqrt{ \wp^2/a(t_0)^2 + r(\wp^2/a(t_0^2)) 
+ {\cal U}''(\vp,R(t_0))/k^2}}\,, 
\ee
with $\wp = p/k$. Let us stress that the instantaneous limit ruins 
the Hadamard property and is in general neither mathematically nor 
physically viable \cite{FRWHamdiag0}. Another caveat is that the 
operations ``taking the instantaneous limit'' and ``selecting a time 
function through a choice of ${n}$'' do not commute because the 
(non-instantaneous) Schwarzian terms arising from the elimination 
of the first order term are missed. 

It is nevertheless instructive to see that the limit produces a flow 
equation of the form (\ref{Mink3}) which is formally applicable to 
generic \FL spacetimes merely by replacing $\wp$ in the integrand 
by $\wp/a(t_0)$ and $U_k$ by ${\cal U}_k = \sum_{n \geq 0} 
{}^nU_k(\vp) R(t_0)^n$. The explicit time dependence of the SLE 
EPA flow disappears and the projection onto the orthonormal 
basis (\ref{ONB1}) is both unnecessary and tautological. Generally, 
if in (\ref{ONB5}) the weight $w(t)$ approaches $\delta(t\!-\!t_0)$,
the coefficients $h_l$ approach $\mathfrak{p}_l(t_0) h(t_0)$, and on 
account of the completeness relation the series resums to $h(t) =
h(t_0) \delta(t\!-\!t_0)$. In the expansion (\ref{sleepa2}) one
finds from the definitions (\ref{sle6}), (\ref{sle7}) 
that $J_{1,\wp}[\overline{\cV}''_k](t_0) \ra 0$ and 
$\varepsilon_{2,\wp}^2[\overline{\cV}''_k] \ra 0$
in the instantaneous limit. The right hand side (\ref{sleepa2}) 
therefore reduces to (\ref{sleinst0}).

In order to compare with the results of Section 3 we fix conformal 
time and consider the SLE solution $\chi_k^{\rm SLE}(\eta,p)$ associated 
with the (homogeneous) wave equation (\ref{fl1l3}).  The shifted 
potential is $W(\eta) - \frac{d-1}{4d} R(\eta)$, and we write 
$\overline{\cV}''(\vp_0(\eta), \scriptr_k(\eta))$ for  its dimensionless 
counterpart, in parallel to (\ref{flrg7}). In both cases, as in 
\eqref{sleinst0}, we omit the subscript $k$ referring to the dynamical 
$k$ dependence. The instantaneous limit 
of the SLE solution in conformal time and $d=3$ then reads 
\begin{equation} 
\label{sleinst1} 
|\chi^{\rm SLE}_k(\eta_0, k \wp)|^2 \rra 
\frac{1}{2 k a(\eta_0)\sqrt{ \overline{\cV}''(\vp_0(\eta_0),\scriptr_k(\eta_0)) 
+ \wp^2/a(\eta_0)^2 + r(\wp^2/a(\eta_0^2)) }} \,,
\end{equation}
with  $\eta_0 \in (\eta_i,\eta_f)$. Restoring the subscript and abbreviating momentarily $\cV_k(\eta_0):= 
\cV_k(\vp_0(\eta_0),\scriptr_k(\eta_0))$, etc., this results in  
\ba
\label{sleinst2}  
&\nspace & k \dd_k \cV_k(\eta_0) + (1\!+\!d) \cV_k(\eta_0) 
- k \dd_k \ln \kappa(k)^2\; \scriptr_k(\eta_0) 
\frac{\dd \cV_k(\eta_0)}{\dd \scriptr_k(\eta_0)} 
- \frac{d\!-\!1}{2} \vp_0(\eta_0) 
\frac{\dd \cV_k(\eta_0)}{\dd  \vp_0(\eta_0)} 
\nonum
&\nspace & \quad = \frac{\hbar}{c_0(4\pi)^{d/2} \Gamma(d/2)} 
\int_0^{\infty} \! d\varrho \varrho^{d-1} 
\dfrac{r(\varrho^2) - \varrho^2 r'(\varrho^2)}%
{ \sqrt{\overline{\cV}_k''(\eta_0) + \varrho^2 + r(\varrho^2)}}\,.
\ea 
The right hand side coincides with that of the spatial 
EPA flow equation in Minkowski space (\ref{Mink5}) up to the replacement 
of $V_k''(\vp_0) = {}^0V''_k(\vp_0)$ by the generalized potential 
$\overline{\cV}_k''(\eta_0)$ depending also on the Ricci scalar 
$\scriptr_k(\eta_0) = R(\eta_0)/\kappa(k)^2$. Upon expansion of 
(\ref{sleinst2}) in powers of $\scriptr_k(\eta_0)$ one obtains 
(\ref{Mink5}) at lowest order and (with $\hbar=1$)
\ba 
\label{sleinst3}
&& k \dd_k {}^1V_k(\eta_0) + \big(1\!+\!d - k \dd_k \ln \kappa(k)^2
\big)  {}^1V_k(\eta_0) 
- \frac{d\!-\!1}{2} \vp_0(\eta_0) 
\frac{\dd \,{}^1V_k(\eta_0)}{ \dd \vp_0(\eta_0)}
\nonum
&& \quad = - \frac{1}{2} q_1\big( {}^0V_k''(\eta_0) \big) 
{}^1\overline{V}''_k(\eta_0) \,, 
\nonum
&& k \dd_k {}^2V_k(\eta_0) + \big(1\!+\!d - 2 k \dd_k \ln \kappa(k)^2\big) 
{}^2V_k(\eta_0) 
- \frac{d\!-\!1}{2} \vp_0(\eta_0) 
\frac{\dd \,{}^2V_k(\eta_0)}{\dd \vp_0(\eta_0)}
\nonum
&& \quad = - \frac{1}{2} q_1\big( {}^0V_k''(\eta_0) \big) {}^2V''(\eta_0)  
+ \frac{3}{8} q_2\big( {}^0V_k''(\eta_0) \big) 
\big({}^1\overline{V}_k''(\eta_0)\big)^2\,,
\ea 
at the next two orders, with $q_{\nu}(v)$ from (\ref{Mink7}). 
Here ${}^n \overline{V}_k''(\vp_0) = {}^n V_k(\vp_0) 
- \frac{d-1}{4d} \delta_{n,1}$.  

It is instructive to compare the UV flow (\ref{sepa8}) 
with the instantaneous limit flow (\ref{sleinst3}). The flow
equations (\ref{sleinst2}), (\ref{sleinst3}) are not limited 
to large $k$ but an expansion in powers of the potential 
can be compared with the dimensionless form of (\ref{sepa8}). 
Since $\kappa(k) \sim k$ for large $k$ the latter is obtained 
via the transition relations ${}^n U_k(\vp) = c_0 k^{d+1 -2n} 
\;{}^nV_k(\vp_0)$. For the comparison of the right hand sides 
we note $q_{\nu}(v) = q_{\nu}(0) - \frac{1}{2}(2\nu\!+\!1) v q_{\nu +1}(0) 
+ \frac{1}{8} (2\nu\!+\!1)(2\nu\!+\!3) v^2 q_{\nu + 2}(0) + O(v^3)$, 
in the notation (\ref{Mink7}). Using 
this in the $d=3$ version of (\ref{sleinst3}) together with 
$1/[(4\pi)^{d/2} \Gamma(d/2)]\big|_{d=3} = 4/(4\pi)^2$,
$4\pi^2 c_0 q_0(0) = q_0$, $4\pi^2 c_0 q_1(0) 
= q_1$, $4 \pi^2 c_0q_2(0) = 1/3$, one
finds for $n=0,2$ an exact match, while for $n=1$ the $\breve{q}$ 
term in (\ref{sepa8}) is missed. Since the latter arises from 
differentiating the time dependence in the modulator, it 
is unsurprising that this term is not seen in the 
instantaneous limit.

\newpage 
\newappendix{Subleading small $k^2$ terms in SLE based EPA flow} 

Here we justify the substitution recipe (\ref{sleepa0}) and 
use it to compute the explicit form of the $O(k^2)$ corrections
displayed in (\ref{sleepa3}). 

As seen in Appendix B, in the instantaneous limit  
$|T^{\rm SLE}_k(t_0, k \wp)|^2$ is a shifted inverse square root in 
the (generalized) potential ${\cal U}_k''$  and it is plain that 
the substitutions (\ref{dimless2}) or (\ref{irdiml2}) transitioning 
to the dimensionless 
formulation do not affect the dependence on the re-interpreted 
(generalized) potential. In the non-instantaneous case, 
however, the dimensionful formulation (\ref{sepa1}) does not allow 
one to isolate the ${\cal U}''_k$ dependence of the Green's function 
explicitly, not even in a series expansion in powers of $k^2$. This is 
because in an expansion like
(\ref{FLhad15}) below the lowest order Green's function of 
$({n}^{-1} \dd_t)^2 + \om_0(t)^2$ needed to solve the 
recursion cannot be obtained explicitly for generic $\om_0(t)^2 = 
a(t)^{2 d} W(t)$. In contrast, the iterative 
solution of (\ref{dimless11}) for the power series expansion  
of the dimensionless Green's function $G_k[\vp_0](t,t';\wp)$ 
is in itself straightforward, as the only a Green's function 
for the trivial $({n}^{-1} \dd_t)^2$ is needed.  
For the SLE this $\om_0 \equiv 0$ expansion 
can be implemented and leads to the explicit result 
(\ref{sleepa2}). The gap to the dimensionful expansion 
is bridged by the following  

{\bf Lemma C:} {\itshape The SLE Green's function $k G_k[\vp](t,t;p)$ 
in (\ref{sepa1}) with dimensionful generalized potential 
${\cal U}''(\vp,t)$ expanded in powers of $k^2$ has upon 
substitution of $\vp = \vp[\vp_0]$, $p = k \wp$, 
${\cal U}''(\vp,t) = k^2 \cV''(\vp_0,t)$, 
and re-expansion in powers of $k^2$ (with $\cV''(\vp_0,t)$ 
held fixed) the {\it same} coefficients as the directly 
expanded dimensionless $k G_k[\vp_0](t,t;\wp)$ leading to 
(\ref{sleepa2}). }

\underline{Remarks:}
\vspace{-4mm}

\begin{itemize}[leftmargin=8mm, rightmargin=-0mm]
\itemsep -1mm
\item[(i)] A result of this form does {\it not} hold for the large $k$ 
expansion (\ref{FLhad12}). One way of seeing this is by noting that 
for $\om_0 \equiv 0$ the coefficients $\bar{G}_n$ vanish 
for constant $w_2$. Since constant $w_2$ occurs in 
Minkowski space (for constant field) this clashes with the 
direct expansion in the dimensionful formulation, where the 
coefficients (\ref{Mink9}) arise. The origin of the clash is
that upon substitution ${\cal U}''(\vp,R)\mapsto k^2 
\cV''(\vp_0, \scriptr_0)$ the $\bar{G}_n$  are $O(k^{2n})$, 
which invalidates the organizing principle in (\ref{Mink9}).   
\item[(ii)] A simple toy model may illustrate the point of the argument. 
Consider the following three series (where $\dd^2$, $W$, $V$, $r$ 
are suggestive notations for positive numbers) 
\ba
\label{sleepa5} 
&& \frac{1}{\dd^2 + W + k^2 r} = \sum_{n \geq 0} (-)^n 
\frac{r^n k^{2n}}{ (\dd^2 + W)^{n+1}}\,, 
\nonum
&& \frac{1}{(\dd^2 + W)^{n+1} }\Big|_{W \ra k^2 V} = \sum_{j \geq 0} 
{ -n\!-\!1 \choose j} \frac{ V^j k^{2 j}}{ (\dd^2)^{n+1 +j} }\,,
\nonum
&& \frac{1}{\dd^2 + k^2( r + V)}= 
\sum_{n \geq 0} (-)^n \frac{(V+r)^n k^{2n}}{(\dd^2)^{n+1}}\,. 
\ea 
Inserting the first two series into each other (re-)produces 
the third, on account of 
$\sum_{j =0}^n (-)^{n-j} { - n - 1 +j \choose j} 
V^j r^{n-j} = (-)^n (V + r)^n $.
\end{itemize}

{\it Proof of Lemma C.} We apply an abstracted version of 
(\ref{sleepa5}) to the fiducial solutions entering the SLE. 
To this end, we search for a fiducial solution $S_k(t,\wp)$ 
of the homogeneous wave equation (\ref{FLhad2}) for fixed 
$\wp = p/k$ in the form of a power series ansatz 
\be 
\label{FLhad14} 
S_k(t,\wp) = S_0(t) + \sum_{n \geq 1} S_n(t,\wp)\,k^{2n}\,, 
\quad \wp = p/k \;\;\mbox{fixed}\,.
\ee
Inserted into (\ref{FLhad2}) one obtains the recursive system 
\ba  
\label{FLhad15} 
\big[ (a^d {N}^{-1} \dd_t)^2 + a^{2d} W \big] S_{0}(t) \is 0\,,
\\[2mm]
\big[ (a^d {N}^{-1} \dd_t)^2 + a^{2d} W \big] S_{n}(t,\wp) \is
- a^{2d-2} \Big( \wp^2 + a^2 r(\wp^2/a^2) \Big) S_{n-1}(t,\wp)\,, 
\quad n \geq 1\,,
\nonumber
\ea  
with $W = {\cal U}''$ is as in (\ref{Diffop}). The coefficients 
$S_n$ are in addition slightly constrained by the expansion of 
the Wronskian condition. 
Since an SLE does not depend on the choice of fiducial solution 
we may begin with a fiducial solution of the dimensionful
wave equation whose initial data are $k,\wp$ independent.
The recursion relations (\ref{FLhad15}) can then be solved 
unambiguously in terms of the (universal) retarded Green's 
functions for $({n}^{-1} \dd_t)^2 + a^{2d} {\cal U}''$, 
$n = N a^{-d}$ \cite{BonusSLE}.   
The resulting series has finite radius of convergence 
and thus yields an exact solutions for $k < k_*$. Since its initial 
data are $k$ independent one can appeal to a standard result 
on ordinary differential equations (see e.g.~\cite{ODEbook}, 
Chapter 5.2) to conclude that the dependence of the exact solution 
on ${\cal U}''$'s overall scale is smooth. The same holds for the  
Green's function of $({n}^{-1} \dd_t)^2 + a^{2d} {\cal U}''$. 
Replacing now ${\cal U}''$ by $k^2 \cV''$ the basic Green's 
function needed for the iteration as well as the exact solution 
can be re-expanded in powers of $k^2$.     
The re-expanded series is a solution of the $\om_0^2=a^{2d}W\equiv 0$ 
recursion relations uniquely characterized by the initial data 
induced by re-expanding the exact solution, and the trivial 
$\om_0^2\equiv 0$ retarded Green's function. It also has finite 
radius of convergence $k<k_{\ast,0}$, see \cite{BonusSLE}.
Within the (nonempty) intersection of the intervals of convergence the 
two-step expanded solution and the solution of the $\om_0^2 
\equiv 0$ recursion will coincide. This justifies the substitution 
recipe (\ref{sleepa0}) for the fiducial solutions. 

Either of these expansions can now be inserted into the SLE 
solution $T_k^{\rm SLE}[S_k](t,p)$, viewed as a functional of $S_k$. 
The functionals also carry an explicit $k,\wp$ dependence 
entering via the $\cE, \cD$ from (\ref{gsle3}). Since 
analyticity in $k^2$ is maintained in the substitution process the
explicit $k$ dependence allows for a power series expansion 
in $k^2$. The same holds for the $k^2$ dependence carried by the 
two distinct series realizations for the fiducial solution.  
The respective expansions are of the form (\ref{sle3}) 
(with $k$ playing the role $p$) but the $\varepsilon_0=0$  
version must by the previous step be the re-expansion of the 
$\varepsilon_0>0$ version. This justifies the substitution 
recipe (\ref{sleepa0}) for  $|T_k^{\rm SLE}[S_k](t,p)|^2$, 
constructed from the above specific fiducial solution $S_k(t,p)$. 
Since  $|T_k^{\rm SLE}[S_k](t,p)|^2$ does not depend on the choice 
of $S_k$ the result follows. \qed 
\medskip 

Application of Lemma C yields (\ref{sleepa2}) so that the 
small $k$ regime of the SLE based FL-sFRG flow equation 
(\ref{dimless12}) is governed by the coefficients $J_{1,\wp}[\cV''](t)$ 
and $\varepsilon_{2,\wp}^2[\cV'']$, where $\cV''(t)$ is short for 
$\cV''(\vp_0,t)$. Using the averages (\ref{sleepa4})  
the right hand side of (\ref{dimless12}) becomes
\ba
\label{C1} 
\mbox{RHS of} \; (\ref{sleepa3}) = 
{}_lQ_0\big(1|\,{}_0\!\cV_k''\big) 
+ \frac{k^2}{J_0} \,{}_lQ_0\big(J_{1,\cdot}|\,{}_0\!\cV_k''\big) 
- \frac{k^2}{2 J_0^2} \,{}_lQ_1\big(\,
\varepsilon_{2,\cdot}^2[\cV''_k] |\,{}_0\!\cV_k''\big) \,.
\ea 
In view of the linearity of the ${}_lQ_m(h|V)$ in the first argument we  
prepare expansions for $J_{1,\wp}(t)$ and $\varepsilon_{2,\wp}^2$. To this end 
the expansion of the commutator function $\Delta_{1,\wp}$ and its derivatives 
is needed. In parallel to (\ref{sleepa1}) we write 
\ba 
\label{Delta1expansion} 
{n}(t)^{-1} \dd_t \Delta_{1,\wp}[\cV''](t',t)  \is 
\int_{t'}^t \! ds \,{n}(s) a(s)^{2d} \big[ \cV''(s) + 
\rf(\wp/a(s)) \big] \Delta_0(s,t') \,, 
\nonum
\Big[{n}(t)^{-1} {n}(t')^{-1} \dd_{t'}\dd_t 
\Delta_{1,\wp}[\cV''](t',t) \Big]^2
\is \int_{t'}^t \! ds ds' {n}(s) {n}(s') a(s)^{2d} a(s')^{2d} 
\\[2mm]
&\times& \big[ \cV''(s) + \rf(\wp/a(s))\big] 
\big[ \cV''(s') + \rf(\wp/a(s'))\big]\,, 
\nonumber
\ea 
with the abbreviation $\rf(\wp/a(t)) := \wp^2/a(t)^2 + r(\wp^2/a(t)^2)$. 
Using (\ref{sle6}), (\ref{sle8}), (\ref{sleepa0}), these enter
\ba
\label{Jexpansion} 
J_{1,\wp}[\cV''](t) \is \frac{1}{2} \int\! dt' {n}(t') f(t') 
\Big\{ 2 {n}(t')^{-1} \dd_{t'} \Delta_{1,\wp}[\cV''](t',t)
\nonum
&+& 
\Delta_0(t,t')^2 a(t')^{2d} \big[\cV''(t') + 
\rf\big(\wp/a(t')\big) \big]\Big\}\,,    
\ea 
and 
\ba 
\label{eps2expansion} 
&\nspace & \varepsilon_{2,\wp}^2[\cV''] = \frac{1}{8} \int dt d t'\,
{n}(t) {n}(t') f(t) f(t') \Big\{ 
\big({n}(t)^{-1}{n}(t')^{-1} \dd_t \dd_{t'} 
\Delta_{1,\wp}[\cV''](t,t') \big)^2 
\nonum
&\nspace & \quad - 4 a(t')^{2d} 
\big[\cV''(t') + \rf\big(\wp/a(t')\big) \big]
{n}(t)^{-1} \dd_t \Delta_{1,\wp}[\cV''](t',t) 
\\[2mm] 
&\nspace & \quad +a(t)^{2d} a(t')^{2d} \Delta_0(t,t')^2  
\big[\cV''(t) + \rf\big(\wp/a(t)\big) \big]
\big[\cV''(t') + \rf\big(\wp/a(t')\big) \big]
\Big\} \,.
\nonumber
\ea 
Next, we expand $\cV''(t) = \cV''(\vp_0, t) = \sum_{l \geq 0} 
{}_l \cV''(\vp_0) \pf_l(t)$ and insert into (\ref{Jexpansion}). 
This gives 
\ba 
\label{Jlexpansion} 
&\nspace & J_{1,\wp}[\cV''](t)\!= \! 
\sum_{l \geq 0} {}_l \cV''(\vp) \jmath_l(t) + \jmath(t,\wp) \,,
\nonum 
&\nspace & \quad \jmath_l(t) \!=\! \int\! dt' n(t') f(t') 
\Big\{ \int_{t'}^t \! ds n(s) a(s)^{2d} \,
\pf_l(s) 
\Delta_0(s,t') + \frac{1}{2} a(t')^{2d} \Delta_0(t,t')^2 \,
\pf_l(t') \Big\}\,,
\\[2mm] 
&\nspace & \jmath(t,\wp) \!=\!\int\!\! dt' n(t') f(t') \Big\{\! \int_{t'}^t 
\!\! ds n(s) a(s)^{2d} \,\rf\big(\wp/a(s)\big)  \Delta_0(s,t') 
+ \frac{1}{2} a(t')^{2d} \Delta_0(t,t')^2 \,\rf\big(\wp/a(t')\big) 
\Big\}. 
\nonumber
\ea 
Finally, 
\ba
\cE_{2,\wp}^2[\cV''] \!\!\is\! \!
\sum_{l,l'\geq 0} {}_l\cV''(\vp_0) \;{}_{l'} \cV''(\vp_0) E_{l,l'} + 
\sum_{l\geq 0} {}_l\cV''(\vp_0) E_l(\wp) + E(\wp)\,,
\\[2mm]
\!E_{l,l'} \!\!\is\!\! \frac{1}{8} \int\! dtdt' n(t)f(t) n(t') f(t') 
\Big\{ \int_{t'}^t \! ds n(s) a(s)^{2d} \,\pf_l(s) 
\int_{t'}^t \! ds' n(s') a(s')^{2d} \,\pf_{l'}(s') 
\nonum
\!&\!-\!&\! 4 a(t')^{2d} \pf_{l'}(t') 
\int_{t'}^t \! ds a(s)^{2d} \,\pf_l(s) \Delta_0(s,t') 
+ a(t)^{2d} a(t')^{2d} \Delta_0(t,t')^2 \,\pf_l(t) \pf_{l'}(t') \Big\}\,,
\nonum
\!E_l(\wp) \!\is\! \frac{1}{4}  \int\! dtdt' n(t) f(t)n(t') f(t') 
\Big\{ \int_{t'}^t \! ds n(s) a(s)^{2d} \,\pf_l(s) 
\!\int_{t'}^t \! ds' n(s') a(s')^{2d} \,\rf\big(\wp/a(s')\big) 
\nonum
\!&\!-\!&\! 2 a(t')^{2d} \,\pf_l(t') \int_{t'}^t \! ds n(s) a(s)^{2d} 
\,\rf\big(\wp/a(s)\big) \Delta_0(s,t')  
\nonum
\!&\!-\!&\! 2 a(t')^{2d} \,\pf_l(t') \rf\big(\wp/a(t')\big) 
\int_{t'}^t \! ds n(s) a(s)^{2d} \,\pf_l(s) \Delta_0(s,t')  
\nonum
\!&\!+\!&\! a(t)^{2d} a(t')^{2d} \Delta_0(t,t')^2 \,\pf_l(t) \,
\rf\big(\wp/a(t')\big) \Big\}\,, 
\nonum
\!E(\wp) \!\is\!  \frac{1}{8} \int\! dtdt' n(t)f(t) n(t')f(t') \Big\{ 
\Big[ \int_{t'}^t \! ds n(s) a(s)^{2d} \rf\big(\wp/a(s)\big) \Big]^2 
\nonum
\!&\!-\!&\! 4 a(t')^{2d} \,\rf(\wp/a(t')\big) 
\int_{t'}^t \! ds n(s) a(s)^{2d} \rf\big(\wp/a(s)\big)  
\Delta_0(s,t') 
\nonum
\!&\!+\!&\!  a(t)^{2d} a(t')^{2d} \Delta_0(t,t')^2\, 
\rf\big(\wp/a(t)\big)\rf\big(\wp/a(t')\big)\Big\}\,.
\ea  
These expansions are now inserted into (\ref{C1}) and 
yield the explicit form of (\ref{sleepa3})'s right hand side.

\newpage


\begin{thebibliography}{99} 

\bibitem{FRGreview} N.~Dupuis, L.~Canet, A.~Eichhorn, W.~Mertzner,
J.~Pawlowski, M.~Tissier, and N.~Wschebor, The nonperturbative 
functional renormalization group and its applications, Phys.\ Rept.\
{\bf 910} (2021) 1. 


\bibitem{Percbook} R.~Percacci, {\it An Introduction to covariant
Quantum Gravity and Asymptotic Safety}, World Scientific, 2017.

\bibitem{RSbook} M.~Reuter and F.~Saueressig, {\it Quantum Gravity and 
the  Functional Renormalization Group}, Cambridge UP, 2018. 

\bibitem{Kopbook} P.~Kopietz, L.~Bartosch, and F.~Sch\"{u}tz,
Introduction to the Functional Renormalization Group, 
Springer, 2010.  

\bibitem{Kaya} A.~Kaya, Exact RG flow in an expanding universe and 
screening of the cosmological constant, Phys.\ Rev.\ {\bf D87} (2013) 
123501. 

\bibitem{Serreau1} J.~Serreau, Renormalization group flow 
and symmetry restauration in de Sitter space, Phys.\ Lett.\ {\bf B730} 
(2014) 271. 

\bibitem{Serreau2} M.~Guilleux and J.~Serreau, Quantum fields 
in de Sitter space from the nonperturbative renormalization group,
Phys.\ Rev.\ {\bf D92} (2015) 084010. 

\bibitem{FRGWick1} S.~Floerchinger, Analytic continuation of 
functional renormalization group equations, JHEP {\bf 05} (2012) 021. 

\bibitem{FRGWick2} J.~Pawlowski and N.~Strodthoff, Real 
time correlation functions and the functional renormalization group, 
Phys.\ Rev.\ {\bf D92} (2015) 094009. 

\bibitem{Graphpaper} R.~Banerjee and M.~Niedermaier,
Graph rules for the linked cluster expansion of the
Legendre effective action, J.\ Math.\ Phys. {\bf 60} (2018)
013504.

\bibitem{Mainzproc} M.~Niedermaier, Anti-Newtonian expansions and the 
Functional Renormalization Group, Universe {\bf 5} (2019) 5. 

\bibitem{FPPT} M.~Niedermaier, Gravitational fixed points from 
perturbation theory, Phys.\ Rev.\ Lett.\ {\bf 101} (2008) 131301. 

\bibitem{Wicked} A.~Baldazzi, R.~Percacci, and V.~Skrinjar, 
Wicked metrics, Class.\ Quant.\ Grav.\ {\bf 36} (2019) 105008.


\bibitem{Moretti} I.~Khavkine and V.~Moretti, Algebraic QFT in 
curves spacetimes and quasifree Hadamard states: an introduction,
in {\it Advances in Algebraic QFT}, R.~Brunetti et al (eds), 
Springer, 2015.

\bibitem{Hackbook} T.~Hack, {\it Cosmological applications of 
algebraic Quantum Field Theory in curved spacetimes}, 
Springer, 2016.

\bibitem{wavebook} C.~B\"{a}r, N.~Ginoux, and F.~Pf\"{a}ffle, 
Wave equations on Lorentzian manifolds, European Mathematical
Society, 2007.  

\bibitem{Olbermann} H.~Olbermann, States of low energy on 
Robertson-Walker spacetimes, Class.\ Quant.\ Grav.\ {\bf 24} 
(2007) 5011. 

\bibitem{Luders} C.~L\"{u}ders and J.~Roberts, 
Local quasi-equivalence and adiabatic vacuum states,
Commun.\ Math,\ Phys.\ {\bf 134} (1990) 29.

\bibitem{BonusSLE} R.~Banerjee and M.~Niedermaier,
Bonus properties of States of Low Energy, J.\ Math.\ Phys. {\bf 61} 
(2020)  103511. 

\bibitem{SLE2} K.~Them and M.~Brum, States of low energy 
in homogeneous and inhomogeneous expanding spacetimes, 
Class.\ Quant.\ Grav. {\bf 30} (2013) 23035. 

\bibitem{SJstates} N.~Ashordi, S.~Aslanbeigi, and R.~Sorkin,
A distingished vacuum state for a quantum field in a curved spacetimes:
formalism, features, and cosmology, 
JHEP {\bf 2012} (2012)~1.

\bibitem{SJHadamard} M.~Brum and K.~Fredenhagen, 
Vacuum-like Hadamard states for quantum fields on curved spacetimes, 
Class.\ Quant.\ Grav.\ {\bf 31} (2014) 025024.  

\bibitem{GlobalHyp} F.~Finster, A.~Much, and K.~Papadopoulos, 
On global hyperbolicity of spacetimes: Topology meets 
Functional Analysis, in: {\it Mathematical Analysis in 
Interdisciplinary Research}, ed.~Parasidi et al., Springer, 2022;
arXiv: 2107.07156. 

\bibitem{ODEbook} T.~Sideris, {\it Ordinary Differential Equations 
and Dynamical Systems}, Springer, 2d edition, 2014. 


\bibitem{Szegobook} G.~Szeg\"{o}, {\it Orthogonal Polynomials}, 
4th edition, AMS 1975.   


\bibitem{Moretti1} V.~Moretti, Proof of the symmetry of the
off-diagonal Hadamard/Seeley-DeWitt's coefficients in 
$C^{\infty}$ Lorentzian manifolds by a `local Wick rotation', 
Commun.\ Math.\ Phys.\ {\bf 212} (2000) 165.  


\bibitem{heatkoffdiag1} Y.~Decanini and A.~Folacci, Off-diagonal 
coefficients of the DeWitt-Schwinger and Hadamard representations 
of the Feynman propagator, Phys.\ Rev.\ {\bf D73} (2006) 044027. 

\bibitem{FRWHamdiag0} S.~Fulling, Remarks on positive energy and 
Hamiltonians in expanding Universes, 
Gen.\ Rel.\ Grav.\ {\bf 10} (1979) 807.  



\bibitem{FRWkindom1} W.~Handley, S.~Brechet, A.~Lasenby, and M.~Hobson,
Kinetic initial conditions for inflation, Phys.\ Rev.\ {\bf D89} (2014) 
063505.  

\bibitem{FRWkindom2} L.~Hergt, W.~Handley, M.~Hobsen, and 
A.~Lasenby, A case for kinetically dominated initial conditions 
for inflation, Phys.\ Rev.\ {\bf D100} (2019) 023502. 


\bibitem{CosmFRGtrad0} C.~Wetterich, Cosmic fluctuations
from the quantum effective action, Phys.\ Rev.\ {\bf D92} 
(2015) 083507.

\bibitem{CosmFRGtrad1} A.~Platania and F.~Saueressig, Functional 
renormalization group flows on Friedmann-Lema\^{i}tre-Robertson-Walker 
backgrounds, Found.\ Phys.\ {\bf 48} (2018) 1291. 


%
\bibitem{ParkerTomsbook} L.~Parker and D.~Toms, {\it Quantum Field Theory 
in curved spacetime}, Cambridge UP, 2009.
%
\bibitem{Phi4onFRWPT1} J.~Baacke, L.~Covi, N.~Kevlishvili, 
Coupled scalar fields on a flat FRW universe: renormalization, 
JCAP {\bf 1008} (2010) 026.  
%
\bibitem{Phi4onFRWPT2} A.~Maroto and F.~Prada, Higgs effective potential 
in perturbed Robertson-Walker background, Phys.\ Rev.\ {\bf D90} 
(2014) 123541.
%
\bibitem{Phi4onFRWPT3} M.~Parvizi, General form of the renormalized,
perturbed energy density via interacting quantum fields in 
cosmological spacetimes, arXiv: 1809.03588.
%
\bibitem{Codelloetal} A.~Codello, R.~Percacci, L.~Rachwal, 
and A.~Tonero, Computing the effective action with the Functional 
Renormalization Group, Eur.\ Phys.\ J.\ {\bf C76} (2016) 4.  
%
\bibitem{FRGfoliated} J.~Biemans, A.~Platania, and F.~Saueressig, 
Quantum gravity on foliated spacetime -- asymptotically safe and sound,
Phys.\ Rev. {\bf D95} (2017) 086013. 
%
\bibitem{CurvedLPA1} I.~Shapiro, P.~Teixeira, 
and A.~Wipf, On the renormalization group for the scalar field on 
curved background with non-minimal interaction,  Eur.\ Phys.\ J.\ 
{\bf C75} (2015) 262.
%
\bibitem{CurvedLPA2} B.~Merzlikin, I.~Shapiro, A.~Wipf, 
and O.~Zanusso, Renormalization group flows and fixed points 
for a scalar field in curved space with nonminimal $F(\phi)R$ coupling,
Phys.\ Rev.\ {\bf D96} (2017) 125007. 

\end{thebibliography}
\end{document}